\documentclass[twocolumn]{aastex631}
\usepackage{newtxtext,newtxmath}
\usepackage[T1]{fontenc}
\usepackage{ae,aecompl}
\usepackage[toc,page]{appendix}
\usepackage{enumitem}
\usepackage{amsmath,bm}
\usepackage{graphicx}	
\usepackage{amssymb}	
\usepackage{tabularx}
\usepackage{subfigure}
\usepackage{float}
\usepackage{array}   
\usepackage{ulem}  
\usepackage{longtable}

\newcommand{\phiref}{\phi_{\rm ref}}
\newcommand{\azily}{azimuthally}

\newcommand{\axi}{axisymmetric}
\newcommand{\epi}{epicyclic}
\newcommand{\snd}{solar neighborhood}
\newcommand*{\firstitacr}[4][]{{{#4#1} (#3)}\expandafter\gdef\csname#2\endcsname{#3}}
\newcommand*{\mw}[1]{\firstitacr[#1]{mw}{MW}{Milky Way}}
\newcommand*{\arl}[1]{\firstitacr[#1]{arl}{ARL}{axisymmetric resonance line}}
\newcommand*{\CR}[1]{\firstitacr[#1]{CR}{CR}{corotation}}
\newcommand*{\ilr}[1]{\firstitacr[#1]{ilr}{ILR}{inner Lindblad resonance}}
\newcommand*{\olr}[1]{\firstitacr[#1]{olr}{OLR}{outer Lindblad resonance}}

\newcommand{\SpFid}{\texttt{SpFid}}
\newcommand{\Spaten}{\texttt{Sp$\alpha$10}}
\newcommand{\Spathirty}{\texttt{Sp$\alpha$30}}
\newcommand{\SpCRsix}{\texttt{SpCR6}}
\newcommand{\SpCRten}{\texttt{SpCR10}}
\newcommand{\SpAhalf}{\texttt{SpA0.5}}
\newcommand{\SpAdouble}{\texttt{SpA2}}
\newcommand{\SpTone}{\texttt{SpT1}}
\newcommand{\SpTfour}{\texttt{SpT4}}
\newcommand{\SpTeight}{\texttt{SpT8}}
\newcommand{\SpWind}{\texttt{SpWind}}

\renewcommand{\sout}[1]{\unskip}

\newcommand{\kdc}[1]{\unskip}
\newcommand{\refedit}[1]{{\color{black} {#1}}}

\shorttitle{Age Trends in Action Space from Transient Spirals}
\shortauthors{Smock et al.}

\begin{document}

\title{Wrinkles in Time. II. Stellar Age Trends in Kinematic Signatures from Transient Spiral Structure}

\author[0000-0002-3041-7822]{Amy Smock}
\affiliation{University of Arizona Department of Astronomy and Steward Observatory \\ 933 N Cherry Ave. \\ Tucson, AZ 85719, USA}
\thanks{Corresponding author: \href{mailto:amysmock@arizona.edu}{amysmock@arizona.edu}}

\author[0000-0003-2594-8052]{Kathryne J. Daniel}
\affiliation{University of Arizona Department of Astronomy and Steward Observatory \\ 933 N Cherry Ave. \\ Tucson, AZ 85719, USA}

\author[0000-0001-7337-5936]{Rayna Rampalli}
\affiliation{Dartmouth College Department of Physics and Astronomy \\ 6127 Wilder Laboratory \\ Hanover, NH 03755, USA}

\author[0000-0003-4150-841X]{Elisabeth R. Newton}
\affiliation{Dartmouth College Department of Physics and Astronomy \\ 6127 Wilder Laboratory \\ Hanover, NH 03755, USA}

\author[0009-0006-1832-6713]{Olivia McAuley}
\affiliation{Bryn Mawr College Department of Physics \\ 101 North Merion Ave. \\ Bryn Mawr, PA 19010, USA}

\author[0000-0002-6036-1858]{S\'{o}ley Hyman}
\affiliation{University of Arizona Department of Astronomy and Steward Observatory \\ 933 N Cherry Ave. \\ Tucson, AZ 85719, USA}

\author[0009-0000-9825-9755]{Lipika Chatur}
\affiliation{University of Arizona Department of Astronomy and Steward Observatory \\ 933 N Cherry Ave. \\ Tucson, AZ 85719, USA}


\begin{abstract}

Spiral arms in the disks of galaxies like the Milky Way can generate kinematic signatures, which appear as ridges or wrinkles in action space. Such signatures have proven difficult to disentangle using kinematic measures alone. In this study, we investigate how including stellar age as an additional dimension for analysis may provide a novel insight into the physical characteristics, timescales, and nature of the progenitors of such perturbations, where these novel insights could contribute to our understanding of the history of spiral arms in the Milky Way. We used a suite of tracer particle simulations that modeled a variety of prescriptions for spiral arms to characterize observable trends.  The Lindblad resonances of nonwinding spirals produce signature overdensities, or wrinkles, in a kinematic space that is typically associated with older stellar populations (high radial action). We find that these  wrinkles are preferentially populated with stars that were initially in nearly circular orbits, kinematics that is generally correlated with younger stellar ages. It follows that the stellar age distribution of wrinkle populations could serve to place constraints on the past passage of a transient spiral pattern in the solar neighborhood. For example, our simulations suggest that a physically motivated spiral pattern could significantly populate a wrinkle with zero-age stars in orbits typically occupied by stars much older than the Sun.

\end{abstract}




\section{Introduction}\label{sec:intro}

Striking substructures have been uncovered in the \mw{} stellar disk thanks to a wealth of kinematic survey data from the Gaia mission \citep{gaiamission,GaiaDR2,GaiaDR3}.  These include the \refedit{prediction \citep{Dehnen98} and} discovery of kinematic ridges, arches, clustering, and a vertical "snail shell" feature \citep[e.g.,][]{Antoja18,BinneySchoenrich18,Kawata18,Koppelman18,Monari18,Quillen18,Ramos18,SchoenrichDehnen18,Bland-Hawthorn19,Trick19,Alves20}.
Despite the plethora of studies on these kinematic structures, the timeline for their genesis, particularly in the plane of the MW's disk, is poorly constrained.

Substructures within the \mw{} disk have many proposed progenitors. Internal drivers include spiral arms \citep{DeSimone04,QuillenMinchev05,Sellwood12,FouvryPichon15b,Hunt18b,Quillen18,Khanna19,Trick21}, the central bar \citep{Raboud98,Dehnen2000,Fux01,MuhlbauerDehnen03,Chakrabarty2007,Monari17,HuntBovy18,Ramos18,Fragkoudi19,Khoperskov19}, or the combined effects of both spirals and the bar \citep{Quillen03,MinchevFamaey10,Grand15,Kawata18,Hunt19,Martinez-Medina19,Cao24}. Interactions with external satellites can also affect disk dynamics and form phase-space substructures \citep{Helmi99,Minchev09,Gomez12}, particularly for motions in the vertical direction. Explanations for observed vertical phase-space substructure in the \mw{} have been attributed to an external satellite \citep{Antoja18,Bland-Hawthorn19,Khanna19,Laporte19,Banik23}, spirals \citep{Khanna19}, the bar \citep{Khoperskov19,Li23}, a dark matter wake \citep{Grand23}, disk noise \citep{TremaineFrankelBovy23}, \refedit{warps induced by gas \citep{Khachaturyants22,Wang26}}, or various combinations of the above \citep{Garcia-Conde24}.

\begin{figure*}
\centering
    \begin{tabular}{@{}cccc@{}}
        \includegraphics{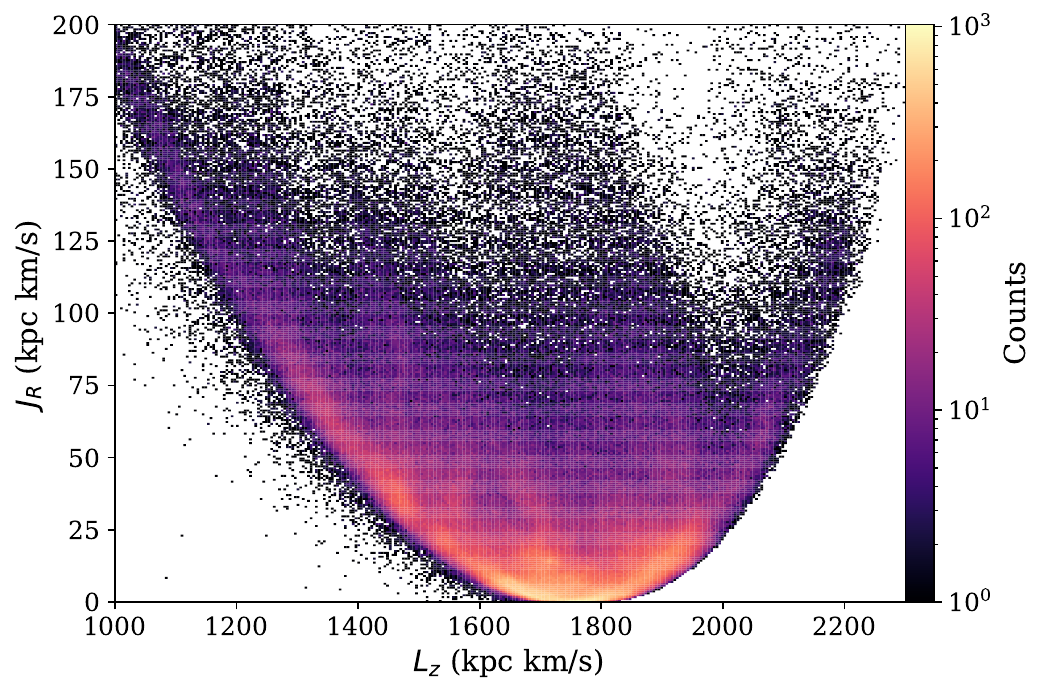} 
    \end{tabular}
    \caption{Distribution in action coordinates ($J_R$,$L_z$) of ${\sim}7\times10^5$ stars that are within 200~pc of the Sun.  This catalog combines 6D kinematics from Gaia Data Release 3 (DR3), GALAH DR3, and APOGEE DR16, as seen in \cite{Rampalli23}.  Wrinkles are the diagonal, extended overdensities in this space and are especially pronounced at high $J_R$.}
    \label{fig:gaia_clustering}
\end{figure*}

In this paper, we take particular interest in the complex, rich clustering in the action space distribution of \snd{} stars.  A more thorough discussion of action space is given in \S\ref{s:actions} of this work \citep[see also][their Section~3.5]{BT08}.
Specifically, we focus on a set of distinct overdensities that appear in high radial action ($J_{\rm R}$) space, a space that corresponds to disklike orbits that have high eccentricity.  These overdensities are termed "wrinkles" in this series of papers \citep[e.g.,][]{Rampalli23} and are likely postresonant signatures from transient spiral arms \citep{Sellwood10b,SellwoodTrick19,Trick19}. 
This sort of clustering was first identified using Gaia Data Release 2 survey data \citep{GaiaDR2}, described by \cite{Trick19}, and can be observed in Figure~\ref{fig:gaia_clustering}.

Transient spiral patterns are critical drivers of disk evolution \citep[see][and references therein]{SellwoodMasters22}. However, the history of spiral structure in the \mw{}, or even in the \snd{} has thus far been a rather intractable question. Attempts to match local kinematic signatures to structures, such as wrinkles caused by spiral arms or the bar, have proven challenging since these signatures have degeneracies, particularly in the case where more than one structure is creating them \citep[see, e.g.,][]{Quillen11,Hunt18b,Hunt19,Fujii19}.

Consequently, disentangling and classifying progenitors of substructure within the action space of the \mw{} disk in the \snd{} has proven challenging \citep[see, e.g.,][]{Hunt19,SellwoodTrick19,Trick19}. \cite{Sellwood10b} proposed that the ages of stars present in resonant signatures in low action space could be used to place a date on the origin of the signature.  However, best estimates for stellar ages in \snd{} stars at the time, namely in the Geneva Copenhagen Survey \citep{Nordstrom04}, were not sufficiently accurate.  

In this paper, we examine postresonant signatures left by a range of isolated transient spiral perturbations in a suite of tracer particle simulations.  
Our approach is to use the kinematics of unperturbed orbits in a \mw-like disk as a proxy for stellar age.  
Specifically, we quantify kinematic temperature using the radial action, $J_{R}$ (see \S\ref{s:actions}~and~\S\ref{s:kinematic age}), and consider the initial values of the radial action in the unperturbed disk, $J_{R_0}$,  as a proxy for age.  In this scheme, kinematically cold (hot) orbits have low (high) values for $J_{R_0}$ and are assumed to be associated on average with younger (older) stars.
The redistribution of these orbits to form wrinkles is used to characterize trends in the age distribution of wrinkle stars.  In turn, this provides a window into the history of spiral structure affecting the \snd{}. We pay particular attention to trends that could provide additional information about possible wrinkle progenitors.

This paper is organized as follows.  We give a background overview of the underlying theory for wrinkle formation in Section \S\ref{sec:background}.  In Section \S\ref{sec:methods}, we describe our suite of tracer particle simulations and the variety of prescriptions for transient spiral arms we superpose on a smooth \mw{} disk potential. In Section \S\ref{sec:analysis}, we describe trends in the orbital response as a function of spiral characteristics and give an analytic explanation for these trends. Following our analysis, we discuss the implications for these trends for future studies of the wrinkles observed in the \snd{} in Section \S\ref{sec:discussion}.

\section{Background} \label{sec:background}
\subsection{Overview of action-space}\label{s:actions}

Under the assumption of a smooth potential, a given star's 6D phase space can be projected into 6D action-angle space, a natural space for describing orbits. Actions provide a measure of orbital shapes and sizes and, when working with disk populations, are best expressed using a cylindrical symmetry with Galactocentric coordinates $(R, \phi, z)$. The resulting actions ($J_{R}$, $J_{\phi}$, $J_{z}$) are integrals of motion in an \axi\ potential.  The associated oscillation frequencies are $\omega_{R}$, $\omega_{\phi}$, and $\omega_{z}$, where the angles $\Theta_R$, $\Theta_\phi$, and $\Theta_z$ give phase information along their orbital trajectories.  

In an \axi\ disk, the radial action $J_{R}$ of an orbit is related to radial motions and is correlated with orbital eccentricity.  For reference,
$J_{R}=E_{\rm nc}/\kappa$ in the \epi{} approximation \citep{BT08}, where $\kappa=\omega_R$ is the \epi{} frequency and $E_{\rm nc}=E-E_{\rm c}(L_z)$ is the energy associated with noncircular \refedit{planar} motions for a star with angular momentum in the $z$-direction $L_z$. This component of the angular momentum is the same as its azimuthal action, $L_z=J_{\phi}$. Hereafter we will use $L_z$. In a disk with a flat rotation curve, $v_c={\rm const}$ and so the $L_{z}$ of an orbit is proportional to its orbital radius. Vertical action, $J_{z}$, describes the vertical orbital motion. 

Since galaxies contain nonaxisymmetric structures, such as bars and spiral arms, the actions are not conserved.  However, they can still be used as effective tools to estimate how nonaxisymmetric perturbations affect orbital dynamics provided a well-fit and slowly varying background model \citep{Binney12,BovyRix13,Trick17,Trick19,Trick21}. Several best-fitting \axi{} disk models have been developed for the \mw{} \citep{McMillan11a,Bovy15,Eilers19}, providing a reasonable basis by which instantaneous actions can be calculated. 
In order to avoid any complications with the estimation of action values \cite{Debattista25}, we \textit{only} calculate actions in an unperturbed potential, either before or after the passage of a transient spiral structure. 

\refedit{The Galactocentric actions shown in Figure~\ref{fig:gaia_clustering} for each star in our catalog were computed using the Python-based Galactic dynamics modeling package \texttt{galpy v.1.7} \citep{Bovy15}.
We assumed the default underlying potential for the \mw{}, \texttt{MWPotential14}, and invoked
the Stäckel–Fudge approximation \citep{Binney12}.  These choices provide a consistent comparison to previous studies of this kind \citep[e.g.,][]{Trick19}.} 


\subsection{postresonant signatures}\label{s:PRsignatures}

Figure~\ref{fig:gaia_clustering} shows the distribution of the radial action ($J_R$) for a given angular momentum ($L_z$) for stars in the \snd{}.  The method for calculating these action space coordinates is described \refedit{\S\ref{s:actions}} and the 6D kinematics are taken from a catalog of 731,961 stars within 200~pc of the Sun.  The catalog combines data from Gaia DR3 \citep{GaiaDR3,GaiaEDR3,GaiaDR3RVs}, GALAH DR3 \citep{deSilva15GALAH}, and APOGEE DR16 \citep{Majewski17APOGEE} and is restricted to stars with measured \refedit{radial velocity} and parallax errors of less than 10\%. 
Further information on this catalog can be found in \cite{Rampalli23}, which is Paper I of this series.
Wrinkles appear in Figure~\ref{fig:gaia_clustering} as diagonal, extended overdensities that are particularly prevalent at higher values of $J_R$ and are likely due to a dynamical response to the Lindblad resonances of transient spiral arms \citep{Sellwood10b,SellwoodTrick19,Trick19}. 

Dynamical resonances occur when the natural frequency of an orbit is equal to or otherwise commensurate with a forcing frequency from a perturbation, such as a spiral or a bar. 
There are three primary resonances of interest in this work.  The \CR{} resonance is where the orbital frequency $\Omega_{0}$ is equal to the pattern speed of a spiral $\Omega_{\rm s}$. There are also two primary Lindblad resonances, which can be defined as occurring when
\begin{equation}\label{eq:lindblad}
    m(\omega_\phi - \Omega_{\rm s}) = \pm \omega_R,
\end{equation}
is satisfied, where $m$ is the integer number of spiral arms, and $\omega_\phi=\Omega_0$ and $\omega_R$ are the azimuthal and radial frequencies, respectively \citep{BT08}. The \ilr{} occurs when the frequency at which a star passes a spiral arm is equal to the radial frequency ($\Omega_R=\kappa$ in the \epi{} approximation) of that orbit, corresponding to the positive form of the expression in Equation \ref{eq:lindblad}. Conversely, the \olr{} is where a spiral passes orbits at the same frequency as the radial oscillation, corresponding to the negative form of Equation \ref{eq:lindblad}.

The orbital response near a dynamical resonance is nonlinear and strong, even if the perturbing field is weak \citep{LBK72}.  Resonant orbits can exchange angular momentum with the perturber causing changes \refedit{to} their orbital sizes. The Lindblad resonances, in particular, can induce irreversible changes in the noncircular motions of an affected population, increasing the average noncircular (or ``random") energy of the orbits \citep{LBK72, Mark74, SellwoodMasters22}.

Orbits that encounter a dynamical resonance with a spiral pattern can have rapid changes in their orbital energy $E$ and angular momentum $L_z$.
During these exchanges, $E$ and $L_z$ are not conserved.  For a spiral with a constant pattern speed $\Omega_{\rm s}$ the Jacobi integral \citep{BT08}, 
\begin{equation}\label{eq:JacobiIntegral}
I_{\rm J} = E - \Omega_{\rm s}L_{z},
\end{equation}
is conserved.
It follows that changes in angular momentum $\Delta L_{z}$ and orbital energy $\Delta E$ are related by \citep{BT87,SB02}
\begin{equation}\label{eq:conservedJacobi}
\Delta E = \Omega_{\rm s}\Delta L_{z}. 
\end{equation}

In the \epi{} approximation, the radial frequency $\omega_{r}$ of an orbit is equivalent to its \epi{} frequency $\kappa$.  It can be shown that the energy associated with radial motions $E_R$ is related to radial action by $E_R = J_{R}/\kappa$ and that a change in the $L_{z}$ of a star at the Lindblad resonances corresponds to change in radial action $J_{R}$ by \citep[see][for a detailed derivation]{SB02,BT08}
\begin{equation}\label{eq:deltaJR}
\Delta J_{R} = \mp\frac{l}{m}\Delta L_{z}. 
\end{equation}
Here $l$ is an integer representing the harmonic of the resonance, and $m$ is the number of spiral arms. If the star is at the \ilr{}, the upper sign is applicable, and conversely for the \olr{}. \refedit{For a Lindblad resonance, l=1, while higher harmonics are given by integers of \textit{l}>1.}

Significant, lasting changes to the orbital energy and angular momentum \sout{from a spiral pattern will} occur at dynamical resonances \citep{LBK72}.  \refedit{In the limit where a spiral pattern grows slowly, the Jacobi integral (Equation~\ref{eq:JacobiIntegral}) is approximately conserved, even for resonant orbits, and so orbital changes will obey Equation~\ref{eq:conservedJacobi}.} 

How actions are redistributed around a resonance depends on the resonance in question. It is expected that the \CR{} resonance primarily effects changes in angular momentum $L_{z}$ without inciting changes in $J_{R}$  \citep{SB02}. In contrast, the Lindblad resonances cause significant changes in both $J_{R}$ as well as changes in $L_{z}$ \citep{BW67,LBK72,SC84,JB90} according to equation~\ref{eq:deltaJR}. Hence, we expect to find strong changes in $J_{R}$ in the vicinity of the Lindblad resonances \citep{LBK72,Mark74,Sellwood12,SellwoodMasters22} with a net trend toward increasing $J_R$.

\refedit{It can be useful to plot lines of constant Jacobi integral in various forms of energy-angular momentum space} as a sufficient first-order guide when \refedit{examining the orbital response to resonances \citep{Sellwood10b}. In this study, we adopt a presentation of lines of constant Jacobi integral similar to that of \cite{Trick21}, which they called self-same \axi{} resonance lines, where we assume the \epi{} approximation and set $J_z=0$. \sout{These lines are shown in Figure~\ref{fig:delta_actions_for_varying_alphas} and similar}.}

In this work, we study the postresonant signatures left in the ($J_R$,$L_z$) action space from the passage of a transient spiral arm.  We primarily focus on changes at the \ilr{}, rather than the \olr{} and \CR{} resonances, since the \ilr{} is expected to have the most prominent changes in $J_R$ \citep{Sellwood12}. Our premise is that the unperturbed value of a star's $J_R$ can on average be correlated with its age.


\section{Methods}\label{sec:methods}

In order to investigate how spiral morphology affects the redistribution of orbits in a disk galaxy, we constructed a suite of tracer particle simulations. This method is ideal since we produce a controlled series of experiments with known potentials, thus enabling straightforward identification of postresonant trends and how they correlate to specific spiral characteristics.

We used the Python-based Galactic dynamics package \texttt{galpy} \citep{Bovy15} v.1.7\footnote{{\url{http://github.com/jobovy/galpy}}} to produce nine \mw{}-like galaxies with distinct models for transient spiral patterns.
The \texttt{galpy} package uses natural units such that the circular velocity $v_o$ equals unity at some nominal scale length $r_o$, which is also set to unity. We assumed \texttt{galpy}'s standard conversion factors so that the natural unit for radius $r_o=8$~kpc and $v_o(r_o)=220$~km~s$^{-1}$, approximating values at the solar circle.  These values are given in Table~\ref{t:FiducialModelParameters}.


\subsection{Potentials}\label{s:potentials}

We adopted the same prescription in all simulations to construct a \mw{}-like underlying axisymmetric disk potential.  Various transient spiral perturbations were then superposed. The bar was not modeled at this time as our focus was to isolate the effects of spiral arm perturbations.


We employed \texttt{galpy}'s default class for a smooth, \mw{}-like disk potential, \texttt{MWPotential2014} \citep{Bovy15}, hereafter referred to as \texttt{MWP}. This potential consists of a 3D Navarro-Frenk-White (NFW) dark matter halo \texttt{NFWPotential} \citep{NFW96}, a 3D power-law spherical bulge with an exponential cutoff, \texttt{PowerSphericalPotentialwCutoff}, and an axisymmetric Miyamoto-Nagai disk, \texttt{MiyamotoNagaiPotential} \citep{MiyamotoNagai75}. 
These potentials are weighted and added to produce a reasonable approximation for the \mw{} disk in the \snd{} \citep[between ${\sim}4~\mathrm{and}~9$ kpc, ][]{BovyRix13, Bovy15}. All scaling parameters and normalization values were the default values assigned for \texttt{MWP}.


\subsubsection{Spiral Pattern Perturbation}\label{ss:spiralpotential}

We adopted an \azily\ sinusoidal prescription for the spiral potential, \texttt{SpiralArmsPotential} \citep{CoxGomez02}. This spiral potential is described by

\begin{align}\label{eq:SpPotential}
\Phi_s(R,\phi,z) =& -A_s H \exp\left[{\frac{R - r_{\rm ref}}{R_{s}}}\right] \\
&\times \sum \frac{C_{n}}{K_{n} D_{n}} \text{cos}(n \gamma) \text{sech} \left(\frac{K_{n}z}{B_{n}}\right)^{B_{n}} \nonumber,
\end{align}

where

\begin{equation}
    K_{n} = \frac{nm}{R \sin{\alpha}},
\end{equation}

\begin{equation}
    B_{n} =  K_{n}H (1 + 0.4 K_{n}),
\end{equation}

\begin{equation}
    D_{n} = \frac{1 + K_{n}H + 0.3 (K_{n}H)^2}{1 + 0.3 K_{n}H}
\end{equation}

and

\begin{equation}
    \gamma = m \left[\phi - \phi_{\rm ref} - \frac{\text{ln}\,(R / r_{\rm ref})}{\text{tan}\alpha}\right].
\end{equation}
Here, $m$ is the number of spiral arms, $\alpha$ is the spiral arm pitch angle, $r_{\rm ref}$ is the fiducial radius, and $\phiref$ is the reference angle $\phi_{p}(r_{0})$ where the spiral density is defined in \cite{CoxGomez02}, $R_{s}$ is the radial scale length for the arm density amplitude, $H$ is the spiral arm scale height, and $C_{n}$ (here set to unity to create a sinusoidal arm profile) is a list of constants multiplying the cos($n \gamma$) terms that modify the density profile of the spiral arms. Default values set by \texttt{galpy} were used, unless otherwise stated as described in \S\ref{s:Spirals}. 

The time-dependent amplitude of each spiral pattern follows a 
Gaussian prescription: 
\begin{equation}\label{eq:spmag}
    A(t) = A_{s} \text{exp}\left(-\frac{[t-\tau_{\rm peak}]^2}{2\sigma_{\rm t}^2}\right),
\end{equation}
which \refedit{replaces $A_{s}$ in Equation~\ref{eq:SpPotential} and}  is implemented \refedit{using}  the \texttt{galpy} function \texttt{GaussianAmplitudeWrapperPotential}. \sout{and
by replacing $A_{s}$ in Equation~\ref{eq:SpPotential} is replaced by $A(t)$.}  Here, $A_s$ is the peak amplitude of the spiral potential reached at time $\tau_{\rm peak}$, and $\sigma_{\rm t}$ is the standard deviation of the Gaussian timescale for growth and decay. The natural units of amplitude adopted here are multiples of the density where $A_s=4\pi G\rho_0$, and $\rho_0$ is the nominal spiral density. 

The spiral prescription used for all models, with the exception of the winding spiral model \SpWind{} (introduced below), are given a constant pattern speed $\Omega_{\rm s}$ that does not shear with radius or evolve over time. In this case, the pattern speed and the radius of \CR{}, $R_{\rm CR}$, are related through the underlying rotation curve by
\begin{equation}\label{eq:omegas}
\Omega_{\rm s}=v_{\rm c}(R_{\rm CR})/R_{\rm CR},    
\end{equation}
where $v_{\rm c}$ is the circular velocity at radius $R$.  In practice, we set $R_{\rm CR}$ and \sout{diagnosed} \refedit{found the pattern speed by identifying} $v_{\rm c}(R_{\rm CR})$ in \texttt{MWP} using the built-in function \texttt{potential.vcirc}. 

The above-described prescription represents a transient density wave such as described by \cite{Grand12a} and \cite{Hunt18b}.
In model \SpWind{}, we modify the transient spiral potential, which has a constant pattern speed $\Omega_{\rm s}$, to a model where the spiral arms corotate at all radii such that $\Omega_p(R)=\Omega(R)$. To do this we used the \texttt{CorotatingRotationWrapperPotential} wrapper.  This is similar to the prescription used by \cite{Hunt19}, which \refedit{is} designed to mimic arms that wind over time. \texttt{CorotatingRotationWrapperPotential} is parameterized by
\begin{equation}\label{eqn:wrappingwrapper}
\phi \rightarrow \phi + \frac{V_{\rm c}(R)}{R} \times \left(t - t_{\rm ref}\right) + \phi_{0}
\end{equation}
where $t_{\rm ref}$ is the reference time when the wrapped potential is equal to the unmodified potential, $\phi_{0}$ is the position angle at the starting time, and $V_{\rm c}(R)$ is the circular velocity curve at radial position $R$ as described by 
\begin{equation}
V_{\rm c}(R) = V_{\rm c,0} \left( \frac{R}{R_0} \right)^\beta \,,
\end{equation}
where we set $V_{\rm c,0} = v_o$, $\beta = 0$, and $\phi_{0}$ = 0.


\subsection{Initial Conditions}\label{s:ICs}

The initial phase-space coordinates for the tracer particles are the same for each model in our suite of simulations. Initial coordinates were sampled for $N=10^5$ particles from a reasonable distribution function (described in \S\ref{ss:DistributionFunction}).
\refedit{The number of particles was chosen to be an order of magnitude greater than the necessary number to clearly resolve the signatures discussed in this paper from noise.
These initial phase-space coordinates were used to initiate orbits that were} evolved through the smooth disk potential (\S\ref{s:QuietPhase}) for 500~Myr. The final phase-space positions from this \refedit{smooth disk potential with no perturbations, which we term the} quiet phase\refedit{,} were used as the initial conditions for each simulation in this study. We here describe this process in detail.

We restricted our initial coordinate space to include only radial coordinates above a minimum initial radius $r_{\rm min}=3$~kpc and less than a maximum initial radii $r_{\rm max}=15$~kpc \refedit{to ensure that the distribution function for our area of interest was appropriately populated.} The maximum initial height above and below the disk $|z_{\text{max},0}|$ was set to 2.4~kpc. These limits allowed us to focus on the region of interest for this work, as the resonant signatures we investigated occurred between approximately $4\lesssim R\lesssim14$~kpc. The radial buffer of on order 1 kpc both minimized the number of highly eccentric orbits that were removed from this sample while also maximizing the efficiency of our computation time given that the central regions of the disk are exponentially more populated.


\subsubsection{Distribution Function}\label{ss:DistributionFunction}

Initial conditions were acquired by sampling a distribution function for a quasi-isothermal \refedit{disk}, which provides a good approximation to the \mw{} disk \citep{Binney10, BM11} and is expressed in terms of action-angle variables using the St\"{a}ckel approximation \citep{DeZeeuw85,Binney12}. Use of this approximation allows for coupling of planar and vertical stellar motion, providing an improvement on the adiabatic approximation. This choice is important for effective determination of actions for orbits with larger vertical excursions  \citep{Binney12,BovyRix13,Trick17}; for a pedagogical approach to actions and action estimates in nonaxisymmetric models see \cite{BT08} Sections 3.5 and 4.6.

In practice, our set of initial conditions were produced using a Markov Chain Monte Carlo sampling scheme to find phase-space positions from the \texttt{quasiisothermaldf} class.
This quasi-isothermal disk distribution function is given by \citep{BM11}
\begin{equation}\label{eq:DistributionFunction}
f(J_{R}, L_{z}, J_{z}) = f_{\sigma_{r}}(J_{R}, L_{z}) \times \dfrac{v_z}{2 \pi \sigma_{z}^2}  e^{-v_{z}J_{z}/\sigma_{z}^2}.
\end{equation}
where \(v_{z}\) is the velocity in the \(z\)-direction, and \(\sigma_{z}\) is the vertical velocity dispersion. The first term in equation~\ref{eq:DistributionFunction}, $f_{\sigma_{R}}(J_{R}, L_{z})$, is given by
\begin{equation}\label{eq:df_first_term}
    \left.f_{\sigma_{R}}\left(J_{R}, L_{z}\right) \equiv \frac{\Omega \Sigma}{\pi \sigma_{R}^{2} \kappa}\right|_{R_{\mathrm{c}}}\left[1+\tanh \left(L_{z} / L_{0}\right)\right] \mathrm{e}^{-\kappa J_{R} / \sigma_{R}^{2}},
\end{equation}
where $\nu$ is the vertical \epi{} frequency and $\Sigma$ is an approximation of the radial surface-density profile, and the variables $\Omega$, $\kappa$, $\nu$, and $\Sigma$ are taken to be the values for a given angular momentum $L_{\rm z}$. The surface-density profile adopted is given by $\Sigma=\Sigma_{0} \mathrm{e}^{-\left(R-R_{\mathrm{c}}\right) / R_{\mathrm{d}}}$, 
where $R_{\mathrm{c}}=R_{\mathrm{c}}\left(L_{z}\right)$ is the radius of a circular orbit with angular momentum $L_{z}$ and $R_{\mathrm{d}}=r_{\mathrm {o}}/3$ is the surface-density scale length. 
The factor $1+\tanh \left(L_{z} / L_{0}\right)$ in equation~\ref{eq:df_first_term} restricts the phase space to include only stars on prograde orbits, where the default value of $L_{\rm 0}$ is chosen to be small ($L_{\rm 0}/L_{z}\ll 1$ for all values of $L_z$ in the range of interest). 

The functional forms for the radial and vertical velocity dispersions are 
\begin{equation}\label{eq:radial vel disp}
    \sigma_{\mathrm{R}}\left(L_{\mathrm{z}}\right)=\sigma_{R_{0}} \mathrm{e}^{\mathrm{q}\left(R_{0}-R_{\mathrm{c}}\right) / \mathrm{h_{\mathrm{R}}}}
\end{equation}
and 
\begin{equation}\label{eq:vertical vel disp}
    \sigma_{\mathrm{z}}\left(L_{\mathrm{z}}\right)=\sigma_{\mathrm{z_{\mathrm{0}}}} \mathrm{e}^{\mathrm{q}\left(R_{0}-R_{\mathrm{c}}\right) / \mathrm{h_{\mathrm{z}}}}, 
\end{equation}
respectively, where $q$=0.45, and $\sigma_{\rm R_{\rm 0}}$ and $\sigma_{\rm z_{\rm 0}}$ are the normalization factors for the radial and vertical velocity dispersions, respectively.

The radial scale length of the exponential surface density was set as $\rm R_{\rm d}$ = $\rm r_{\rm o}/3$ (${\sim}2.7$~kpc, similar to values adopted by \cite{Bovy15,DanielWyse18}). The scale of the radial velocity dispersion was set to equal $\sigma_{\rm R_{\rm o}}=0.16v_o$ (${\sim}$35.2 km/s) at $\rm r_o$ (similar to values adopted by \cite{DanielWyse18,Khrapov21}), and the vertical velocity dispersion was set to equal $\sigma_{\rm z_{\rm o}}=\rm \sigma_{\rm R_{\rm o}} /2$  (${\sim}17.6$~km/s) at $\rm r_o$ (similar to values adopted by \cite{BT08,AumerBinney09,Binney10}). These choices approximate the average value for stars in the \snd{}. The radial scale lengths $h_{\rm R}$ and $h_{\rm z}$ for the radial and vertical velocity dispersion profiles are both set to \sout{to} 3$R_{\rm d}$ \citep[similar to values adopted by][]{Minchev12,Bovy15}. These scale lengths were chosen in order to obtain an exponential surface-density profile that mimics the \mw{} \citep{Bovy15}.

\begin{longtable}[htb]{>{\raggedright\arraybackslash}m{5cm} >
{\centering\arraybackslash}m{2cm}}
\caption{Underlying Model Variables}\label{t:FiducialModelParameters}\\
\hline
\textbf{Variable} & \textbf{Value} \\
\hline
\endfirsthead

\multicolumn{2}{c}%
{{\tablename\ \thetable{} -- continued from previous page}} \\
\hline
\textbf{Variable} & \textbf{Value} \\
\hline
\endhead

\hline \multicolumn{2}{r}{{Continued on next page}} \\
\endfoot

\hline
\endlastfoot
Number of particles ($N$) & 100,000 \\
Time step length ($\delta t, \mathrm{Myr}$) & $\sim$0.377 \\
\hline
Scale radius ($r_o, \mathrm{kpc}$) & 8 \\
Scale velocity ($v_o, \mathrm{km}\,s^{-1}$) & 220 \\
Minimum initial radius ($r_{\text{min},0}, \mathrm{kpc}$) & 3 \\
Maximum initial radius ($r_{\text{max},0}, \mathrm{kpc}$) & 15 \\
Maximum initial height ($|z_{\text{max},0}|, \mathrm{kpc}$) & 2.4 \\
Radial vel. disp. at $r_o$ ($\sigma_{R_{o}}, \mathrm{km}\,s^{-1}$) & 35.2 \\
Vertical vel. disp. at $r_o$ ($\sigma_{z_{o}}, \mathrm{km}\,s^{-1}$) & 17.6 \\
\hline
Surf. density radial scale length ($R_{d}$) & $r_{\rm o}/3$ \\
Radial vel. disp. scale length ($h_{R}$) & $3R_{d}$ \\ 
Vertical vel. disp. scale length ($h_{z}$) & $3R_{d}$ \\

\end{longtable}

\tablecomments{All simulations utilize the \texttt{galpy} underlying \mw{} potential \texttt{MWPotential2014} and the \texttt{galpy} St\"{a}ckel action-angle approximation \texttt{stklAA}.}
\label{table:Table 1}


\subsection{Orbit Integration}\label{ss:OrbitIntMethod}

Orbit integration was performed using \texttt{galpy}'s symplectic C integration method \texttt{symplec4\_c}. The integration was split into two phases: a quiet phase where the initial tracer populations were allowed to evolve in a smooth disk potential and a transient spiral phase where various prescriptions for spiral structure grew and decayed over a prescribed time frame.


\subsubsection{Orbit Integration: Quiet Phase}\label{s:QuietPhase}

The initial orbital phase-space coordinates obtained by sampling \texttt{quasiisothermaldf} described in \S\ref{ss:DistributionFunction} were evolved in the smooth disk, \texttt{MWP}, for 0.5~Gyr.  
The length of a single time step was chosen to be one-tenth of an orbital time for a circular orbit at \sout{at} $r_{\rm min}$, so $\delta t \approx.377$~Myr.  
This phase allowed for equilibration and ensured that our model choices did not lead to any significant nonadiabatic response.
\refedit{The final coordinates from this quiet phase were used as the}\sout{The} initial phase-space coordinates for all spiral phase simulations described in Sections \S\ref{s:Spirals}-\ref{ss:winding spirals}\refedit{.}\sout{, were equal to the final coordinates from this quiet phase.}


\subsubsection{Orbit Integration: Spirals}\label{s:Spirals}

Eleven simulated \mw{} analogs were produced with a systematic \sout{variety} \refedit{variation} of the following transient spiral properties: pitch angle, amplitude, pattern speed, lifetime, and prescription for the radial dependence of the pattern speed (i.e.,~density-wave-like or winding). Each model and its parameters are described in Table~\ref{t:SpiralVariables}. The total integration time for each simulation was set to be 4 times the orbital period at the radius of \CR{}, $\Delta t=4 T_{\rm dyn}(R_{\rm CR})$. We emphasize here that it is not the goal of this paper to perfectly reproduce the spirals in the \mw{}, but instead to examine postresonant trends from a variety of transient spiral prescriptions.

\begin{deluxetable*}{lccccc}
\tablenum{2}
\tablecaption{Parameters \refedit{values for each spiral model}\label{t:SpiralVariables}}
\tablewidth{0pt}
\tablehead{
\colhead{Model} & \colhead{Pitch Angle $\alpha$} & \colhead{Corotation radius $R_{\rm CR}$} & \colhead{Spiral Amplitude $A_s(4\pi G\rho_{o}$)} & \colhead{Spiral Timescale $\tau_{\rm peak}$ ($T_{\rm Dyn}(R_{\rm CR})$)} & \colhead{Spiral Type: Winding?}
\\
\nocolhead{Model} & \colhead{(deg)} & \colhead{(kpc)} & \colhead{} & \colhead{} & \colhead{} 
}
\startdata 
{\SpFid{}} & 20 & 8 & 1.0 & 2 & no
\\
{\Spaten{}} & 10 & 8 & 1.0 & 2 & no
\\
{\Spathirty{}} & 30 & 8 & 1.0 & 2 & no
\\
{\SpCRsix{}} & 20 & 6 & 1.0 & 2 & no
\\
{\SpCRten{}} & 20 & 10 & 1.0 & 2 & no
\\
{\SpAhalf{}} & 20 & 8 & 0.5 & 2 & no
\\
{\SpAdouble{}} & 20 & 8 & 2.0 & 2 & no
\\
{\SpTone{}} & 20 & 8 & 1.0 & 1 & no
\\
{\SpTfour{}} & 20 & 8 & 1.0 & 4 & no
\\
{\SpTeight{}} & 20 & 8 & 1.0 & 8 & no
\\
{\SpWind{}} & 20 & 8 & 1.0 & 2 & yes
\\ 
\enddata 
\label{table:Table 2}
\end{deluxetable*}


\subsubsection{Spiral pitch angle}\label{ss:spiral PA}

The openness of a spiral arm can be defined by its pitch angle $\alpha$.  Observationally, spiral arm pitch angles can be approximated fairly well by either logarithmic or hyperbolic functional fitting forms. Spiral arm pitch angles often range from a few degrees to~30$^{\circ}$--~40$^{\circ}$ \citep[e.g.,][]{Kennicutt81,Diaz-Garcia19,Masters19,YuHo20,Wisz26} and typically vary only few degrees within a given spiral arm.

We adopt a logarithmic form for our spiral models where the pitch angle, defined as the angle between the arm and a circle about the Galactic center \citep{BT87}, is the same at all radii.  As such, a tight winding spiral will have $\alpha\rightarrow 0$ and a bar has $\alpha=90^\circ$.  

In this work we use \texttt{galpy}'s logarithmic spiral arm prescription, \texttt{SpiralArmsPotential} \citep{ConsidereAthanassoula82}.  Using polar coordinates $(r,\theta)$ the peak amplitude of the arms is traced by a curve obeying
\begin{equation}
    r = r_{0}\exp{\left[-\frac{m}{p}\theta\right]},
\end{equation}
where $r_{0}$ sets the rotational position of the spiral, $m$ is the number of spiral arms, and $p$ is related to pitch angle $\alpha$ by
\begin{equation}
    \tan{\alpha} = -\dfrac{m}{p}.
\end{equation}
 
The fiducial pitch angle adopted for spiral arms in the simulation suite is $\alpha=20^{\circ}$. Pitch angles of 10$^{\circ}$ and 30$^{\circ}$ are implemented for models \Spaten{} and \Spathirty{} respectively. As can be seen in Table~\ref{t:SpiralVariables}, all other parameters for these two models reflect the same values as in the fiducial model.


\subsubsection{Spiral amplitude}\label{ss:spiral amp}

The parameter governing the amplitude of a spiral pattern, $A_s$, is unitless and described by the relation $A_s=4\pi G\rho_0$, where $\rho_0$ is the mass density. The amplitude of each spiral perturbation is modified by the Gaussian prescription given in Equation~\ref{eq:spmag} to model the growth and decay of a transient pattern (further described in \S\ref{ss:spiral lifetime}).

The fiducial maximum amplitude of spiral perturbations to the potential is a 20\% overdensity with respect to the underlying disk. Amplitudes of one-half and twice the fiducial value were modeled by \SpAhalf{} (10\%) and \SpAdouble{} (40\%), respectively.  The fiducial amplitude was used as the default spiral strength for all other models.
Table~\ref{t:SpiralVariables} documents the amplitudes used for each model.


\subsubsection{Spiral pattern speed}\label{ss:corotation loc}

The pattern speed of each spiral can be quantified by the radius of \CR{}, $R_{\rm CR}$, through Equation~\ref{eq:omegas}.
The \ilr{} and \olr{} are similarly related to the pattern speed through Equation~\ref{eq:lindblad} and are located at smaller and larger radii, respectively, than the \CR{} radius.  In this suite of simulations a selected $R_{\rm CR}$ was used to prescribe the pattern speed, which in turn specifies the locations of the Lindblad resonance.

The $R_{\rm CR}$ for the fiducial \mw{} model (\SpFid{}) was set to 8~kpc to approximate the radius of the Sun. Model \SpCRsix{} implemented a $R_{\rm CR}$ of 6~kpc, while model \SpCRten{} set the $R_{\rm CR}$ to 10~kpc.


\subsubsection{Spiral lifetime}\label{ss:spiral lifetime}

Transient spirals grow and decay through their evolution cycles over lifetimes which are often poorly constrained \citep{Sellwood11}. Simulated spirals have lifetimes that span on order of $\sim$100~Myr to multiple gigayears \citep[e.g.,][]{Weinberg04,Grand12a}. 

In this work, we base the spiral lifetime on each given model's dynamical time $T_{\rm dyn}$, which we define as equaling the orbital period at the \CR{}.
Spiral amplitudes follow the Gaussian time dependence prescription given by Equation~\ref{eq:spmag}. Each model reached peak amplitude at $\tau_{\rm peak}=2\sigma_t$, where the fiducial model uses $\sigma_t=2T_{\rm dyn}$.

The spiral lifetime for the fiducial model (\SpFid{}), where $T_{\rm dyn}\simeq0.22$~Gyr, is set by \refedit{$\sigma_t=2T_{\rm dyn}\sim0.45$~Gyr). In practice, we set the spiral potential to be zero at $t=\tau_{\rm peak}\pm2T_{\rm dyn}$ so that} the total spiral lifetime was $4T_{\rm dyn}$ ($\sim$0.89~Gyr). 

In order to identify trends related to the way spiral lifetime could affect the amplitude and character of the signatures in action space, we simulated a spiral lifetime shorter than that of \SpFid{} in model \SpTone{}, in addition to simulations of successively longer spiral lifetimes in models \SpTfour{} and \SpTeight{}. \SpTone{} incorporated spirals with one-half the lifetime of \SpFid{} \refedit{with $\sigma_t=T_{\rm dyn}$ and the full length of the simulation spanning the nonzero spiral lifetime of} $2T_{\rm dyn}$ ($\sim$0.45~Gyr). \SpTfour{} modeled spirals with double the lifetime of \SpFid{} \refedit{such that $\sigma_t=4T_{\rm dyn}$} ($\sim$0.89~Gyr), with total spiral lifetime equivalent to $8T_{\rm dyn}$ ($\sim$1.79~Gyr). Model \SpTeight{} was quadruple the lifetime of \SpFid{} \refedit{with $\sigma_t=8T_{\rm dyn}$} ($\sim$1.79~Gyr) and a total lifetime of $16T_{\rm dyn}$ ($\sim$3.57~Gyr). 

\refedit{To test the lifetimes of wrinkles, we allowed the orbits in \SpFid{} to evolve an additional $2T_{\rm dyn}$ ($\sim$0.45~Gyr) after the complete dissipation of the spiral potential. We found no discernible obfuscation of the wrinkles created by the spirals even after the spirals were gone.  This indicates that phase mixing does not play a significant role in wrinkle dissolution. Since tracer particle simulations do not have any perturbing forces after the spiral amplitude goes to zero, we cannot test the expectation that wrinkles would blur on a diffusion timescale, as one might expect in physical or \textit{N}-body systems.}


\subsubsection{Winding spirals}\label{ss:winding spirals}

Along with the models thus far discussed, which qualitatively well represent transient density waves, we also modeled a winding transient spiral arm perturbation \SpWind{}. This model is similar to what was used by \cite{Hunt18b} and motivated by studies of \textit{N}-body and smoothed particle hydrodynamics simulations of disk galaxies \citep[e.g.,][]{Grand12a}, which have recurring transient material spiral arms that corotate at all radii. For this simulation, all model parameters are the same as in \SpFid{} except for the inclusion of the additional wrapper described by Equation~\ref{eqn:wrappingwrapper}. Here the \sout{the} nominal pitch angle $\alpha$ is set to be achieved when the spiral reaches its peak amplitude at $\tau_{\rm peak}$.


\section{Analysis}\label{sec:analysis}

In this section we analyze trends in the \sout{unique} signatures in action space produced by the passage of the transient spiral patterns described Table~\ref{t:SpiralVariables}. 
It is necessary to invoke perturbation theory to retrieve a full time evolution in action-angle coordinates for a galaxy containing nonaxisymmetric structures such as spirals \citep{Binney18}. However, given a well-constrained and slowly varying potential, reasonable approximations for final actions may be calculated within such a potential \citep{Binney12,BovyRix13,Trick17,Trick19}. Accordingly, the action space values for disk stars are here evaluated before and after the growth and decay of each transient spiral pattern in our suite of simulations.  The signatures we investigate are the changes to the distribution of action space coordinates from before the onset of the spiral pattern to after it has dissipated. In other words, the action space coordinate values are evaluated only when the underlying potential is smooth. 
This approach isolates the overall effect of a transient spiral pattern to the disk kinematics while ensuring each measure is using the same smooth, axisymmetric potential.

\refedit{The \mw{} hosts nonaxisymmetric structures such as a bar and spiral arms and so does not have a smooth potential. }
\sout{For real galaxies like the \mw{}, which do not have} 
\sout{smooth potentials and host nonaxisymmetric structures} 
\sout{like bars or spiral arms,} 
Actions \refedit{for \mw{} orbits} can be estimated with \refedit{reasonable} accuracy using the St\"ackel--Fudge algorithm \citep[][and described in \S\ref{ss:DistributionFunction}]{Binney12}.  We adopt this technique which is available within the \texttt{galpy} package \citep{Bovy15}. \refedit{We note here that, as discussed in \citep{Debattista25}, there can be correlated errors associated with use of the St\"ackel--Fudge algorithm, which may cause blurring of wrinkles and other kinematic features in real data.}

\begin{figure*}
\centering

    \includegraphics[width=\textwidth]{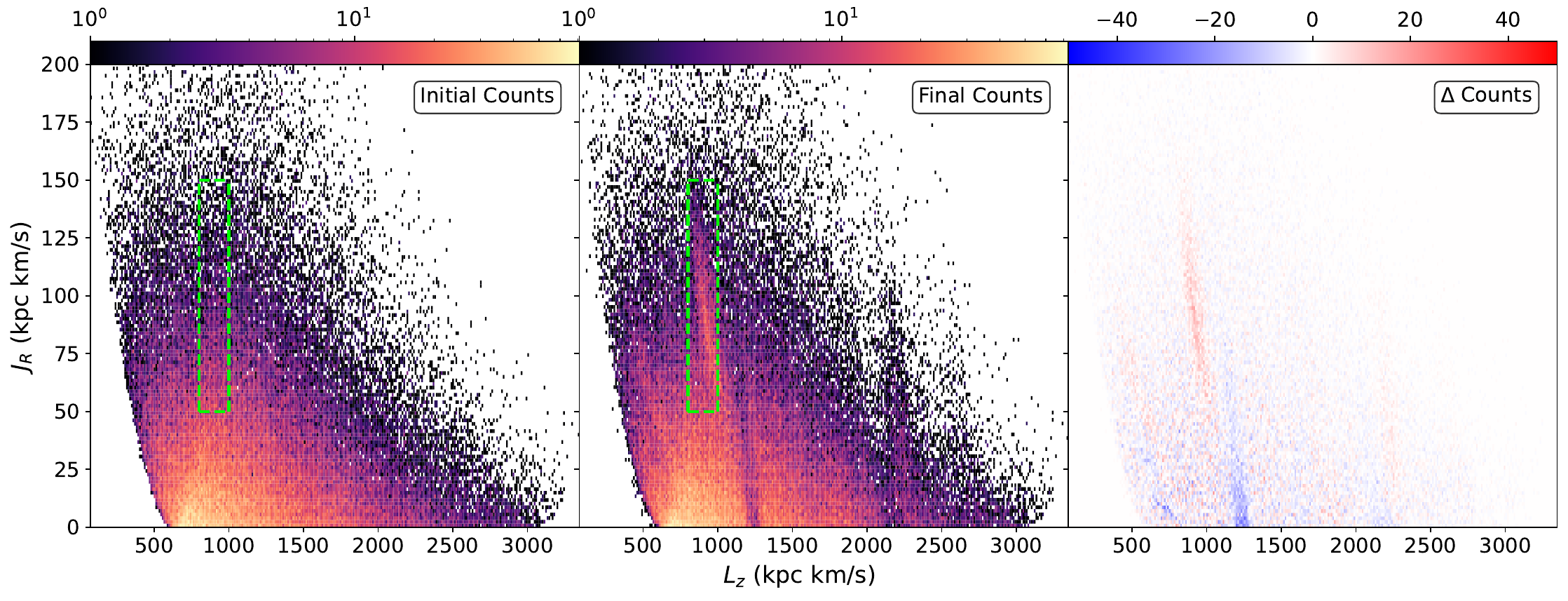} \\

  \caption{Distribution of actions in the $J_{\rm R}-L_{\rm z}$ plane for model \Spathirty{}. The color bars indicate the \refedit{particle} density before the introduction of spirals \textbf{(left)}, the \refedit{particle} density after the occurrence of a spiral pattern \textbf{(middle)}, and the \refedit{average} change in \refedit{particle} density after the passage of the spiral pattern \textbf{(right)}. We \refedit{outline} \sout{highlight in green} the \refedit{approximate} region where the wrinkle \refedit{(particle overdensity in high action space that formed as a response to the \ilr{} of a transient spiral pattern)} forms \refedit{with a dashed, green rectangle}.  We place an identical rectangle over the smooth distribution in the left plot and will compare the enclosed populations in the discussion.}
  \label{fig:spAlpha30_init_fin_delta_counts}
\end{figure*}

Figure~\ref{fig:spAlpha30_init_fin_delta_counts} demonstrates stellar density counts in $J_{\rm R}-L_{\rm z}$ space before (left) and after (middle) the passage of a spiral pattern.  The \refedit{percent} change in counts \sout{per} over time are also shown (right). \refedit{A green rectangle is placed roughly around the postresonant wrinkle in the middle panel and the same area is shown in the smooth distribution on the left.  These populations are compared in the discussion of this paper (\S\ref{s:kinematic age}).}
A visual inspection indicates that an overdensity, or wrinkle, forms when compared to the smooth distribution. This feature is visibly prominent in \Spathirty{}, but the amplitude and shape of the wrinkle signature depend on the prescription for the perturbing spiral. All other models, with exception of \SpWind{} (see \S\ref{ss:model type trends} for details), produce similar signatures.


\subsection{Trends with kinematic temperature}\label{s:trends}

It is well established that out of all the resonances, the strongest changes in $J_{\rm R}$ occur at the \ilr{} \citep{SB02,Weinberg04,SellwoodTrick19}. This study builds on those results by showing that at the \ilr{} the orbits with the lowest initial $J_{R}$ tend to exhibit the largest increase in $J_{R}$.  This can be seen as a reorganization of orbits near the \ilr{} ($L_z{\sim}900-1300\,{\rm kpc\, km\, s^{-1}}$) in Figure~\ref{fig:spAlpha30_init_fin_delta_counts}, where orbits with the lowest initial values for $J_R$ are moved to higher $J_R$ values after the passage of a spiral and form a wrinkle (discussed in \S\ref{sec:discussion}). 
While all of our models demonstrated this trend, we observed additional trends with pitch angle $\alpha$ and spiral amplitude that appeared to strengthen this behavior more substantially than changes in other model parameters.


\subsection{Trends with pitch angle}\label{ss:PA trends}

\begin{figure}
    \includegraphics[width=\columnwidth]{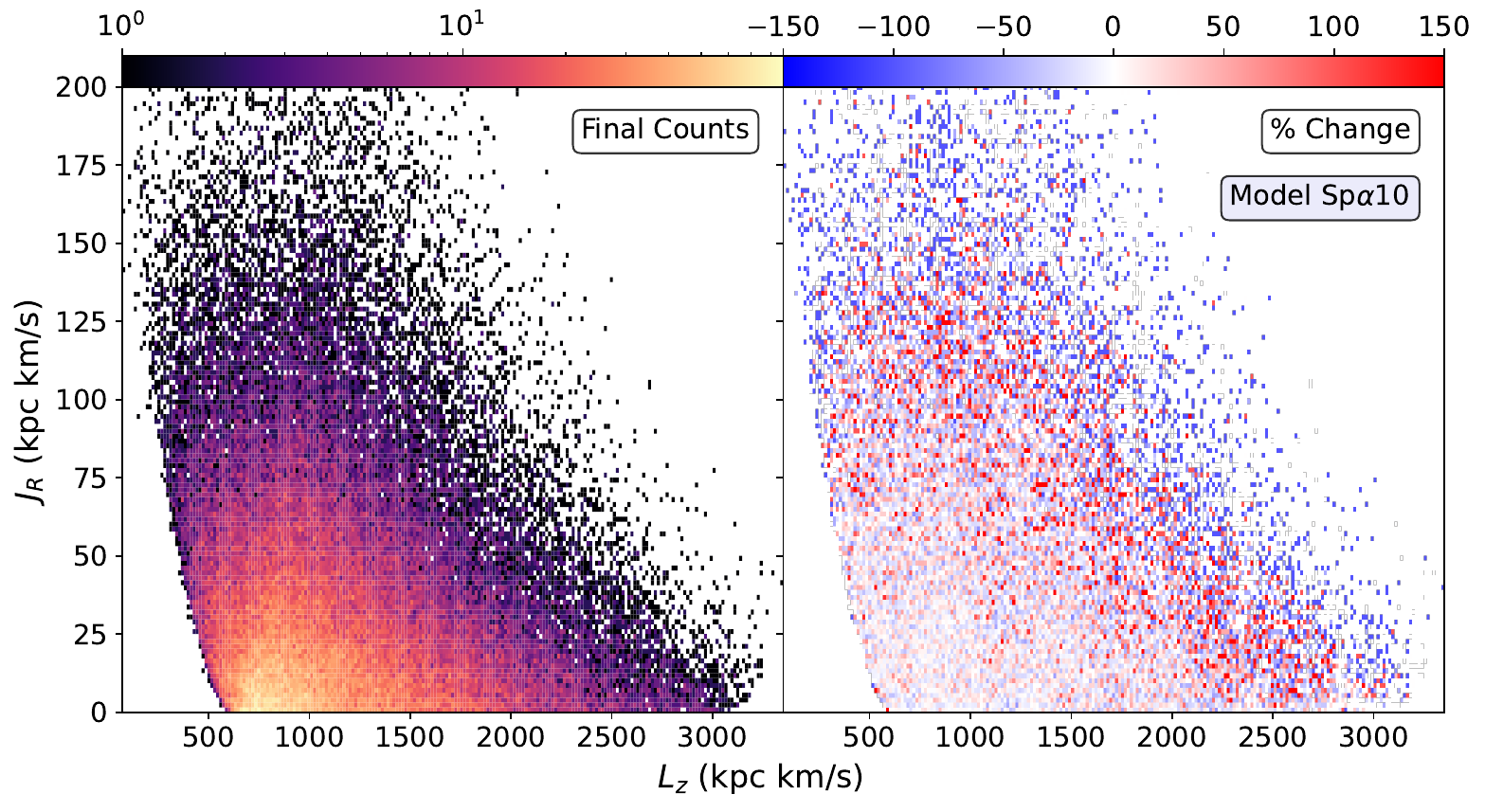} \\
    \includegraphics[width=\columnwidth]{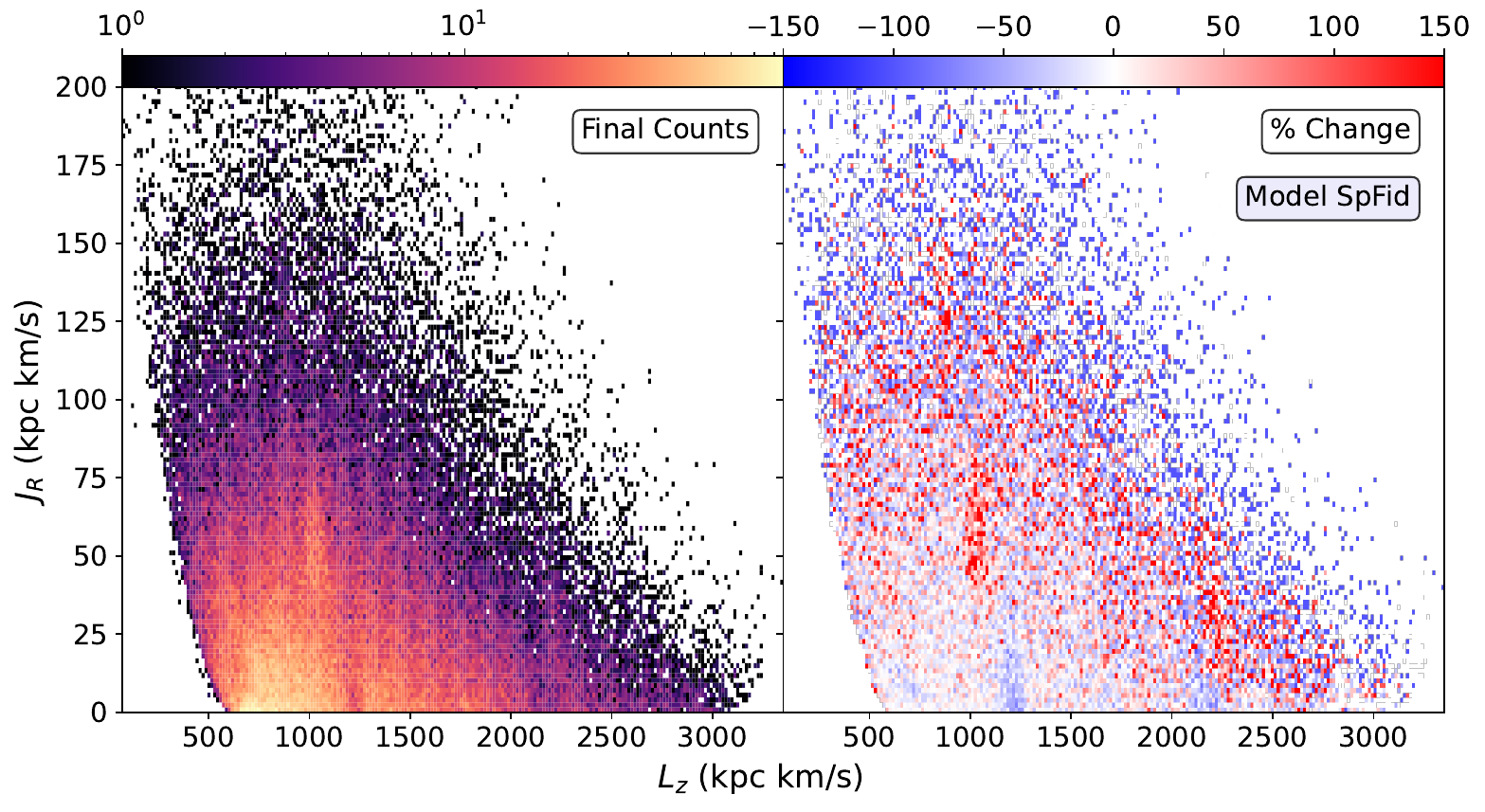}
    \caption{\refedit{Similar to Figure~\ref{fig:spAlpha30_init_fin_delta_counts}, showing stellar density counts in $J_{\rm R}-L_{\rm z}$ space after \sout{before} the occurrence of a spiral pattern (left) and the percent change in counts after the occurrence of the spiral pattern \sout{over time} (right) for models \Spaten{} (top) and \SpFid{}  (bottom), with pitch angles $\alpha=10^\circ$ and $20^\circ$, respectively}}
    \label{fig:spAlpha1020_fin_delta_counts}
\end{figure}

\begin{figure*}
\centering
  \begin{tabular}{@{}cccc@{}}
    \includegraphics{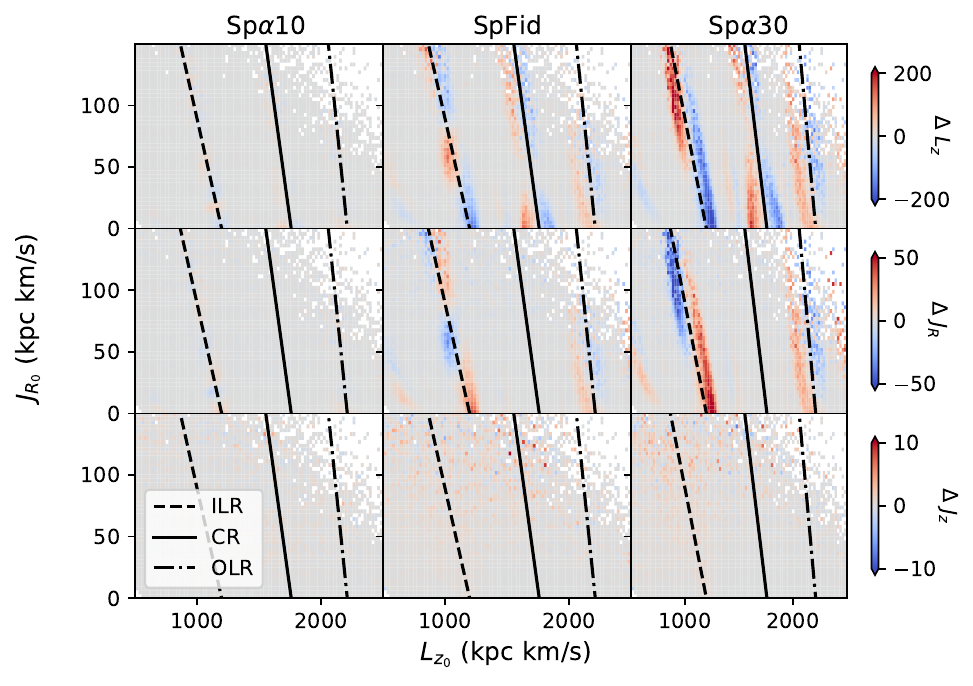} 
  \end{tabular}
  \caption{Average changes in actions $L_z$ (top), $J_R$ (middle), and $J_z$ (bottom) after the passage of a transient spiral pattern as a function of initial action for models differing only in spiral pitch angle.  The horizontal and vertical axes are the initial $J_{R_0}$ and initial angular momentum $L_{z,0}$, respectively, for models \Spaten{} (left), \SpFid{} (middle), and \Spathirty{} (right). The colorbar illustrates a positive change in action \refedit{as} red and negative change as blue. The locations of the \ilr{} (dashed line), the \CR{} (solid line), and \olr{} (dot-dashed line) are shown.
  }
  \label{fig:delta_actions_for_varying_alphas}
\end{figure*}

\refedit{In this subsection we compare the trends from models \Spaten{}, \SpFid{}, and \Spathirty{}.  These have pitch angle $\alpha=10^\circ,20^\circ,$~and~$30$\textdegree{}, respectively. Other than pitch angle all other parameters are held the same between these models.} 
\refedit{Figure~\ref{fig:spAlpha1020_fin_delta_counts} shows the particle density and its change from equilibrium and Figure~\ref{fig:delta_actions_for_varying_alphas} shows the} average change in $L_z$ (top), $J_R$ (middle), and $J_z$ (bottom) from the passage of a transient spiral pattern as a function of initial coordinate value in $L_{z_0}-J_{R_0}$ space. Resonance lines are plotted using the methods discussed in \S\ref{s:PRsignatures}. 
\sout{From left to right, this figure illustrates trends with pitch }
\sout{angle $\alpha=10^\circ,20^\circ,$~and~$30$\textdegree{}, from the corresponding models }
\sout{\Spaten{}, \SpFid{}, and \Spathirty{}, respectively.  Other than pitch }
\sout{angle all other parameters are held the same between these}
\sout{models.} 
The amplitude of the wrinkle signature increases with increasing pitch angle, \sout{and} \refedit{but the signatures} are otherwise relatively similar. 

In every model, orbits near the \ilr{} with the smallest values for the initial radial action $J_{R_0}$ had the most significant average decrease in $L_{z}$ and increase in $J_R$. The opposite trend appears more mildly for higher initial values of $J_{R_0}$. Particles with a slightly lower (higher) $L_{z}$ than the \olr{} experienced a mild increase (decrease) in $J_{R}$. No change in $J_{R}$ occurs around the CR, where there appears to be a symmetric exchange in $L_{z}$ about the \CR{} resonance, as predicted by \cite{SB02}. 
The direction of change in these maps was predicted by \cite{SellwoodTrick19}, however we demonstrate an additional trend where orbits with the lowest $J_{R_0}$ have the strongest increase in $J_{R}$ at the \ilr{}. 

\refedit{We note the existence of slight gaps in the amplitude of \sout{$\Delta J_x$} $\Delta J_R$ along the resonance lines, particularly around 75~kpc~km/s in the fiducial model (\SpFid{}). 
We note that these gaps trend toward higher corresponding value for $J_R$ with decreasing pattern speed (\S\ref{ss:CR trends}) but do not change with spiral amplitude or spiral duration.  This suggests that harmonics exist at a higher $J_R$ as orbits diverge from the epicyclic approximation. A deeper exploration of this trend is deferred to a future paper.}

There are no discernible trends in $\Delta J_{z}$ \refedit{that could be associated with a change in the spiral pitch angle.} 

\subsection{Trends with spiral amplitude}\label{ss:amp trends}

\begin{figure}
    \includegraphics[width=\columnwidth]{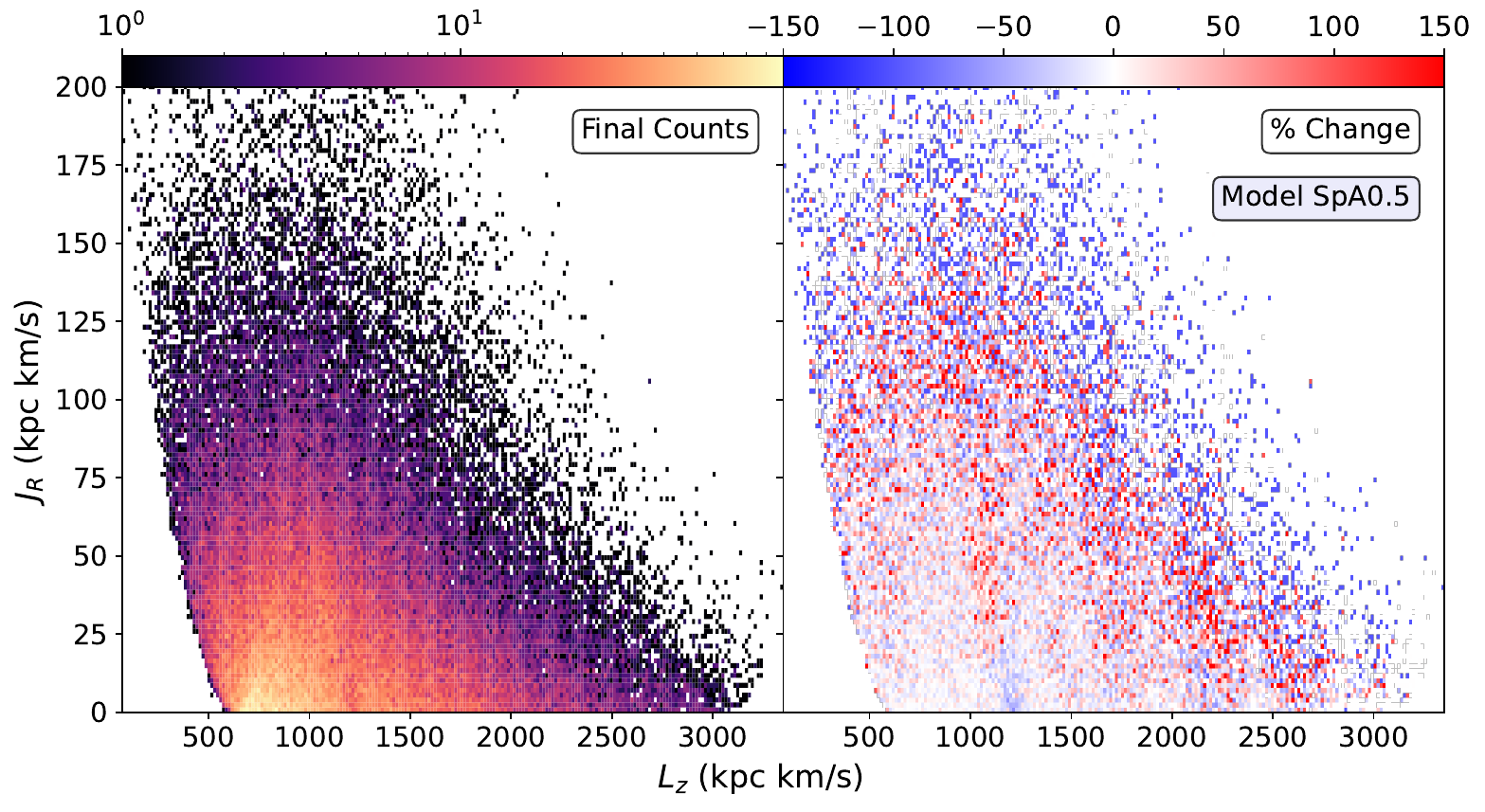} \\
    \includegraphics[width=\columnwidth]{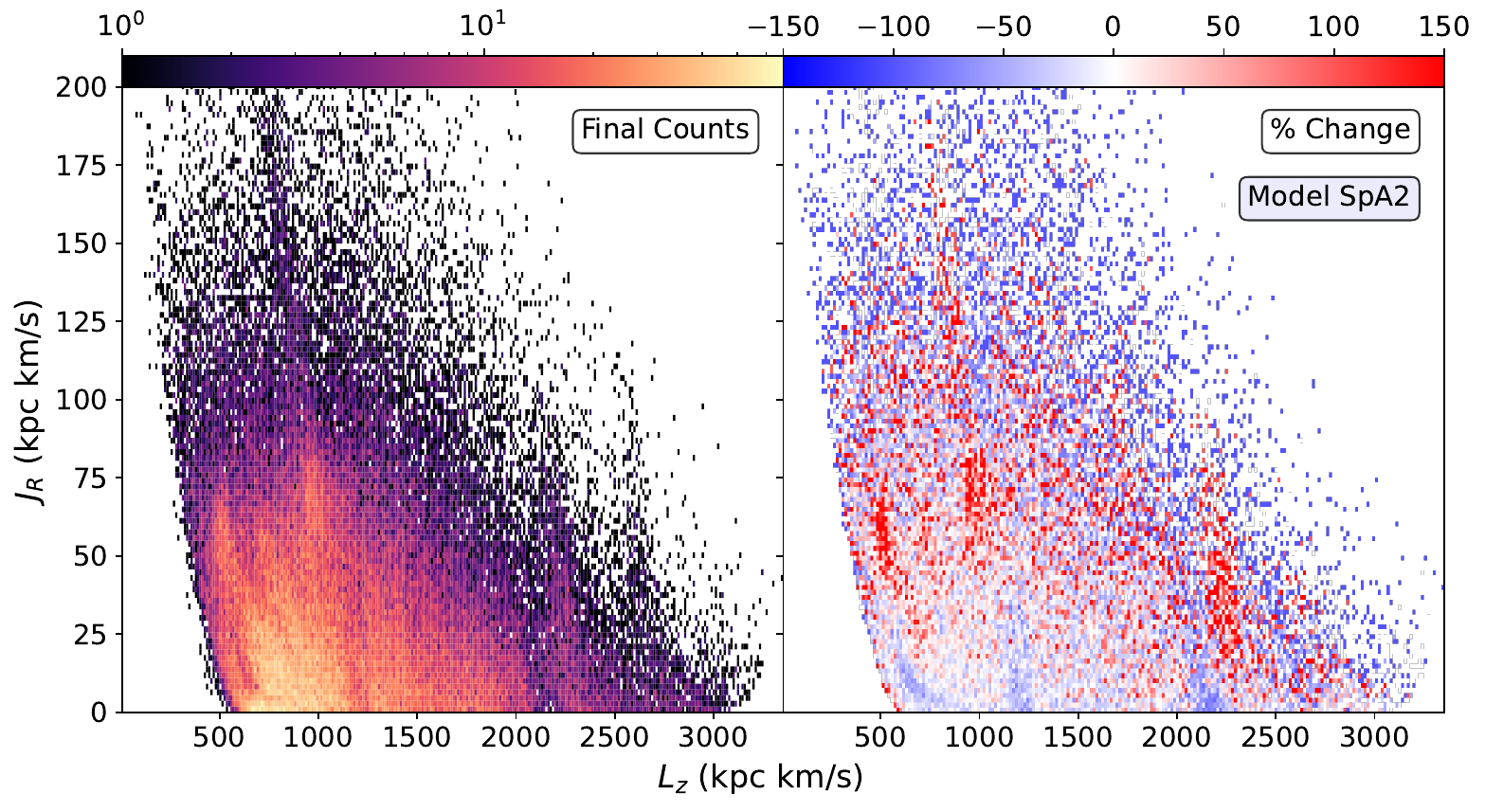}
    \caption{\refedit{Similar to Figure~\ref{fig:spAlpha30_init_fin_delta_counts}, showing stellar density counts in $J_{\rm R}-L_{\rm z}$ space after \sout{before} the occurrence of a spiral pattern (left) and the percent change in counts after the occurrence of the spiral pattern \sout{over time} (right) for models \SpAhalf{} (top) and \SpAdouble{} (bottom), with peak amplitudes one-half and twice, respectively, the fiducial model.}}
    \label{fig:spAmpHalfDouble_fin_delta_counts}
\end{figure}

\begin{figure*}
\centering
    \includegraphics{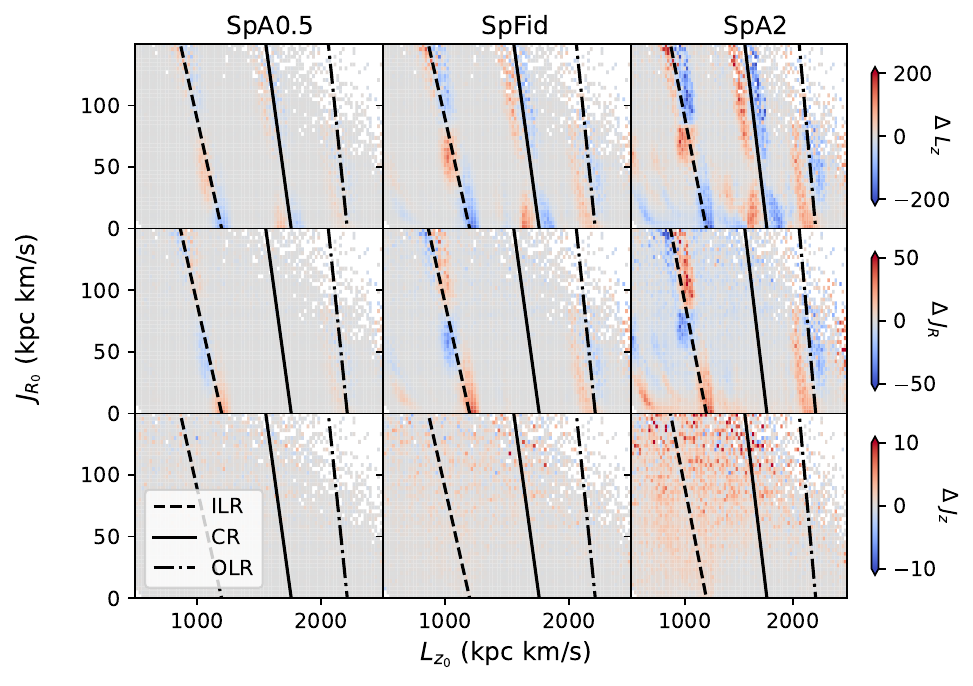} 
   \caption{Average changes in actions $L_z$ (top), $J_R$ (middle), and $J_z$ (bottom) after the passage of a transient spiral pattern as a function of initial action for models differing only in spiral amplitude. Axes are as shown in Figure~\ref{fig:delta_actions_for_varying_alphas}.
  }
  \label{fig:delta_actions_for_varying_amps}
\end{figure*}

\refedit{Spirals in our fiducial model, \SpFid{}, had a maximum amplitude that was a 20\% mass overdensity with respect to the underlying disk. We here compare to the signatures from spirals with maximum amplitudes of 10\% in model \SpAhalf{}, one-half the fiducial value, and 40\% in model \SpAdouble{}, twice the fiducial value.}
Figure~\ref{fig:spAmpHalfDouble_fin_delta_counts} \refedit{shows the trends in density after the passage of a transient spiral, while Figure~\ref{fig:delta_actions_for_varying_amps}} illustrates signatures in the average change in action for a given initial value in $L_{z_0}-J_{R_0}$ from the passage of a transient spiral pattern of a given amplitude. \sout{Spirals in our fiducial model, \SpFid{} (middle column),} 
\sout{had a maximum amplitude that was a 20\% mass overdensity}
\sout{ with respect to the underlying disk. At their respective} 
\sout{maximum amplitudes, model \SpAhalf{} (left column) constitutes} 
\sout{a spiral overdensity of 10\%, half the fiducial value, and \SpAdouble{}}
\sout{(right column) an overdensity of 40\%, twice the fiducial}
\sout{value.}

The trends in $\Delta J_{R}$ and $\Delta L_{z}$ at the resonances are similar to those described in \S\ref{ss:PA trends}, where here the response is weaker for \SpAhalf{} than for \SpFid{}. The signatures from \SpAdouble{} are stronger with an additional response that is particularly apparent in $\Delta L_z$ in the space between the primary resonances, where a nonlinear response might be expected for such a strong spiral amplitude. 

As with all models considered thus far, there exists a fairly even exchange of $L_{z}$ around the \CR{}, with the overall magnitude of the change significantly decreased (enhanced) for a spiral with halved (doubled) amplitudes. Orbits with the lowest initial $J_{R_0}$ have larger average decreases in $L_z$ in the vicinity of the \ilr{}, while overall increases in $L_z$ occur near the \olr{}.

Orbits do not experience changes in $J_{R}$ near the \CR{}, but orbits with the lowest initial $J_{R_0}$ show the strongest average increase in $J_{R}$ at the Lindblad resonances, especially at the \ilr{}. This trend is enhanced with increasing spiral amplitude. 

There are no strong trends in $\Delta J_{z}$.

\subsection{Trends with pattern speed}\label{ss:CR trends}

\begin{figure}[h]
    \includegraphics[width=\columnwidth]{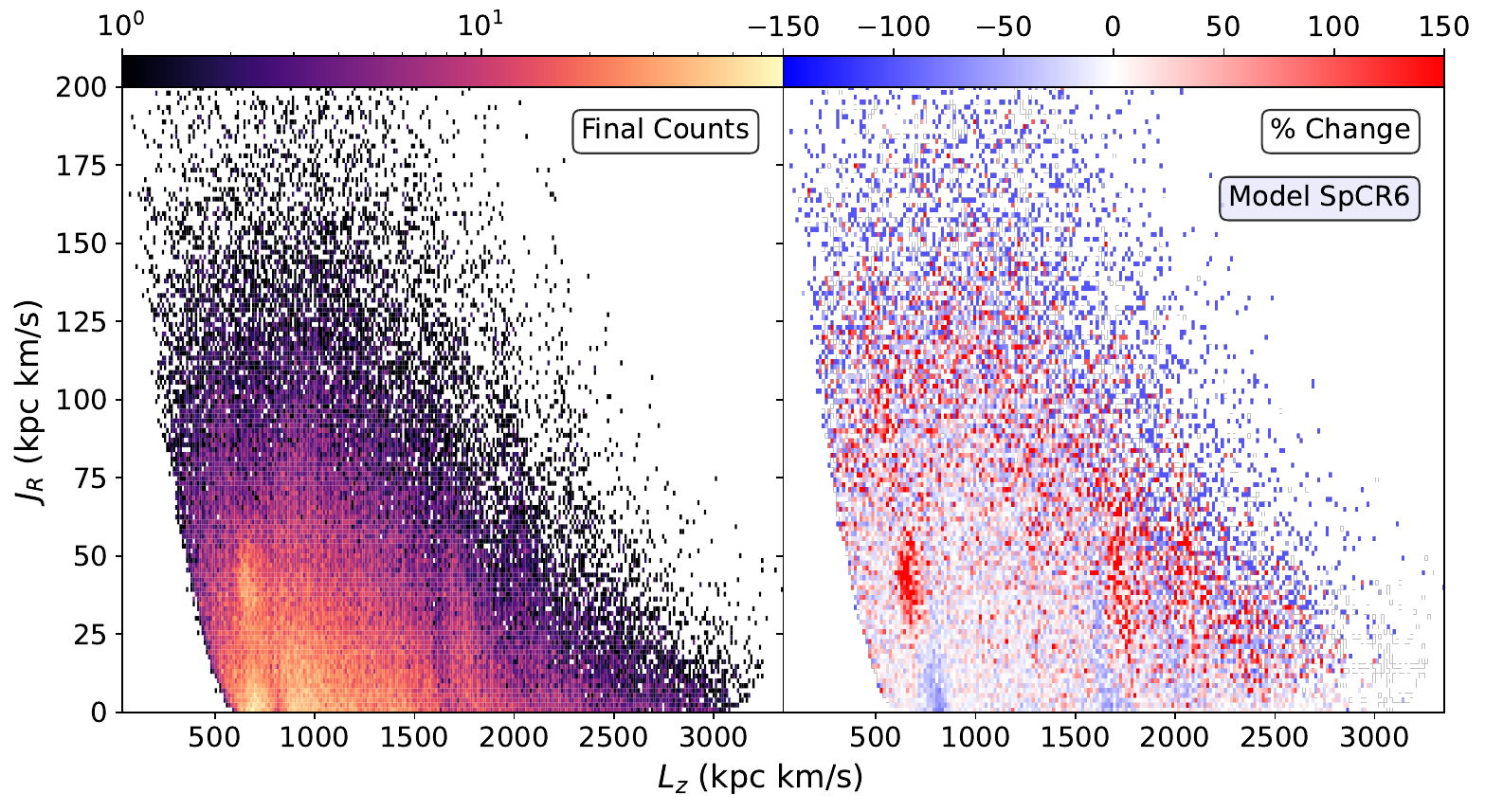} \\
    \includegraphics[width=\columnwidth]{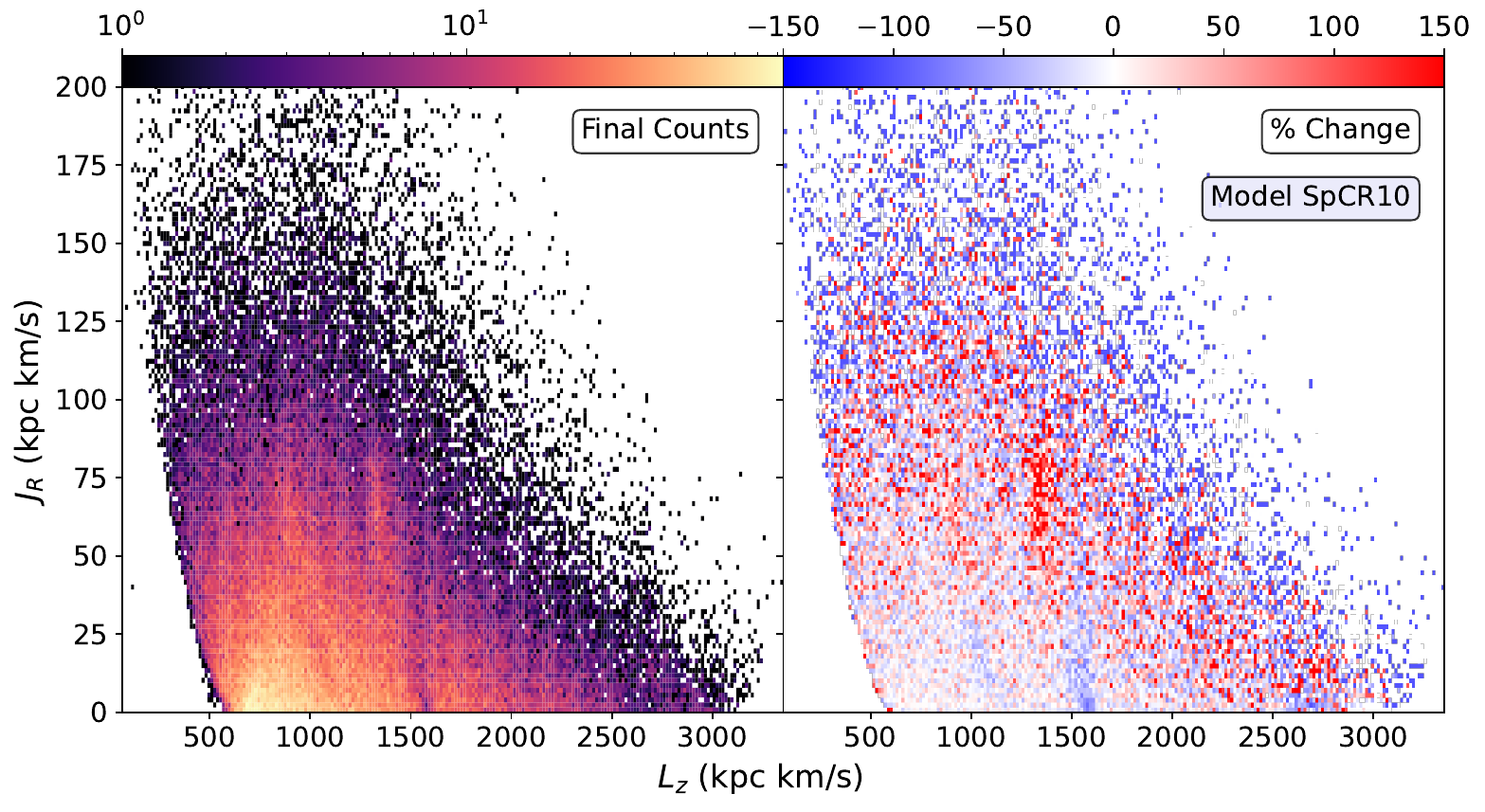}
    \caption{\refedit{Similar to Figure~\ref{fig:spAlpha30_init_fin_delta_counts}, showing stellar density counts in $J_{\rm R}-L_{\rm z}$ space after \sout{before} the occurrence of a spiral pattern (left) and the percent change in counts after the occurrence of the spiral pattern \sout{over time} (right) for models \SpCRsix{} (top) and \SpCRten{} (bottom), with the \CR{} set at 6 kpc and 10 kpc, respectively.}}
    \label{fig:spCR6CR10_fin_delta_counts}
\end{figure}

\begin{figure*}
\centering
    \includegraphics{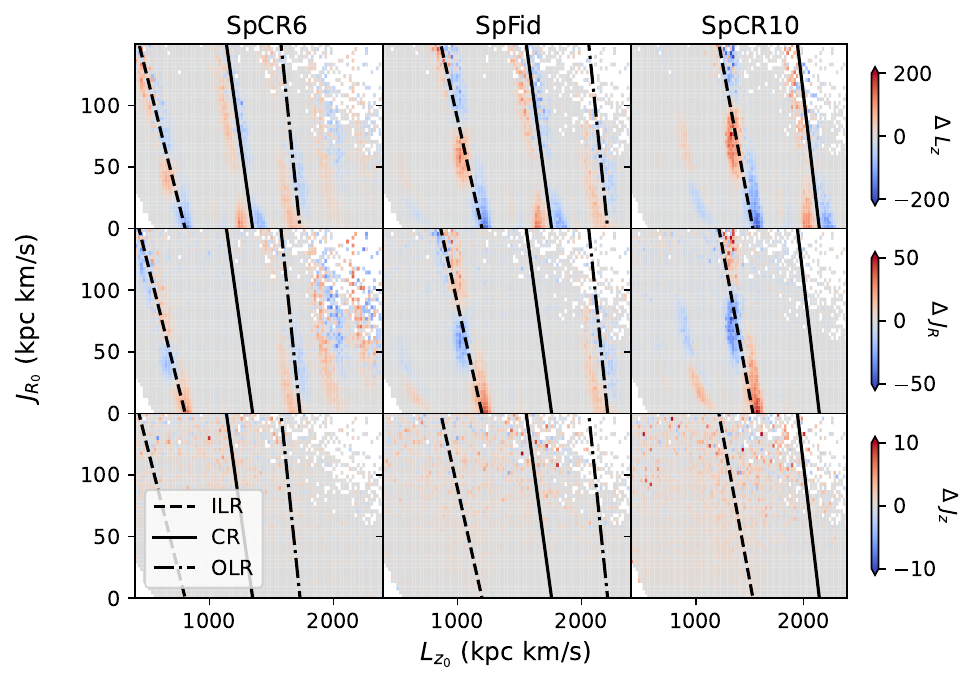}
  \caption{
  Average changes in actions $L_z$ (top), $J_R$ (middle), and $J_z$ (bottom) after the passage of a transient spiral pattern as a function of the initial action for models differing only in the pattern speed $\Omega_s$, which is related to the radius at which the spiral pattern corotates with disk orbits.  Axes are the same as in Figure~\ref{fig:delta_actions_for_varying_alphas}.
  }
  \label{fig:delta_actions_for_varying_CR}
\end{figure*}

\refedit{The spiral pattern speed $\Omega_s$ is related to the \CR{} radius $R_{\rm CR}$ through equation~\ref{eq:omegas}.  Model \SpCRsix{} has $R_{\rm CR}=6$~kpc, \SpFid{} has $R_{\rm CR}=8$~kpc, and \SpCRten{} has $R_{\rm CR}=10$~kpc. 
The peak amplitude of each spiral models is each set to be a 20\%  overdensity with respect to the underlying disk at the prescribed \CR{} radius. }
Figure~\ref{fig:spCR6CR10_fin_delta_counts} \refedit{shows signatures in the overdensity after a transient spiral passage, where Figure~\ref{fig:delta_actions_for_varying_CR}} illustrates signature trends with the spiral pattern speed.
\sout{$\Omega_s$, which is related to} 
\sout{the \CR{} radius $R_{\rm CR}$  through equation~\ref{eq:omegas}. Model \SpCRsix{} has} 
\sout{$R_{\rm CR}=6$~kpc, \SpFid{} has $R_{\rm CR}=8$~kpc, and \SpCRten{} has} 
\sout{$R_{\rm CR}=10$~kpc. 
The peak amplitude of each spiral models is} 
\sout{each set to be a 20\%  overdensity with respect to the underlying} 
\sout{ disk at the prescribed \CR{} radius.}

Trends in $\Delta J_{R}$, $\Delta L_{z}$, and $\Delta J_z$ are similar to those in \S\ref{ss:PA trends}. All models show a fairly even exchange of $L_{z}$ in the region surrounding the \CR{}, as expected. Orbits with the lowest initial $J_{R_0}$ values had more significant decreases in $L_{z}$ at the \ilr{}, with the opposite trend appearing much more mildly near the \olr{}.
We highlight once again that orbits with the lowest initial values for $J_{R_0}$ have the largest increases in $J_{R}$ in the vicinity of the \ilr{}. Orbits just near the \olr{} experience a very mild overall increase in $J_{R}$, while no discernible $\Delta J_{R}$ occurs around the \CR{}. 

The model with the slowest pattern speed, \SpCRten{}, has faint features at $L_z$ less than the \ilr{}, likely corresponding to a harmonic of the \ilr{}. Similarly, the model with the highest pattern speed, \SpCRsix{}, has features at $L_z$ higher than the \olr{}, likely corresponding to a harmonic of the \olr{}.  These features are also visible in the corresponding plots for \SpAdouble{} and \Spathirty.
No notable trends are seen in $\Delta J_{z}$.

\subsection{Trends with spiral lifetime}\label{ss:time trends}

\begin{figure}
    \includegraphics[width=\columnwidth]{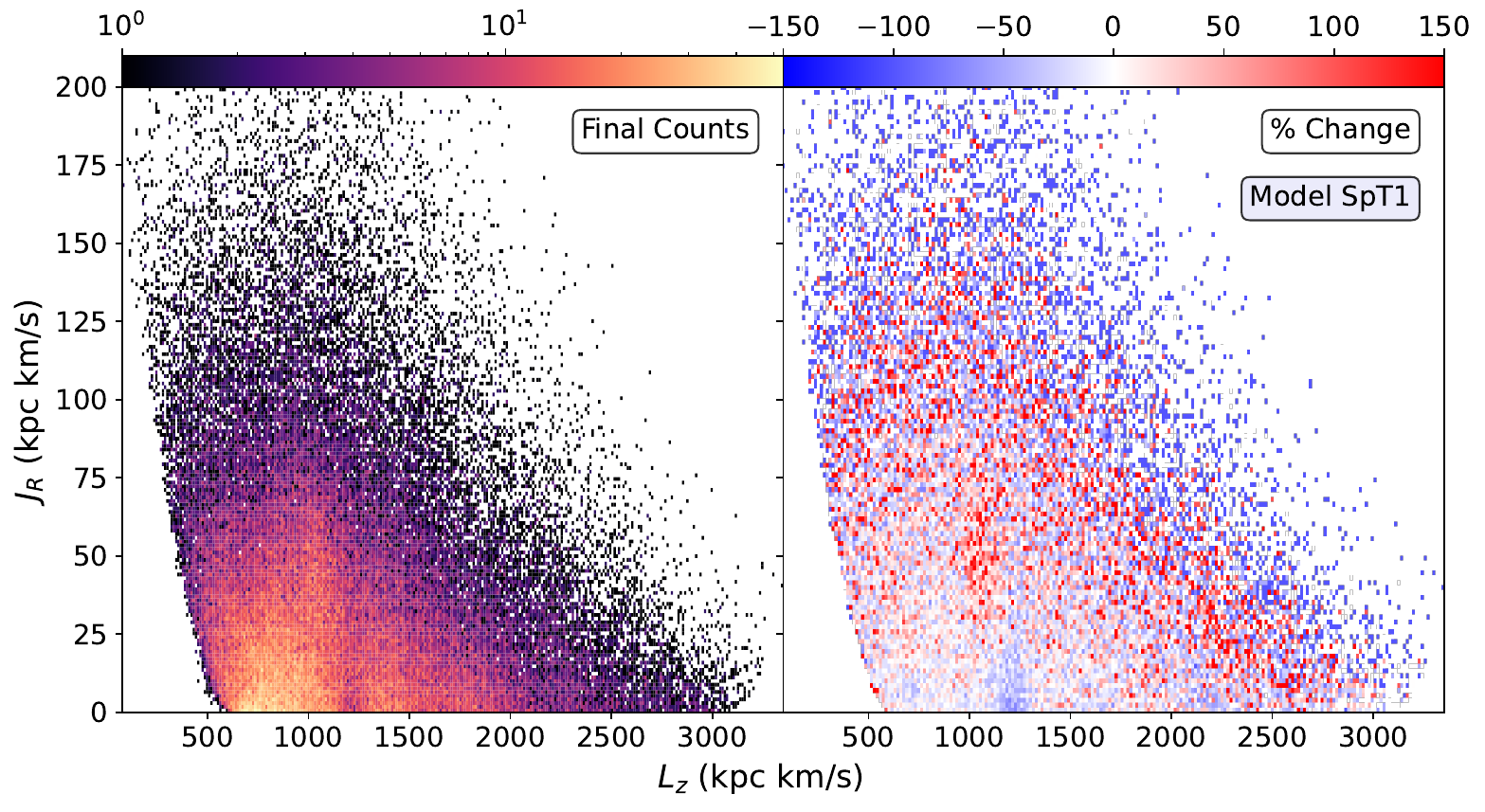} \\
    \includegraphics[width=\columnwidth]{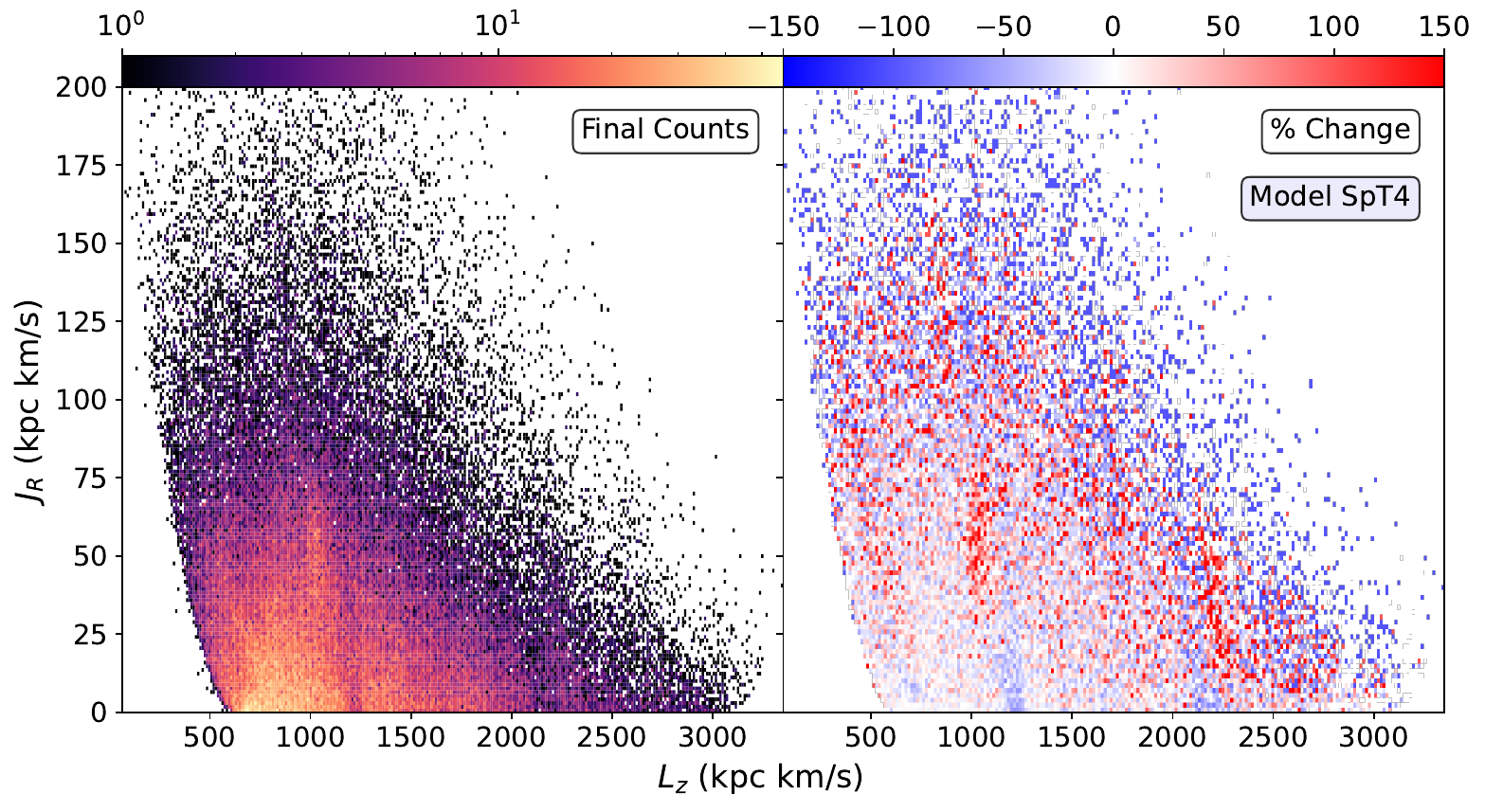}
    \caption{\refedit{Similar to Figure~\ref{fig:spAlpha30_init_fin_delta_counts}, showing stellar density counts in $J_{\rm R}-L_{\rm z}$ space after \sout{before} the occurrence of a spiral pattern (left) and the percent change in counts after the occurrence of the spiral pattern \sout{over time} (right) for models \SpTone{} (top) and \SpTfour{} (bottom), with spiral lifetimes one-half and twice that of the fiducial model, respectively.}}
    \label{fig:spTimeHalfDouble_fin_delta_counts}
\end{figure}

\begin{figure*}
\centering
  \begin{tabular}{@{}cccc@{}}
    \includegraphics{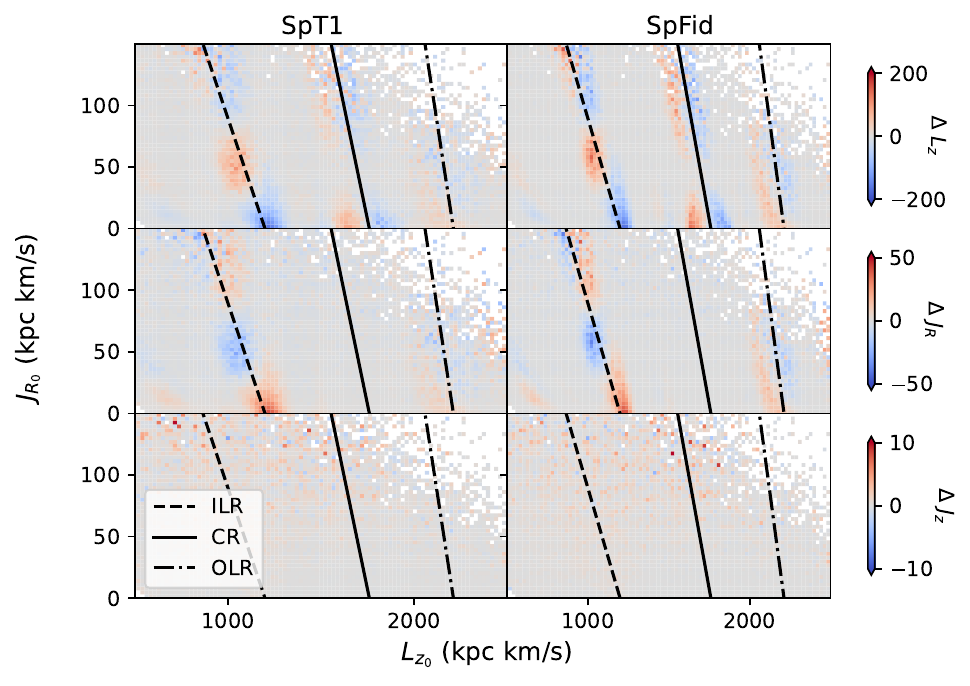} 
  \end{tabular}
  \caption{Average changes in actions $L_z$ (top), $J_R$ (middle), and $J_z$ (bottom) after the passage of a transient spiral pattern for \refedit{\SpFid{} and \SpTone{}, the model with spiral lifetime equal to one-half that of \SpFid{}.  Models \SpTfour{} and \SpTeight{} (not shown) had spiral lifetimes that are 2 and 4 times the lifetime of \SpFid{}, respectively.} Transient spirals with lifetimes longer than two dynamical times, such as \SpFid{}, showed no discernible differences from the fiducial model.  Axes are as shown in Figure~\ref{fig:delta_actions_for_varying_alphas}.
  }
  \label{fig:delta_actions_for_varying_time}
\end{figure*}

The signatures in action are more pronounced with longer transient spiral lifetime \refedit{up to that of \SpFid{}, which has a total spiral lifetime of $\sim$0.89 Gyr, or $4T_{\rm Dyn}$.  Longer spiral lifetimes leave signatures that qualitatively resemble those from \SpFid{}}.
\refedit{Figure~\ref{fig:spTimeHalfDouble_fin_delta_counts} demonstrates the signatures in density after the occurrence of a transient spiral pattern while Figure~\ref{fig:delta_actions_for_varying_time} compares trends in actions for} \sout{for model \SpFid{}, which has a total spiral lifetime of $\sim$0.89 Gyr }
\sout{or $4T_{\rm Dyn}$, alongside} models \SpTone{}, which has a total spiral lifetime of $\sim$0.45 Gyr, or $2T_{\rm Dyn}$, \refedit{and \SpFid{}, which has total spiral lifetime of $\sim$0.89 Gyr, or $4T_{\rm Dyn}$}.

Notable trends include the relatively even exchange of $L_{z}$ around the CR as expected, along with the largest increase in $J_{R}$ occurring for particles with lowest initial $J_{R}$ near the Lindblad resonances. These trends are enhanced as the lifetime of the spiral is increased. 

There are no noteworthy trends in $J_{z}$ with change in spiral lifetime.

\subsection{Density wave versus winding spiral}\label{ss:model type trends}

\begin{figure*}
\centering
  \begin{tabular}{@{}cccc@{}}
    \includegraphics{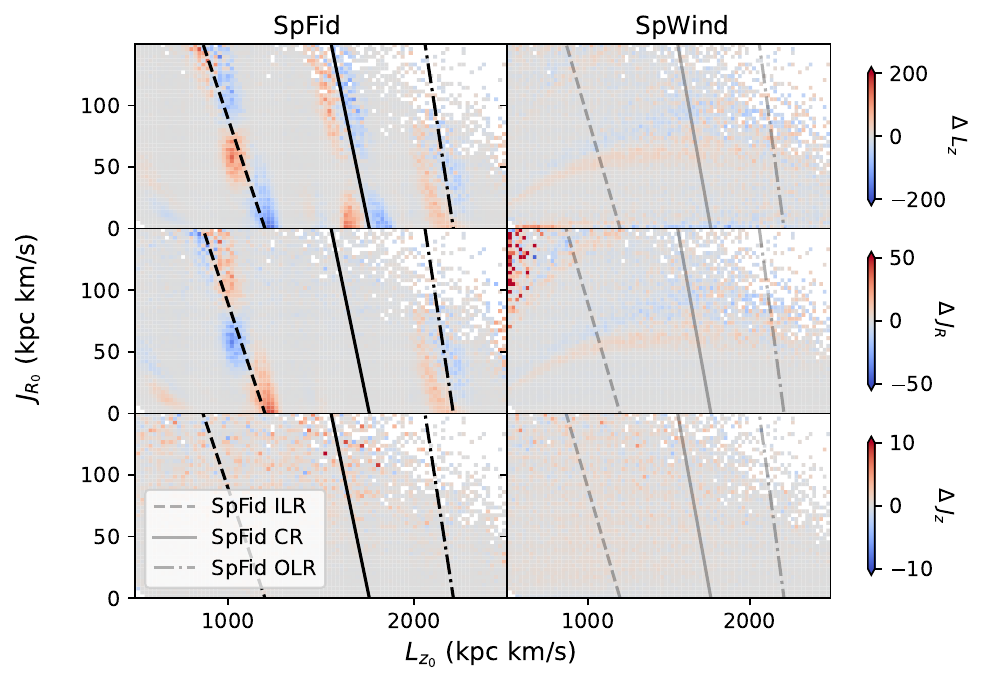} 
  \end{tabular}
  \caption{Average changes in actions $L_z$ (top), $J_R$ (middle), and $J_z$ (bottom) after the passage of a transient spiral pattern as a function of initial action for the fiducial density-wave-like model (\SpFid{}) compared to a model of a corotating or winding spiral pattern (\SpWind{}). Axes are as shown in Figure~\ref{fig:delta_actions_for_varying_alphas}. The resonance lines for \SpFid{} are superposed (gray) on \SpWind{} data for visual reference.
  }
  \label{fig:delta_actions_for_varying_SpType}
\end{figure*}

Figure~\ref{fig:delta_actions_for_varying_SpType} compares the signature trends in action space for the fiducial nonwinding spiral pattern (\SpFid{}) and a transient corotating spiral pattern (\SpWind{}). Since the winding spiral corotates at all radii the location of the resonances cannot meaningfully be plotted in $L_z-J_R$ space.  However, the resonances for \SpFid{} are superposed onto the plots for \SpWind{} for reference.
\refedit{There were no discernible postspiral wrinkle signatures in the density of star particles in the $L_z-J_R$ plane.}

The \SpWind{} model did not show any significant trends in $\Delta J_{R}$, $\Delta L_{z}$, or $\Delta J_{z}$ other than a weak arc across the space for trends in increasing and decreasing $L_z$ and $J_R$.  Significantly, \texttt{\SpWind{}} did not produce a wrinkle.
\refedit{However, \cite{SellwoodTrick19} note that when a winding perturbation is sufficiently strong and long-lived, broad features can appear (see Figure~12).  Our spiral is different in character, simulating the trailing phase only, but neither produces a signature in $J_R-L_z$ space that we would \refedit{characterize} \sout{characterized} as a wrinkle that could clearly be identified in observational data within 200~pc of the Sun.}

\refedit{In an experiment identical to \SpWind{} except with twice the spiral amplitude, we found trends similar to those shown in Figure~\ref{fig:delta_actions_for_varying_SpType} but more pronounced. There were no signatures resembling a wrinkle. The features produced, however, are interesting, and warrant future investigation.}

\section{Discussion}\label{sec:discussion}


The distribution of stars in $L_z-J_R$ space in the local \mw{} (Figure~\ref{fig:gaia_clustering}) \sout{demonstrates} \refedit{exhibits} a variety of overlapping features that are the combined signatures from multiple progenitors \citep{SellwoodTrick19}. \refedit{Resonant signatures from transient spiral patterns in local kinematics have been explored in several studies \citep[e.g.,][]{Dehnen98, SB02, Antoja09, Hunt19, SellwoodTrick19, Trick19}. In this study, we limit our discussion to focus mainly on the kinematic response at the \ilr{}, given that changes in $J_R$ from the \ilr{} are strongest \citep[\refedit{as} predicted by][]{Sellwood10b} and produce a characteristic wrinkle that continues to exist many dynamical times after a transient spiral pattern dissipates.} \sout{the wrinkles discussed in this work are the particular signature produced at the \ilr{} of a transient spiral pattern \citep[predicted by][]{Sellwood10b}}. Examinations of local \mw{} kinematics have indicated that disentangling the progenitors of such features can be challenging at best \citep{Hunt19}. 
The following discussion explores how novel insight into the history of perturbations to the \mw{} may be possible when also considering stellar age.


\subsection{Kinematic age}\label{s:kinematic age}

Young stars are, by \sout{in} \refedit{and} large, born on nearly circular orbits \citep{Jeans1915, Chandrasekhar42Book, Shu77}. Observational studies have also show\refedit{n} that local young stars in the \mw{} are born on nearly circular orbits \cite{StarkBrand89,Kuhn19}. It follows that a selection of young disk stars would have a relatively low velocity dispersion, often described as being kinematically cold. Over time, stellar populations tend to experience a steady increase in velocity dispersion due to many interactions with perturbations to the smooth disk, such as giant molecular clouds (GMCs) and spirals arms, over their lifetimes \citep{Stromberg46,Roman50a,Roman50b,Wielen77,Nordstrom04,SeabrokeGilmore07,Soubiran08,Casagrande11,SandersDas18,Mackareth19,TingRix19}. 
The resulting increase in the velocity dispersion of a stellar population is often referred to as kinematic heating.  


There is mounting evidence that stellar populations were born with higher initial velocity dispersions (i.e., with higher kinematic temperatures) at early times than is the case in the current epoch.\footnote{\refedit{Given that the \mw{} potential has presumably evolved significantly over the last ${\sim}10~Gyr$, including a considerable increase in total mass, it is not obvious that higher initial velocity dispersion equates with higher initial $J_R$.}} This population would then be further heated after birth. At later times, young populations are born with relatively lower initial velocity dispersions (i.e. lower kinematic temperatures), which then increase monotonically with age \citep{Bird21,McCluskey24}.  Whether or not there is a time dependence for birth velocity dispersion, older populations \sout{will} will have a higher velocity dispersion than younger populations when looking at current kinematics in the \mw{}.

Many mechanisms for kinematic heating have been identified.
Scattering from lumps in the mass distribution of the disk, such as GMCs, kinematically heat disk populations in all directions \citep{SpitzerSchwarzchild51,SpitzerSchwarzchild53,Lacey84,JB90}. External satellite bombardment \citep{Quinn93,VW99,Kazantzidis08,Bird12} can also increase the vertical and in-plane velocity dispersions of disk populations \citep{Bird12}, but the primary contributor to kinematic heating in the radial direction is expected to be transient spiral arms \citep{SC84,CS85,JB90,SC14,SellwoodMasters22}.
Repeated exposure to internal perturbers, such as bars and spirals \citep{BW67,LBK72,SC84,CS85,Weinberg94,Dehnen2000,SB02,Roskar08a,Roskar08b,MinchevFamaey10,Loebman11,Hunt19}, along with the intersection of their resonances \citep{Minchev11, Daniel19}, drive vigorous kinematic heating. 
While the overall \sout{affect} \refedit{effect} of kinematic heating is generally a slow and steady process \citep{Chandrasekhar42Book,SpitzerSchwarzchild51,Wielen77,Mackareth19}, permanent orbital changes, such as those explored in this paper, can occur over much shorter timescales.

As stellar populations age and undergo kinematic heating in the radial direction, their radial actions on average increase \citep{DehnenBinney98b,Beane18}. It is instructive to adopt the following toy model in order to gain insight into an approximate scaling relation between radial action and stellar age. 

In the \epi{} approximation \citep[see, e.g.,][their Sections \S3.2.3 and \S3.5.3]{BT08}, $J_{R}$ can be expressed as
\begin{equation}
    J_{\mathrm{R}} = \dfrac{E_{\mathrm{nc}}}{\kappa}.
\end{equation}
The noncircular energy of a population of stars with age $\tau$ can be approximated as 
\begin{equation}
    E_{\rm nc} \approx \frac{1}{2} \sigma_{\mathrm{R}}(\tau)^2 .
\end{equation}
Thus, the probable radial action of a star from a population born some time $\tau$ before present can be estimated to scale as 
\begin{equation}\label{eqn:JrVsigmaR}
    J_{\mathrm{R}} \approx \dfrac{\sigma_{\mathrm{R}}(\tau)^2 }{2\kappa} .
\end{equation}

\cite{SandersBinney15} used this approach to develop a model for a \mw{}-like thin disk (see their \S5 and Table~5 for a full description and adopted values).
They found that the age of a stellar population in the \mw{}'s thin disk is related to its radial velocity dispersion by \citep{Binney10,SandersBinney15} 
\begin{equation}\label{eqn:AVR}
    \sigma_{\rm R}(\tau) = \left(\dfrac{\tau + \tau_1}{\tau_m + \tau_1}\right)^{\beta_{\rm R}} \sigma_{\rm R}(R_{\rm c}),
\end{equation}
where $\tau_{1}=110~\rm{Myr}$, $\tau_{\rm m}$ is the age of the Galaxy, and $\beta_{\rm R}=0.33$ describes the efficiency of stochastic heating. Their form for the radial profile of the radial velocity dispersion was defined to be
\begin{equation}\label{eqn:AVR_initial}
    \sigma_{\rm R}(R_{\rm c}) = \sigma_{\rm R0} \mathrm{e}^{(R_{0}-R_{\rm c})\mathbin{/}R_{\sigma_{\rm R}}} ,
\end{equation}
where the scale length is $R_{\sigma_{\rm R}}\sim 2R_{\rm d}$. Here $\sigma_{\rm R0}$ indicates the radial velocity dispersion at the solar radius.  

Using the above model, \cite{Frankel20} obtained the relation
\begin{equation}\label{eqn:FrankelAVR}
    \sigma_{\rm R}(\tau, R_{\rm c}(L_{\rm z})) = \sigma_{\rm R0} \left(\dfrac{\tau + \tau_1}{\tau_m + \tau_1}\right)^{\beta_{\rm R}} \mathrm{e}^{(R_{0}-R_{\rm c})\mathbin{/}R_{\sigma_{\rm R}}} .
\end{equation}
Their parameterized model, which was fit to APOGEE red clump stars in the \snd{} ($R_{0}=8$~kpc), found that the rms radial action for a population of given age $\tau$ can be expressed as \citep{Frankel20}

\begin{equation}\label{eqn:FrankeljR}
    \sqrt{\langle \Delta J_{\rm R} \rangle^2} \approx 63 \text{ kpc km s}^{-1}  \left(\dfrac{\tau}{6 \text{ Gyr}}\right)^{0.6}.
\end{equation}

\begin{figure*}
\centering
  \begin{tabular}{@{}cccc@{}}
    \includegraphics{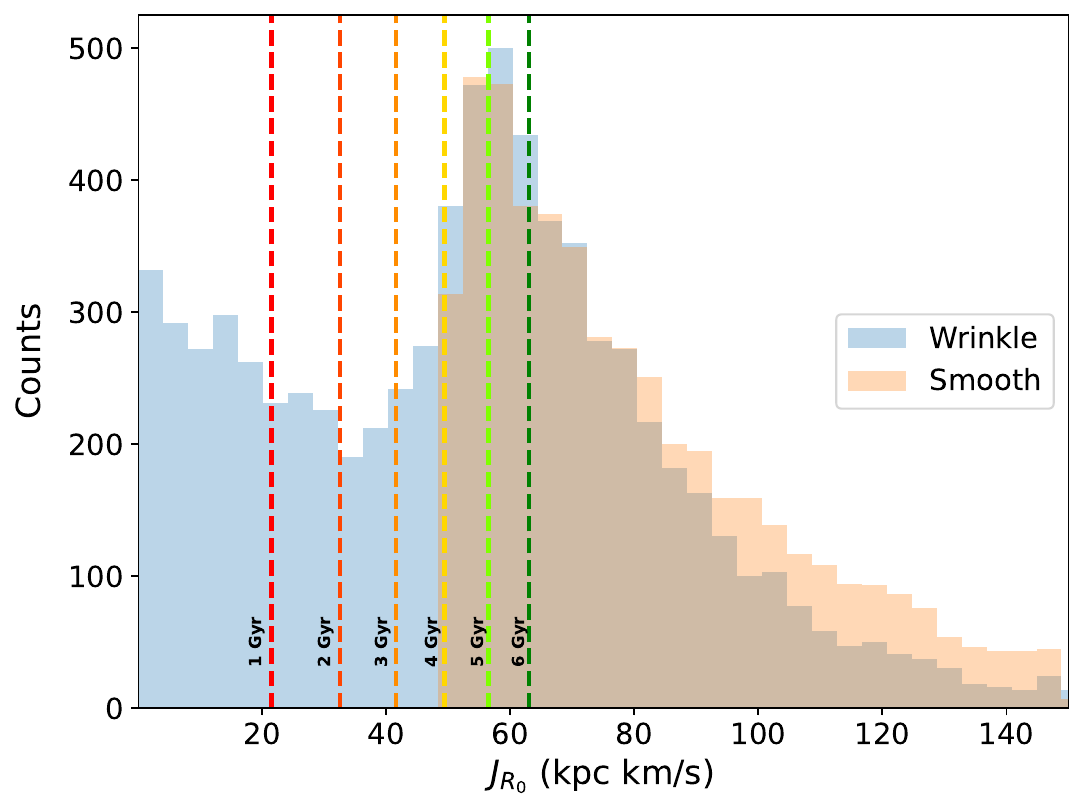}
  \end{tabular}
  \caption{Histogram showing the \refedit{distribution of initial radial actions $J_{\rm R_{0}}$ in the \refedit{region of $L_z-J_R$ space} \sout{of interest} where a wrinkle forms (indicated by the green box in Figure~\ref{fig:spAlpha30_init_fin_delta_counts}) in model \Spathirty{}. The orange distribution is from the unperturbed, smooth population prior to the introduction of a spiral perturbation. The blue distribution is from the same region in $L_z-J_R$ space after a wrinkle formed from the growth and decay of a transient spiral pattern.} 
  Vertical lines indicate the associated ages from Equation~\ref{eqn:FrankeljR} \citep{Frankel20}.  The \refedit{postresonant} wrinkle population has a significant young stellar population (blue), where this region would otherwise be dominated by intermediate-age stars (orange).}
  \label{fig:histogram_for_WrinkleVsSmooth}
\end{figure*}

\refedit{Figure~\ref{fig:histogram_for_WrinkleVsSmooth} shows a histogram distribution of $J_{\rm R_0}$ for the stars within the \sout{wrinkle} $L_z-J_R$ space region demarcated by the green rectangles in Figure~\ref{fig:spAlpha30_init_fin_delta_counts}.}  The values for $J_{\rm R_0}$ can be seen as a proxy for the ages of stars from before the passage of a spiral.  
We overplot these approximate ages for a given radial action based on equation~\ref{eqn:FrankeljR}. 
While stellar populations are born with some small initial velocity dispersion, most orbits will have $J_{\rm R_{0}}$ close to zero at birth and so we approximate the value for $\sqrt{\langle \Delta J_{\rm R} \rangle^2}$ at birth to be zero. 

\refedit{Our toy model implies that near-zero-age stars could significantly contribute to the population of stars contributing to the wrinkle overdensity, where orbits with similar values for $J_R$ are otherwise dominated by intermediate-age stars.}
\sout{Our toy model implies that the \refedit{composition of} a wrinkle could have a significant contribution from near-zero-age stars in a space where the population is otherwise dominated by intermediate age stars.} 
\refedit{The wrinkle signatures in this study are produced on relatively short timescales (e.g., just two dynamical times, 2$T_{\rm Dyn}=0.89~{\rm Gyr}$, for most models) compared to the time} 
\sout{which is significantly less than the time} 
\refedit{it would take a near-zero age population to otherwise reach the same $\langle \Delta J_{\rm R} \rangle$ (e.g., it takes longer the 5~Gyr for an average stellar population to have $\langle \Delta J_{\rm R} \rangle=60~kpc~km~s^{-1}$). 
The \ilr{} of a transient spiral pattern thus provides a mechanism by which dynamically younger stars can have values for $J_R$ that are higher than expected for their age.  Presumably, young stars would therefore preferentially fill the wrinkles in the observed solar neighborhood kinematics (see Figure~\ref{fig:gaia_clustering}).}

\begin{figure*}
\centering
  \begin{tabular}{@{}cccc@{}}
    \includegraphics{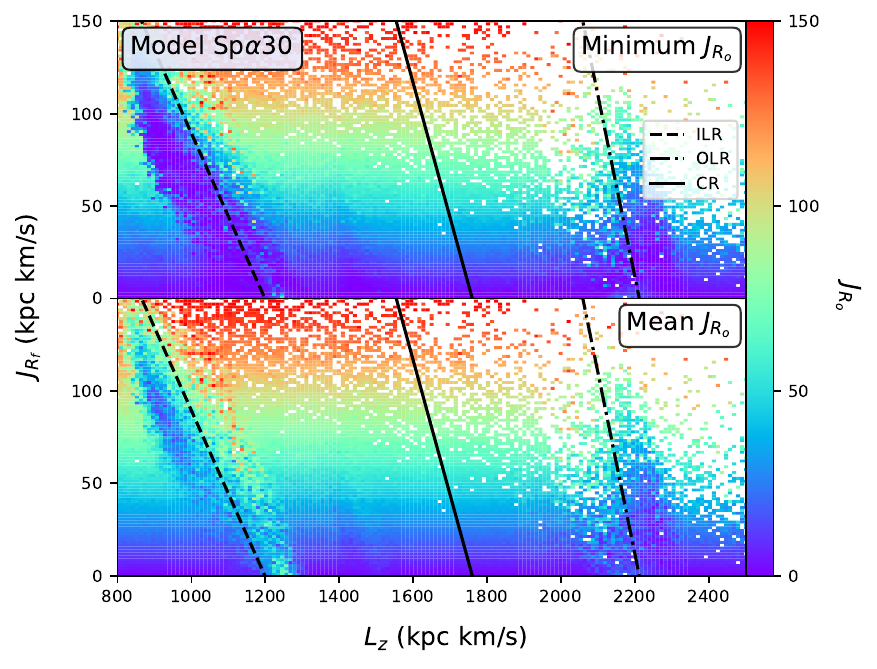} 
  \end{tabular}
  \caption{Final $J_{\rm R}$ vs. final $L_{\rm z}$ distribution from the \Spathirty{} simulation. Colors indicate minimum {(top)} and mean {(bottom)} values of $J_{\rm R_{0}}$. The locations of the \ilr{} (dashed line), the \CR{} (solid line), and the \olr{} (dot-dashed line) are also shown.  Wrinkle populations at the \ilr{} are dominated by stars with initial values of $J_{\rm R_{0}}\sim0$, a good proxy for populations with age $\tau\sim0$.
  }
  \label{fig:MinMean_JR0_SpAlpha30}
\end{figure*}

Figure~\ref{fig:MinMean_JR0_SpAlpha30} shows the wrinkle that formed in the \Spathirty{} model with color indicating the minimum and mean values of $J_{\rm R_{0}}$ in a given bin of the density distribution. The wrinkle is dominated, by both measures, by stars with initial $J_{\rm R_{0}}$ near zero.

In combination with Figure~\ref{fig:histogram_for_WrinkleVsSmooth}, this demonstrates wrinkles are primarily populated by stars with significantly younger ages than those of field stars at similar $J_{\rm R_{0}}$. We should therefore expect to find dynamically young stars within \mw{} wrinkles in the \snd{}.

\subsection{Wrinkles from Winding Spirals}\label{s:winding wrinkles}
\refedit{Several studies have investigated phase-space perturbations from a winding spiral. For example, \cite{SellwoodTrick19} found that a swing-amplified spiral produced broad features in $L_z-J_R$ space. The spiral adopted in that study was different in character from what is used for \SpWind{} since \cite{SellwoodTrick19} followed the spiral from the fully leading to fully trailing phases. It is not obvious at which stage the perturbations in $L_z-J_R$ grew. In this study, the winding spiral only has significant amplitude from a trailing pitch angle starting at about $35^\circ$ and winding down to about $15^\circ$.}

\refedit{\cite{Hunt19} used a similar prescription for a winding spiral to that used for \SpWind{}, but their disk also included various prescriptions for a bar. The nonlinear response from these overlapping structures and their resonances produced signatures that are not present in our experiment, which has only one perturbation. We also note that the physics driving the redistribution of orbits in the case of a winding spiral are fundamentally different from a density wave, where a density wave produces a distinct signature at the \ilr{} (and other resonances) in $L_z-J_R$ space.}

\refedit{Since this paper is focused on identifying clear wrinkle-like signatures in $L_z-J_R$ space that are potentially correlated with the ages of its stellar members, we have not further pursued the winding spiral case and its signatures. Rather, we show that such a spiral does not alone produce a clear signature resembling a wrinkle.  An extension of the work by \cite{SellwoodTrick19} that more closely examines the differences in how orbits change near winding spirals, multiple overlapping modes, density waves, or any of these overlapping would be quite interesting, though beyond the scope of this paper. For this reason, we also refrain from commenting on the likelihood (or not) that the observed distribution of stars within 200 pc of the Sun sufficiently rules out (or reveals) signatures from past winding spirals.}

\subsection{Observable Signatures}\label{s:observable signatures}

The above models predict that a significant number of young stars in highly eccentric orbits could be found in wrinkles formed by spiral patterns. Since wrinkles must have finished forming \textit{after} their youngest stellar members are born, the youngest stellar members could place a constraint on the formation time of a wrinkle. A smoking gun signature for this formation mechanism is that the ages of stars in wrinkles are younger by a few gigayears than surrounding populations with the same $J_{\rm R}$. 

In the first paper in this series, Paper~1 \citep{Rampalli23} used gyrochronology to find that that stars as young as a few hundred million years can be found proximate to wrinkles with $J_{\rm R}\gtrsim 120$~kpc~km~s$^{-1}$. This finding is consistent with this work. 
In forthcoming papers in this series, we will use clustering methods to isolate wrinkle stars in the \snd{} and use stellar ages to explore potential constraints that can be placed on the history of spiral structure in the \snd{}.


\section{Conclusions}\label{sec:conclusions}

This work made use of spiral galaxy models to investigate trends in postresonant signatures from transient spirals.  Our results are as follows.

\begin{itemize}
  \item We created a set of tracer particle models to investigate the impact of various spiral parameters on postresonant signatures in action space.
  \item We confirm the findings first proposed by \cite{Sellwood10b} and further explored in \cite{SellwoodTrick19} that wrinkles can form at the \ilr{} of a spiral pattern. \sout{We also find that the winding spiral arm model had minimal impact on action space and did not produce a wrinkle.} \refedit{We also find that the winding spiral arm model did not produce signatures in action space that resembled what we identified as wrinkles.}
  \item We find that dynamically younger particles, characterized by having the lowest initial $J_{R_0}$, preferentially \sout{from} \refedit{populate} the wrinkles at the \ilr{} of a transient spiral.
  \item High-$J_{R}$ wrinkles can form from a relatively short-lived perturbation and persist indefinitely. 
  \item The inclusion of stellar age as an additional dimension in observational analysis is a promising new avenue for constraining the dynamical history in the \snd{}. 
  \item We predict that finding a significant number of young stars within observed wrinkles could strengthen the claim for a recent occurrence of a transient spiral \ilr{}.
\end{itemize}


\section*{Acknowledgements} \label{sec:Acknowledgements}

\refedit{We thank the anonymous referee for their review of this work, which has significantly improved its presentation.}
This work was funded by the NASA ADAP (21-ADAP21-0134).
A.S. thanks Shambhavi Srivastava for assistance with simulations, and Jason Hunt for constructive conversation on modeling and analysis.
K.J.D. acknowledges that this work was performed in part at Aspen Center for Physics, which is supported by National Science Foundation grant PHY-2210452.
K.J.D. and E.N. thank the Research Corporation for Scialog.
R.R. was supported by the NSF Graduate Research fellowship (DGE-2236868). 
K.J.D., A.S., \refedit{and L.C.} acknowledge support from the Heising-Simons Foundation grant \# 2022-3927. 

We respectfully acknowledge the University of Arizona is on the land and territories of Indigenous peoples. Today, Arizona is home to 22 federally recognized tribes, with Tucson being home to the O’odham and the Yaqui. The University strives to build sustainable relationships with sovereign Native Nations and Indigenous communities through education offerings, partnerships, and community service. 
Dartmouth is situated upon the ancestral and unceded lands of the Abenaki people. This acknowledgment reminds us of the significance of place, the continued existence of Indigenous people, and Dartmouth's commitment to building respectful relationships with those who call these lands home today. 
We respect and honor the ancestral caretakers of the land, from time immemorial until now, and into the future.

\vspace{5mm}
\facilities{Gaia \citep{gaiamission}, AAT GALAH Survey \citep{deSilva15GALAH}, Sloan APOGEE Survey \citep{Majewski17APOGEE}, TESS \citep{Ricker15}}

\software{this work made use of \texttt{astropy}:\footnote{{\url{http://www.astropy.org}}} a community-developed core Python package and an ecosystem of tools and resources for astronomy \citep{astropy:2013, astropy:2018, astropy:2022}. All tracer particle models were built using \texttt{galpy}{\footnote{{\url{http://github.com/jobovy/galpy}}}} (v.1.7, \citealt{Bovy15}), a Python package used for the study of Galactic dynamics. This work also utilized \texttt{scipy} \citep{scipy}, \texttt{numpy} \citep{numpy}, and \texttt{matplotlib} \citep{matplotlib}. 
\sout{The code to locate and plot the \refedit{lines near} spiral resonances \refedit{(similar to \arl s)} was developed by Olivia McAuley for \cite{McAuley23} and can be accessed at http://github.com/oliviamcauley/Resonance\_Lines. \footnote{{\url{http://github.com/oliviamcauley/Resonance_Lines}}}}
}





\setlength{\abovecaptionskip}{2pt}
\setlength{\belowcaptionskip}{2pt}
\setlength{\floatsep}{4pt}
\setlength{\textfloatsep}{4pt}
\setlength{\intextsep}{4pt}

\begin{figure*}[p]
\centering

\newcommand{\panelheightA}{0.44\textheight}

\setlength{\tabcolsep}{2pt}

\begin{tabular}{cc}
  \includegraphics[width=0.49\textwidth,height=\panelheightA,keepaspectratio]
    {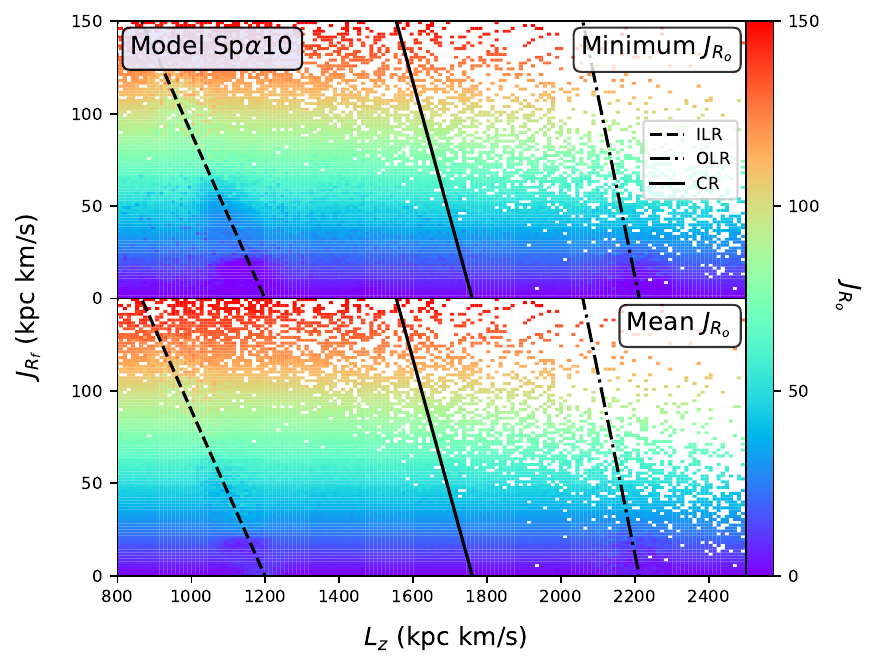} &
  \includegraphics[width=0.49\textwidth,height=\panelheightA,keepaspectratio]
    {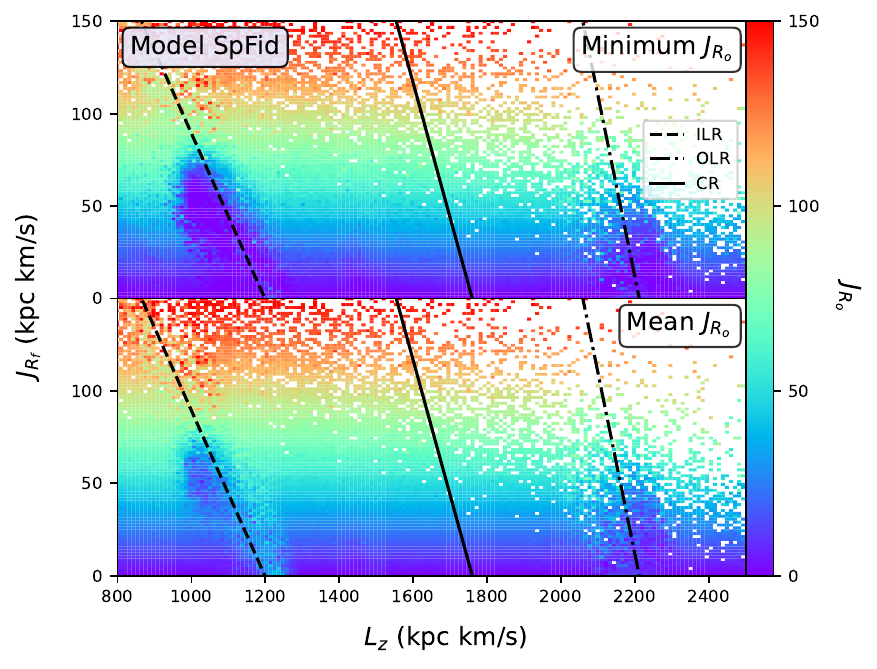} \\[-2pt]
  \includegraphics[width=0.49\textwidth,height=\panelheightA,keepaspectratio]
    {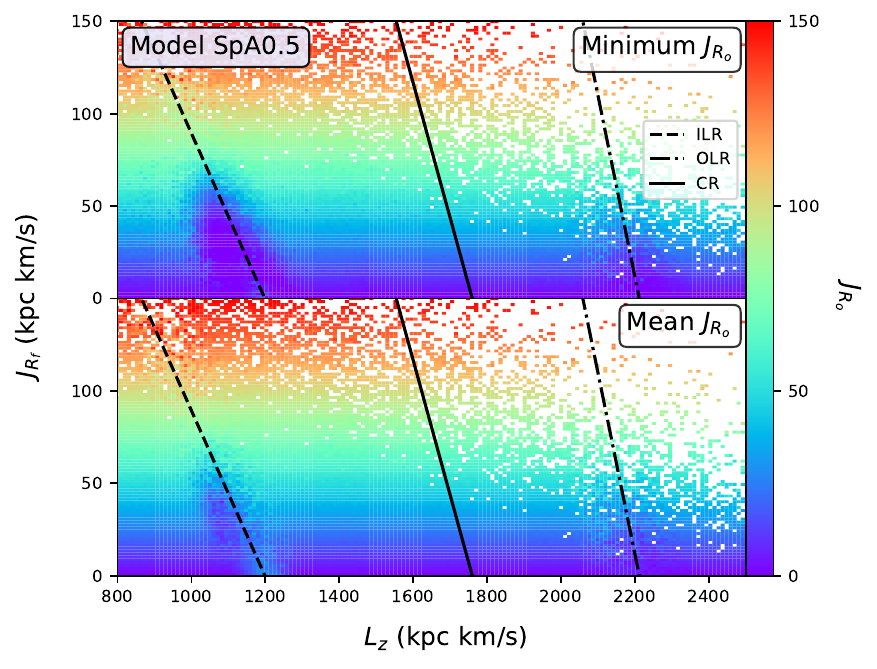} &
  \includegraphics[width=0.49\textwidth,height=\panelheightA,keepaspectratio]
    {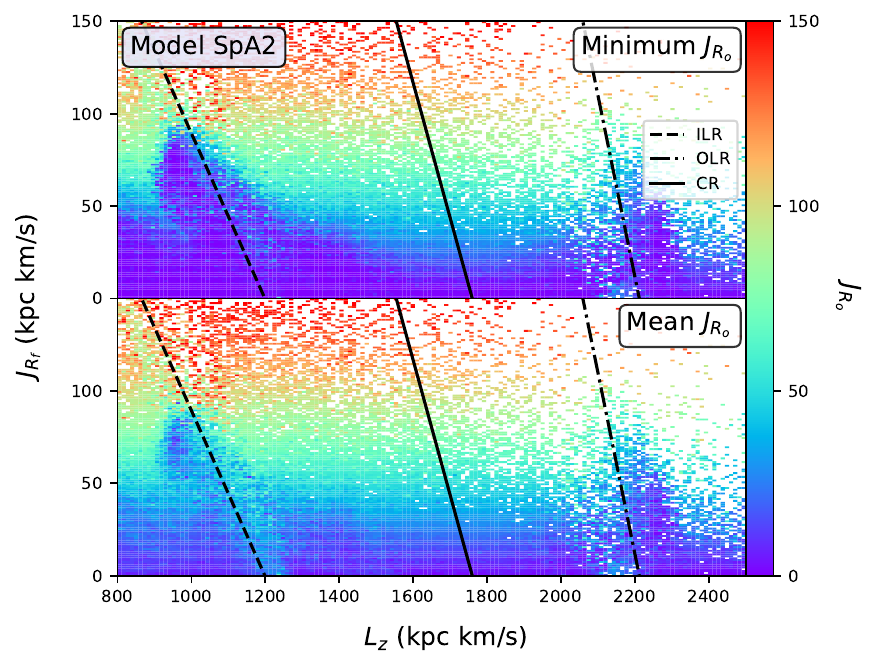}
\end{tabular}

\caption{Final $J_{\rm R}$ vs. final $L_{\rm z}$ distribution from \sout{all} \refedit{select} simulations. 
Colors indicate the minimum \textbf{(top)} and mean \textbf{(bottom)} values of $J_{{\rm R}_{0}}$. 
The locations of the \ilr{} (dashed line), the \CR{} (solid line), and the \olr{} (dot-dashed line) are also shown.}
\label{fig:MinMean_JR0_page1}
\end{figure*}

\begin{figure*}[p]
\centering

\newcommand{\panelheightB}{0.29\textheight}

\setlength{\tabcolsep}{2pt}

\begin{tabular}{cc}
  \includegraphics[width=0.49\textwidth,height=\panelheightB,keepaspectratio]
    {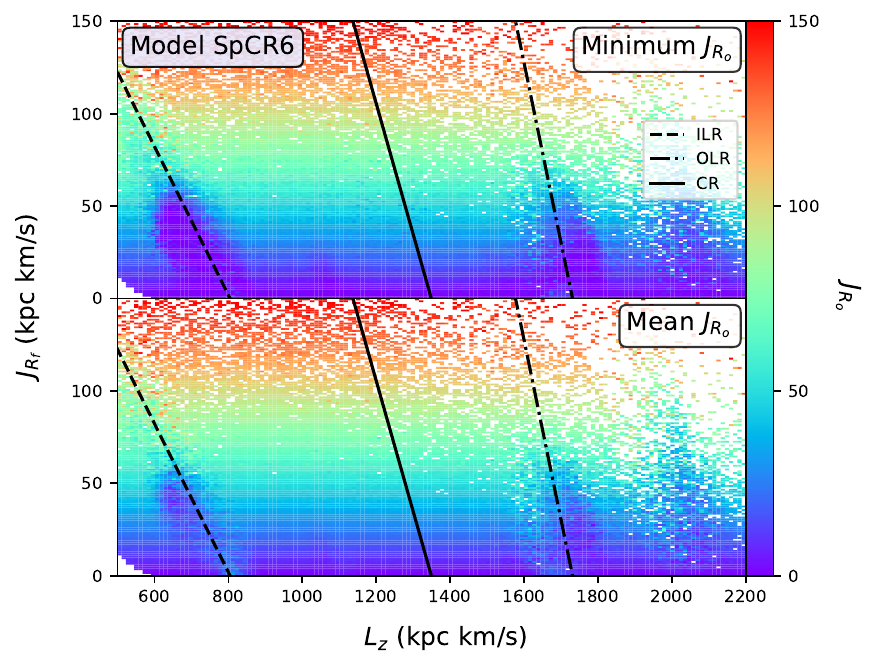} &
  \includegraphics[width=0.49\textwidth,height=\panelheightB,keepaspectratio]
    {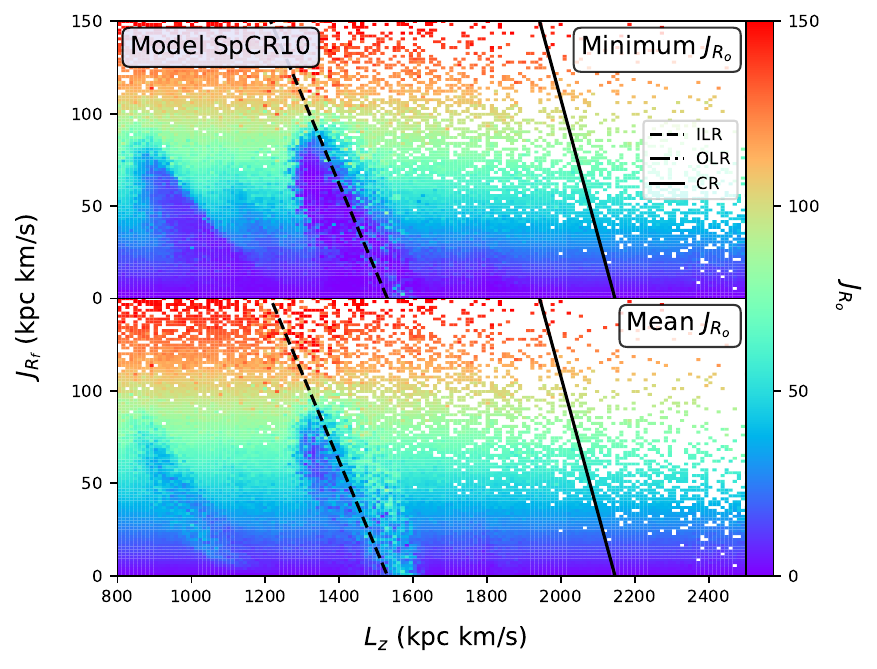} \\[-2pt]
  \includegraphics[width=0.49\textwidth,height=\panelheightB,keepaspectratio]
    {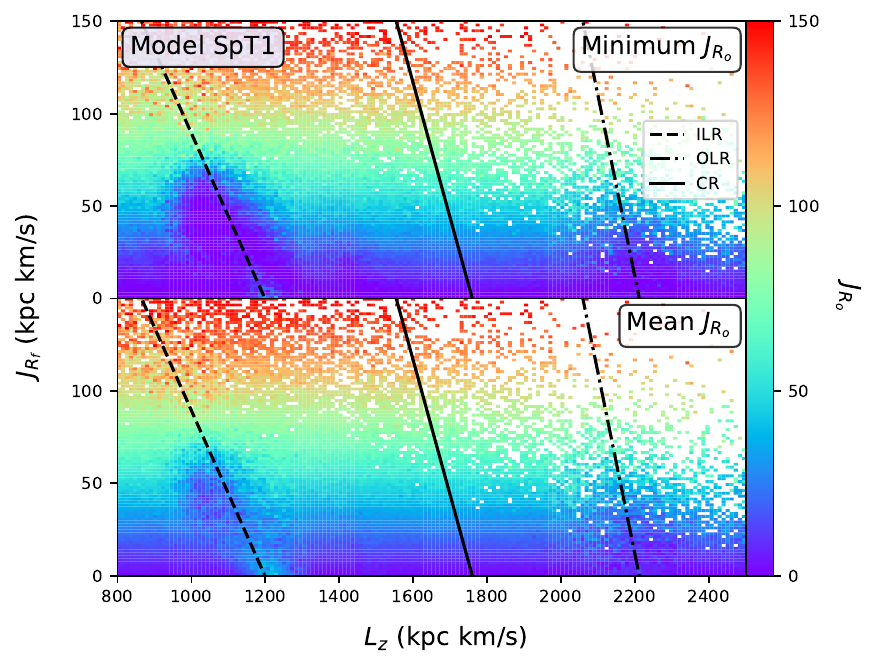} &
  \includegraphics[width=0.49\textwidth,height=\panelheightB,keepaspectratio]
    {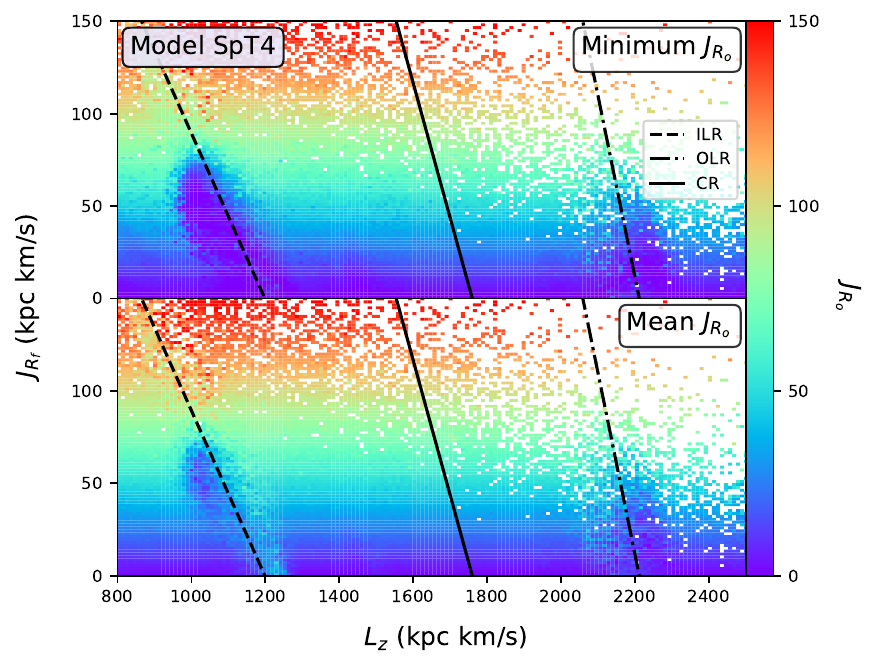} \\[-2pt]
  \multicolumn{2}{c}{
    \includegraphics[width=0.49\textwidth,height=\panelheightB,keepaspectratio]
      {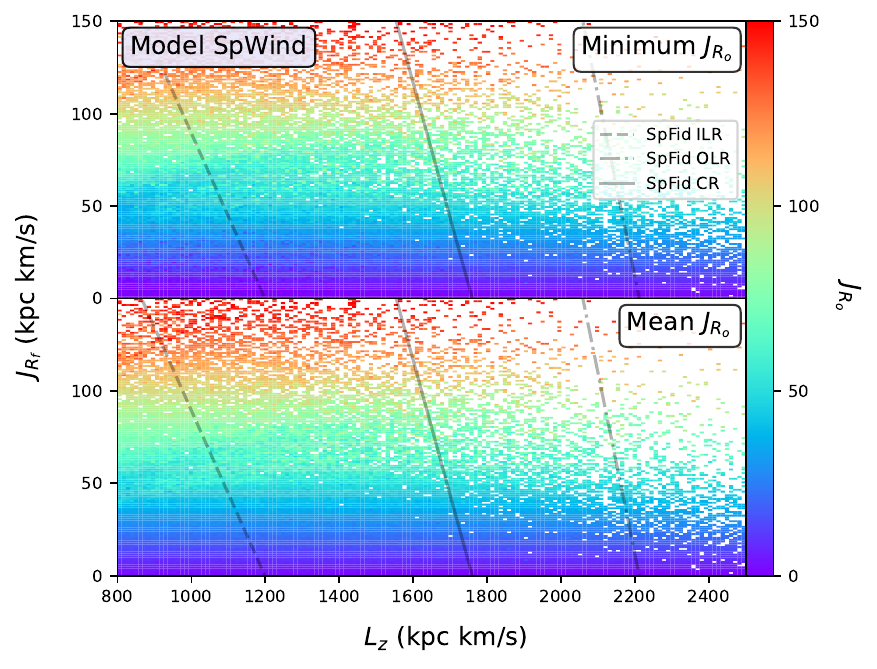}
  }
\end{tabular}

\caption{Final $J_{\rm R}$ vs. final $L_{\rm z}$ distribution from \sout{all} \refedit{select} simulations. 
Colors indicate the minimum \textbf{(top)} and mean \textbf{(bottom)} values of $J_{{\rm R}_{0}}$. 
The locations of the \ilr{} (dashed line), the \CR{} (solid line), and the \olr{} (dot-dashed line) are also shown.}
\label{fig:MinMean_JR0__page2}
\end{figure*}

\appendix
\refedit{In this appendix we include Figures~\ref{fig:MinMean_JR0_page1} and \ref{fig:MinMean_JR0__page2}, which are similar to Figure~\ref{fig:MinMean_JR0_SpAlpha30}, showing the minimum and mean initial radial action ($J_{R_0}$) in $J_{\rm R}-L_{\rm z}$ space at the end of each of our models.}

\bibliography{Main.bib}{}

@article{astropy:2013,
    Adsurl = {http://adsabs.harvard.edu/abs/2013A%26A...558A..33A},
    Archiveprefix = {arXiv},
    Author = {{Astropy Collaboration} and {Robitaille}, T.~P. and {Tollerud}, E.~J. and {Greenfield}, P. and {Droettboom}, M. and {Bray}, E. and {Aldcroft}, T. and {Davis}, M. and {Ginsburg}, A. and {Price-Whelan}, A.~M. and {Kerzendorf}, W.~E. and {Conley}, A. and {Crighton}, N. and {Barbary}, K. and {Muna}, D. and {Ferguson}, H. and {Grollier}, F. and {Parikh}, M.~M. and {Nair}, P.~H. and {Unther}, H.~M. and {Deil}, C. and {Woillez}, J. and {Conseil}, S. and {Kramer}, R. and {Turner}, J.~E.~H. and {Singer}, L. and {Fox}, R. and {Weaver}, B.~A. and {Zabalza}, V. and {Edwards}, Z.~I. and {Azalee Bostroem}, K. and {Burke}, D.~J. and {Casey}, A.~R. and {Crawford}, S.~M. and {Dencheva}, N. and {Ely}, J. and {Jenness}, T. and {Labrie}, K. and {Lim}, P.~L. and {Pierfederici}, F. and {Pontzen}, A. and {Ptak}, A. and {Refsdal}, B. and {Servillat}, M. and {Streicher}, O.},
    Doi = {10.1051/0004-6361/201322068},
    Eid = {A33},
    Eprint = {1307.6212},
    Journal = {\aap},
    Month = oct,
    Pages = {A33},
    Primaryclass = {astro-ph.IM},
    Title = {{Astropy: A community Python package for astronomy}},
    Volume = 558,
    Year = 2013}

@ARTICLE{astropy:2018,
       author = {{Astropy Collaboration} and {Price-Whelan}, A.~M. and
         {Sip{\H{o}}cz}, B.~M. and {G{\"u}nther}, H.~M. and {Lim}, P.~L. and
         {Crawford}, S.~M. and {Conseil}, S. and {Shupe}, D.~L. and
         {Craig}, M.~W. and {Dencheva}, N. and {Ginsburg}, A. and {Vand
        erPlas}, J.~T. and {Bradley}, L.~D. and {P{\'e}rez-Su{\'a}rez}, D. and
         {de Val-Borro}, M. and {Aldcroft}, T.~L. and {Cruz}, K.~L. and
         {Robitaille}, T.~P. and {Tollerud}, E.~J. and {Ardelean}, C. and
         {Babej}, T. and {Bach}, Y.~P. and {Bachetti}, M. and {Bakanov}, A.~V. and
         {Bamford}, S.~P. and {Barentsen}, G. and {Barmby}, P. and
         {Baumbach}, A. and {Berry}, K.~L. and {Biscani}, F. and {Boquien}, M. and
         {Bostroem}, K.~A. and {Bouma}, L.~G. and {Brammer}, G.~B. and
         {Bray}, E.~M. and {Breytenbach}, H. and {Buddelmeijer}, H. and
         {Burke}, D.~J. and {Calderone}, G. and {Cano Rodr{\'\i}guez}, J.~L. and
         {Cara}, M. and {Cardoso}, J.~V.~M. and {Cheedella}, S. and {Copin}, Y. and
         {Corrales}, L. and {Crichton}, D. and {D'Avella}, D. and {Deil}, C. and
         {Depagne}, {\'E}. and {Dietrich}, J.~P. and {Donath}, A. and
         {Droettboom}, M. and {Earl}, N. and {Erben}, T. and {Fabbro}, S. and
         {Ferreira}, L.~A. and {Finethy}, T. and {Fox}, R.~T. and
         {Garrison}, L.~H. and {Gibbons}, S.~L.~J. and {Goldstein}, D.~A. and
         {Gommers}, R. and {Greco}, J.~P. and {Greenfield}, P. and
         {Groener}, A.~M. and {Grollier}, F. and {Hagen}, A. and {Hirst}, P. and
         {Homeier}, D. and {Horton}, A.~J. and {Hosseinzadeh}, G. and {Hu}, L. and
         {Hunkeler}, J.~S. and {Ivezi{\'c}}, {\v{Z}}. and {Jain}, A. and
         {Jenness}, T. and {Kanarek}, G. and {Kendrew}, S. and {Kern}, N.~S. and
         {Kerzendorf}, W.~E. and {Khvalko}, A. and {King}, J. and {Kirkby}, D. and
         {Kulkarni}, A.~M. and {Kumar}, A. and {Lee}, A. and {Lenz}, D. and
         {Littlefair}, S.~P. and {Ma}, Z. and {Macleod}, D.~M. and
         {Mastropietro}, M. and {McCully}, C. and {Montagnac}, S. and
         {Morris}, B.~M. and {Mueller}, M. and {Mumford}, S.~J. and {Muna}, D. and
         {Murphy}, N.~A. and {Nelson}, S. and {Nguyen}, G.~H. and
         {Ninan}, J.~P. and {N{\"o}the}, M. and {Ogaz}, S. and {Oh}, S. and
         {Parejko}, J.~K. and {Parley}, N. and {Pascual}, S. and {Patil}, R. and
         {Patil}, A.~A. and {Plunkett}, A.~L. and {Prochaska}, J.~X. and
         {Rastogi}, T. and {Reddy Janga}, V. and {Sabater}, J. and
         {Sakurikar}, P. and {Seifert}, M. and {Sherbert}, L.~E. and
         {Sherwood-Taylor}, H. and {Shih}, A.~Y. and {Sick}, J. and
         {Silbiger}, M.~T. and {Singanamalla}, S. and {Singer}, L.~P. and
         {Sladen}, P.~H. and {Sooley}, K.~A. and {Sornarajah}, S. and
         {Streicher}, O. and {Teuben}, P. and {Thomas}, S.~W. and
         {Tremblay}, G.~R. and {Turner}, J.~E.~H. and {Terr{\'o}n}, V. and
         {van Kerkwijk}, M.~H. and {de la Vega}, A. and {Watkins}, L.~L. and
         {Weaver}, B.~A. and {Whitmore}, J.~B. and {Woillez}, J. and
         {Zabalza}, V. and {Astropy Contributors}},
        title = "{The Astropy Project: Building an Open-science Project and Status of the v2.0 Core Package}",
      journal = {\aj},
         year = 2018,
        month = sep,
       volume = {156},
       number = {3},
          eid = {123},
        pages = {123},
          doi = {10.3847/1538-3881/aabc4f},
archivePrefix = {arXiv},
       eprint = {1801.02634},
 primaryClass = {astro-ph.IM},
       adsurl = {https://ui.adsabs.harvard.edu/abs/2018AJ....156..123A}
}

@ARTICLE{astropy:2022,
       author = {{Astropy Collaboration} and {Price-Whelan}, Adrian M. and {Lim}, Pey Lian and {Earl}, Nicholas and {Starkman}, Nathaniel and {Bradley}, Larry and {Shupe}, David L. and {Patil}, Aarya A. and {Corrales}, Lia and {Brasseur}, C.~E. and {N{"o}the}, Maximilian and {Donath}, Axel and {Tollerud}, Erik and {Morris}, Brett M. and {Ginsburg}, Adam and {Vaher}, Eero and {Weaver}, Benjamin A. and {Tocknell}, James and {Jamieson}, William and {van Kerkwijk}, Marten H. and {Robitaille}, Thomas P. and {Merry}, Bruce and {Bachetti}, Matteo and {G{"u}nther}, H. Moritz and {Aldcroft}, Thomas L. and {Alvarado-Montes}, Jaime A. and {Archibald}, Anne M. and {B{'o}di}, Attila and {Bapat}, Shreyas and {Barentsen}, Geert and {Baz{'a}n}, Juanjo and {Biswas}, Manish and {Boquien}, M{'e}d{'e}ric and {Burke}, D.~J. and {Cara}, Daria and {Cara}, Mihai and {Conroy}, Kyle E. and {Conseil}, Simon and {Craig}, Matthew W. and {Cross}, Robert M. and {Cruz}, Kelle L. and {D'Eugenio}, Francesco and {Dencheva}, Nadia and {Devillepoix}, Hadrien A.~R. and {Dietrich}, J{"o}rg P. and {Eigenbrot}, Arthur Davis and {Erben}, Thomas and {Ferreira}, Leonardo and {Foreman-Mackey}, Daniel and {Fox}, Ryan and {Freij}, Nabil and {Garg}, Suyog and {Geda}, Robel and {Glattly}, Lauren and {Gondhalekar}, Yash and {Gordon}, Karl D. and {Grant}, David and {Greenfield}, Perry and {Groener}, Austen M. and {Guest}, Steve and {Gurovich}, Sebastian and {Handberg}, Rasmus and {Hart}, Akeem and {Hatfield-Dodds}, Zac and {Homeier}, Derek and {Hosseinzadeh}, Griffin and {Jenness}, Tim and {Jones}, Craig K. and {Joseph}, Prajwel and {Kalmbach}, J. Bryce and {Karamehmetoglu}, Emir and {Ka{l}uszy{'n}ski}, Miko{l}aj and {Kelley}, Michael S.~P. and {Kern}, Nicholas and {Kerzendorf}, Wolfgang E. and {Koch}, Eric W. and {Kulumani}, Shankar and {Lee}, Antony and {Ly}, Chun and {Ma}, Zhiyuan and {MacBride}, Conor and {Maljaars}, Jakob M. and {Muna}, Demitri and {Murphy}, N.~A. and {Norman}, Henrik and {O'Steen}, Richard and {Oman}, Kyle A. and {Pacifici}, Camilla and {Pascual}, Sergio and {Pascual-Granado}, J. and {Patil}, Rohit R. and {Perren}, Gabriel I. and {Pickering}, Timothy E. and {Rastogi}, Tanuj and {Roulston}, Benjamin R. and {Ryan}, Daniel F. and {Rykoff}, Eli S. and {Sabater}, Jose and {Sakurikar}, Parikshit and {Salgado}, Jes{'u}s and {Sanghi}, Aniket and {Saunders}, Nicholas and {Savchenko}, Volodymyr and {Schwardt}, Ludwig and {Seifert-Eckert}, Michael and {Shih}, Albert Y. and {Jain}, Anany Shrey and {Shukla}, Gyanendra and {Sick}, Jonathan and {Simpson}, Chris and {Singanamalla}, Sudheesh and {Singer}, Leo P. and {Singhal}, Jaladh and {Sinha}, Manodeep and {Sip{H{o}}cz}, Brigitta M. and {Spitler}, Lee R. and {Stansby}, David and {Streicher}, Ole and {{{S}}umak}, Jani and {Swinbank}, John D. and {Taranu}, Dan S. and {Tewary}, Nikita and {Tremblay}, Grant R. and {Val-Borro}, Miguel de and {Van Kooten}, Samuel J. and {Vasovi{'c}}, Zlatan and {Verma}, Shresth and {de Miranda Cardoso}, Jos{'e} Vin{'i}cius and {Williams}, Peter K.~G. and {Wilson}, Tom J. and {Winkel}, Benjamin and {Wood-Vasey}, W.~M. and {Xue}, Rui and {Yoachim}, Peter and {Zhang}, Chen and {Zonca}, Andrea and {Astropy Project Contributors}},
        title = "{The Astropy Project: Sustaining and Growing a Community-oriented Open-source Project and the Latest Major Release (v5.0) of the Core Package}",
      journal = {apj},
         year = 2022,
        month = aug,
       volume = {935},
       number = {2},
          eid = {167},
        pages = {167},
          doi = {10.3847/1538-4357/ac7c74},
archivePrefix = {arXiv},
       eprint = {2206.14220},
 primaryClass = {astro-ph.IM},
       adsurl = {https://ui.adsabs.harvard.edu/abs/2022ApJ...935..167A}
}

@ARTICLE{Rampalli23,
       author = {{Rampalli}, Rayna and {Smock}, Amy and {Newton}, Elisabeth R. and {Daniel}, Kathryne J. and {Curtis}, Jason L.},
        title = "{Wrinkles in Time. I. Rapid Rotators Found in High-eccentricity Orbits}",
      journal = {\apj},
         year = 2023,
        month = nov,
       volume = {958},
       number = {1},
          eid = {76},
        pages = {76},
          doi = {10.3847/1538-4357/acff69},
archivePrefix = {arXiv},
       eprint = {2310.02305},
 primaryClass = {astro-ph.SR},
       adsurl = {https://ui.adsabs.harvard.edu/abs/2023ApJ...958...76R}
}

@article{GaiaDR3,
	author = {{Gaia Collaboration} and {Vallenari, A.} and {Brown, A. G. A.} and {Prusti, T.} and {de Bruijne, J. H. J.} and {Arenou, F.} and {Babusiaux, C.} and {Biermann, M.} and {Creevey, O. L.} and {Ducourant, C.} and {Evans, D. W.} and {Eyer, L.} and {Guerra, R.} and {Hutton, A.} and {Jordi, C.} and {Klioner, S. A.} and {Lammers, U. L.} and {Lindegren, L.} and {Luri, X.} and {Mignard, F.} and {Panem, C.} and {Pourbaix, D.} and {Randich, S.} and {Sartoretti, P.} and {Soubiran, C.} and {Tanga, P.} and {Walton, N. A.} and {Bailer-Jones, C. A. L.} and {Bastian, U.} and {Drimmel, R.} and {Jansen, F.} and {Katz, D.} and {Lattanzi, M. G.} and {van Leeuwen, F.} and {Bakker, J.} and {Cacciari, C.} and {Castañeda, J.} and {De Angeli, F.} and {Fabricius, C.} and {Fouesneau, M.} and {Frémat, Y.} and {Galluccio, L.} and {Guerrier, A.} and {Heiter, U.} and {Masana, E.} and {Messineo, R.} and {Mowlavi, N.} and {Nicolas, C.} and {Nienartowicz, K.} and {Pailler, F.} and {Panuzzo, P.} and {Riclet, F.} and {Roux, W.} and {Seabroke, G. M.} and {Sordo, R.} and {Thévenin, F.} and {Gracia-Abril, G.} and {Portell, J.} and {Teyssier, D.} and {Altmann, M.} and {Andrae, R.} and {Audard, M.} and {Bellas-Velidis, I.} and {Benson, K.} and {Berthier, J.} and {Blomme, R.} and {Burgess, P. W.} and {Busonero, D.} and {Busso, G.} and {Cánovas, H.} and {Carry, B.} and {Cellino, A.} and {Cheek, N.} and {Clementini, G.} and {Damerdji, Y.} and {Davidson, M.} and {de Teodoro, P.} and {Nuñez Campos, M.} and {Delchambre, L.} and {Dell’Oro, A.} and {Esquej, P.} and {Fernández-Hernández, J.} and {Fraile, E.} and {Garabato, D.} and {García-Lario, P.} and {Gosset, E.} and {Haigron, R.} and {Halbwachs, J.-L.} and {Hambly, N. C.} and {Harrison, D. L.} and {Hernández, J.} and {Hestroffer, D.} and {Hodgkin, S. T.} and {Holl, B.} and {Janßen, K.} and {Jevardat de Fombelle, G.} and {Jordan, S.} and {Krone-Martins, A.} and {Lanzafame, A. C.} and {Löffler, W.} and {Marchal, O.} and {Marrese, P. M.} and {Moitinho, A.} and {Muinonen, K.} and {Osborne, P.} and {Pancino, E.} and {Pauwels, T.} and {Recio-Blanco, A.} and {Reylé, C.} and {Riello, M.} and {Rimoldini, L.} and {Roegiers, T.} and {Rybizki, J.} and {Sarro, L. M.} and {Siopis, C.} and {Smith, M.} and {Sozzetti, A.} and {Utrilla, E.} and {van Leeuwen, M.} and {Abbas, U.} and {Ábrahám, P.} and {Abreu Aramburu, A.} and {Aerts, C.} and {Aguado, J. J.} and {Ajaj, M.} and {Aldea-Montero, F.} and {Altavilla, G.} and {Álvarez, M. A.} and {Alves, J.} and {Anders, F.} and {Anderson, R. I.} and {Anglada Varela, E.} and {Antoja, T.} and {Baines, D.} and {Baker, S. G.} and {Balaguer-Núñez, L.} and {Balbinot, E.} and {Balog, Z.} and {Barache, C.} and {Barbato, D.} and {Barros, M.} and {Barstow, M. A.} and {Bartolomé, S.} and {Bassilana, J.-L.} and {Bauchet, N.} and {Becciani, U.} and {Bellazzini, M.} and {Berihuete, A.} and {Bernet, M.} and {Bertone, S.} and {Bianchi, L.} and {Binnenfeld, A.} and {Blanco-Cuaresma, S.} and {Blazere, A.} and {Boch, T.} and {Bombrun, A.} and {Bossini, D.} and {Bouquillon, S.} and {Bragaglia, A.} and {Bramante, L.} and {Breedt, E.} and {Bressan, A.} and {Brouillet, N.} and {Brugaletta, E.} and {Bucciarelli, B.} and {Burlacu, A.} and {Butkevich, A. G.} and {Buzzi, R.} and {Caffau, E.} and {Cancelliere, R.} and {Cantat-Gaudin, T.} and {Carballo, R.} and {Carlucci, T.} and {Carnerero, M. I.} and {Carrasco, J. M.} and {Casamiquela, L.} and {Castellani, M.} and {Castro-Ginard, A.} and {Chaoul, L.} and {Charlot, P.} and {Chemin, L.} and {Chiaramida, V.} and {Chiavassa, A.} and {Chornay, N.} and {Comoretto, G.} and {Contursi, G.} and {Cooper, W. J.} and {Cornez, T.} and {Cowell, S.} and {Crifo, F.} and {Cropper, M.} and {Crosta, M.} and {Crowley, C.} and {Dafonte, C.} and {Dapergolas, A.} and {David, M.} and {David, P.} and {de Laverny, P.} and {De Luise, F.} and {De March, R.} and {De Ridder, J.} and {de Souza, R.} and {de Torres, A.} and {del Peloso, E. F.} and {del Pozo, E.} and {Delbo, M.} and {Delgado, A.} and {Delisle, J.-B.} and {Demouchy, C.} and {Dharmawardena, T. E.} and {Di Matteo, P.} and {Diakite, S.} and {Diener, C.} and {Distefano, E.} and {Dolding, C.} and {Edvardsson, B.} and {Enke, H.} and {Fabre, C.} and {Fabrizio, M.} and {Faigler, S.} and {Fedorets, G.} and {Fernique, P.} and {Fienga, A.} and {Figueras, F.} and {Fournier, Y.} and {Fouron, C.} and {Fragkoudi, F.} and {Gai, M.} and {Garcia-Gutierrez, A.} and {Garcia-Reinaldos, M.} and {García-Torres, M.} and {Garofalo, A.} and {Gavel, A.} and {Gavras, P.} and {Gerlach, E.} and {Geyer, R.} and {Giacobbe, P.} and {Gilmore, G.} and {Girona, S.} and {Giuffrida, G.} and {Gomel, R.} and {Gomez, A.} and {González-Núñez, J.} and {González-Santamaría, I.} and {González-Vidal, J. J.} and {Granvik, M.} and {Guillout, P.} and {Guiraud, J.} and {Gutiérrez-Sánchez, R.} and {Guy, L. P.} and {Hatzidimitriou, D.} and {Hauser, M.} and {Haywood, M.} and {Helmer, A.} and {Helmi, A.} and {Sarmiento, M. H.} and {Hidalgo, S. L.} and {Hilger, T.} and {Hładczuk, N.} and {Hobbs, D.} and {Holland, G.} and {Huckle, H. E.} and {Jardine, K.} and {Jasniewicz, G.} and {Jean-Antoine Piccolo, A.} and {Jiménez-Arranz, Ó.} and {Jorissen, A.} and {Juaristi Campillo, J.} and {Julbe, F.} and {Karbevska, L.} and {Kervella, P.} and {Khanna, S.} and {Kontizas, M.} and {Kordopatis, G.} and {Korn, A. J.} and {Kóspál, Á} and {Kostrzewa-Rutkowska, Z.} and {Kruszyńska, K.} and {Kun, M.} and {Laizeau, P.} and {Lambert, S.} and {Lanza, A. F.} and {Lasne, Y.} and {Le Campion, J.-F.} and {Lebreton, Y.} and {Lebzelter, T.} and {Leccia, S.} and {Leclerc, N.} and {Lecoeur-Taibi, I.} and {Liao, S.} and {Licata, E. L.} and {Lindstrøm, H. E. P.} and {Lister, T. A.} and {Livanou, E.} and {Lobel, A.} and {Lorca, A.} and {Loup, C.} and {Madrero Pardo, P.} and {Magdaleno Romeo, A.} and {Managau, S.} and {Mann, R. G.} and {Manteiga, M.} and {Marchant, J. M.} and {Marconi, M.} and {Marcos, J.} and {Marcos Santos, M. M. S.} and {Marín Pina, D.} and {Marinoni, S.} and {Marocco, F.} and {Marshall, D. J.} and {Martin Polo, L.} and {Martín-Fleitas, J. M.} and {Marton, G.} and {Mary, N.} and {Masip, A.} and {Massari, D.} and {Mastrobuono-Battisti, A.} and {Mazeh, T.} and {McMillan, P. J.} and {Messina, S.} and {Michalik, D.} and {Millar, N. R.} and {Mints, A.} and {Molina, D.} and {Molinaro, R.} and {Molnár, L.} and {Monari, G.} and {Monguió, M.} and {Montegriffo, P.} and {Montero, A.} and {Mor, R.} and {Mora, A.} and {Morbidelli, R.} and {Morel, T.} and {Morris, D.} and {Muraveva, T.} and {Murphy, C. P.} and {Musella, I.} and {Nagy, Z.} and {Noval, L.} and {Ocaña, F.} and {Ogden, A.} and {Ordenovic, C.} and {Osinde, J. O.} and {Pagani, C.} and {Pagano, I.} and {Palaversa, L.} and {Palicio, P. A.} and {Pallas-Quintela, L.} and {Panahi, A.} and {Payne-Wardenaar, S.} and {Peñalosa Esteller, X.} and {Penttilä, A.} and {Pichon, B.} and {Piersimoni, A. M.} and {Pineau, F.-X.} and {Plachy, E.} and {Plum, G.} and {Poggio, E.} and {Prša, A.} and {Pulone, L.} and {Racero, E.} and {Ragaini, S.} and {Rainer, M.} and {Raiteri, C. M.} and {Rambaux, N.} and {Ramos, P.} and {Ramos-Lerate, M.} and {Re Fiorentin, P.} and {Regibo, S.} and {Richards, P. J.} and {Rios Diaz, C.} and {Ripepi, V.} and {Riva, A.} and {Rix, H.-W.} and {Rixon, G.} and {Robichon, N.} and {Robin, A. C.} and {Robin, C.} and {Roelens, M.} and {Rogues, H. R. O.} and {Rohrbasser, L.} and {Romero-Gómez, M.} and {Rowell, N.} and {Royer, F.} and {Ruz Mieres, D.} and {Rybicki, K. A.} and {Sadowski, G.} and {Sáez Núñez, A.} and {Sagristà Sellés, A.} and {Sahlmann, J.} and {Salguero, E.} and {Samaras, N.} and {Sanchez Gimenez, V.} and {Sanna, N.} and {Santoveña, R.} and {Sarasso, M.} and {Schultheis, M.} and {Sciacca, E.} and {Segol, M.} and {Segovia, J. C.} and {Ségransan, D.} and {Semeux, D.} and {Shahaf, S.} and {Siddiqui, H. I.} and {Siebert, A.} and {Siltala, L.} and {Silvelo, A.} and {Slezak, E.} and {Slezak, I.} and {Smart, R. L.} and {Snaith, O. N.} and {Solano, E.} and {Solitro, F.} and {Souami, D.} and {Souchay, J.} and {Spagna, A.} and {Spina, L.} and {Spoto, F.} and {Steele, I. A.} and {Steidelmüller, H.} and {Stephenson, C. A.} and {Süveges, M.} and {Surdej, J.} and {Szabados, L.} and {Szegedi-Elek, E.} and {Taris, F.} and {Taylor, M. B.} and {Teixeira, R.} and {Tolomei, L.} and {Tonello, N.} and {Torra, F.} and {Torra, J.} and {Torralba Elipe, G.} and {Trabucchi, M.} and {Tsounis, A. T.} and {Turon, C.} and {Ulla, A.} and {Unger, N.} and {Vaillant, M. V.} and {van Dillen, E.} and {van Reeven, W.} and {Vanel, O.} and {Vecchiato, A.} and {Viala, Y.} and {Vicente, D.} and {Voutsinas, S.} and {Weiler, M.} and {Wevers, T.} and {Wyrzykowski, Ł.} and {Yoldas, A.} and {Yvard, P.} and {Zhao, H.} and {Zorec, J.} and {Zucker, S.} and {Zwitter, T.}},
	title = {Gaia Data Release 3 - Summary of the content and survey properties},
	DOI= "10.1051/0004-6361/202243940",
	url= "https://doi.org/10.1051/0004-6361/202243940",
	journal = {A\&A},
	year = 2023,
	volume = 674,
	pages = "A1",
}

@article{GaiaEDR3,
	author = {{Gaia Collaboration} and {Brown, A. G. A.} and {Vallenari, A.} and {Prusti, T.} and {de Bruijne, J. H. J.} and {Babusiaux, C.} and {Biermann, M.} and {Creevey, O. L.} and {Evans, D. W.} and {Eyer, L.} and {Hutton, A.} and {Jansen, F.} and {Jordi, C.} and {Klioner, S. A.} and {Lammers, U.} and {Lindegren, L.} and {Luri, X.} and {Mignard, F.} and {Panem, C.} and {Pourbaix, D.} and {Randich, S.} and {Sartoretti, P.} and {Soubiran, C.} and {Walton, N. A.} and {Arenou, F.} and {Bailer-Jones, C. A. L.} and {Bastian, U.} and {Cropper, M.} and {Drimmel, R.} and {Katz, D.} and {Lattanzi, M. G.} and {van Leeuwen, F.} and {Bakker, J.} and {Cacciari, C.} and {Castañeda, J.} and {De Angeli, F.} and {Ducourant, C.} and {Fabricius, C.} and {Fouesneau, M.} and {Frémat, Y.} and {Guerra, R.} and {Guerrier, A.} and {Guiraud, J.} and {Jean-Antoine Piccolo, A.} and {Masana, E.} and {Messineo, R.} and {Mowlavi, N.} and {Nicolas, C.} and {Nienartowicz, K.} and {Pailler, F.} and {Panuzzo, P.} and {Riclet, F.} and {Roux, W.} and {Seabroke, G. M.} and {Sordo, R.} and {Tanga, P.} and {Thévenin, F.} and {Gracia-Abril, G.} and {Portell, J.} and {Teyssier, D.} and {Altmann, M.} and {Andrae, R.} and {Bellas-Velidis, I.} and {Benson, K.} and {Berthier, J.} and {Blomme, R.} and {Brugaletta, E.} and {Burgess, P. W.} and {Busso, G.} and {Carry, B.} and {Cellino, A.} and {Cheek, N.} and {Clementini, G.} and {Damerdji, Y.} and {Davidson, M.} and {Delchambre, L.} and {Dell’Oro, A.} and {Fernández-Hernández, J.} and {Galluccio, L.} and {García-Lario, P.} and {Garcia-Reinaldos, M.} and {González-Núñez, J.} and {Gosset, E.} and {Haigron, R.} and {Halbwachs, J.-L.} and {Hambly, N. C.} and {Harrison, D. L.} and {Hatzidimitriou, D.} and {Heiter, U.} and {Hernández, J.} and {Hestroffer, D.} and {Hodgkin, S. T.} and {Holl, B.} and {Janßen, K.} and {Jevardat de Fombelle, G.} and {Jordan, S.} and {Krone-Martins, A.} and {Lanzafame, A. C.} and {Löffler, W.} and {Lorca, A.} and {Manteiga, M.} and {Marchal, O.} and {Marrese, P. M.} and {Moitinho, A.} and {Mora, A.} and {Muinonen, K.} and {Osborne, P.} and {Pancino, E.} and {Pauwels, T.} and {Petit, J.-M.} and {Recio-Blanco, A.} and {Richards, P. J.} and {Riello, M.} and {Rimoldini, L.} and {Robin, A. C.} and {Roegiers, T.} and {Rybizki, J.} and {Sarro, L. M.} and {Siopis, C.} and {Smith, M.} and {Sozzetti, A.} and {Ulla, A.} and {Utrilla, E.} and {van Leeuwen, M.} and {van Reeven, W.} and {Abbas, U.} and {Abreu Aramburu, A.} and {Accart, S.} and {Aerts, C.} and {Aguado, J. J.} and {Ajaj, M.} and {Altavilla, G.} and {Álvarez, M. A.} and {Álvarez Cid-Fuentes, J.} and {Alves, J.} and {Anderson, R. I.} and {Anglada Varela, E.} and {Antoja, T.} and {Audard, M.} and {Baines, D.} and {Baker, S. G.} and {Balaguer-Núñez, L.} and {Balbinot, E.} and {Balog, Z.} and {Barache, C.} and {Barbato, D.} and {Barros, M.} and {Barstow, M. A.} and {Bartolomé, S.} and {Bassilana, J.-L.} and {Bauchet, N.} and {Baudesson-Stella, A.} and {Becciani, U.} and {Bellazzini, M.} and {Bernet, M.} and {Bertone, S.} and {Bianchi, L.} and {Blanco-Cuaresma, S.} and {Boch, T.} and {Bombrun, A.} and {Bossini, D.} and {Bouquillon, S.} and {Bragaglia, A.} and {Bramante, L.} and {Breedt, E.} and {Bressan, A.} and {Brouillet, N.} and {Bucciarelli, B.} and {Burlacu, A.} and {Busonero, D.} and {Butkevich, A. G.} and {Buzzi, R.} and {Caffau, E.} and {Cancelliere, R.} and {Cánovas, H.} and {Cantat-Gaudin, T.} and {Carballo, R.} and {Carlucci, T.} and {Carnerero, M. I} and {Carrasco, J. M.} and {Casamiquela, L.} and {Castellani, M.} and {Castro-Ginard, A.} and {Castro Sampol, P.} and {Chaoul, L.} and {Charlot, P.} and {Chemin, L.} and {Chiavassa, A.} and {Cioni, M.-R. L.} and {Comoretto, G.} and {Cooper, W. J.} and {Cornez, T.} and {Cowell, S.} and {Crifo, F.} and {Crosta, M.} and {Crowley, C.} and {Dafonte, C.} and {Dapergolas, A.} and {David, M.} and {David, P.} and {de Laverny, P.} and {De Luise, F.} and {De March, R.} and {De Ridder, J.} and {de Souza, R.} and {de Teodoro, P.} and {de Torres, A.} and {del Peloso, E. F.} and {del Pozo, E.} and {Delbo, M.} and {Delgado, A.} and {Delgado, H. E.} and {Delisle, J.-B.} and {Di Matteo, P.} and {Diakite, S.} and {Diener, C.} and {Distefano, E.} and {Dolding, C.} and {Eappachen, D.} and {Edvardsson, B.} and {Enke, H.} and {Esquej, P.} and {Fabre, C.} and {Fabrizio, M.} and {Faigler, S.} and {Fedorets, G.} and {Fernique, P.} and {Fienga, A.} and {Figueras, F.} and {Fouron, C.} and {Fragkoudi, F.} and {Fraile, E.} and {Franke, F.} and {Gai, M.} and {Garabato, D.} and {Garcia-Gutierrez, A.} and {García-Torres, M.} and {Garofalo, A.} and {Gavras, P.} and {Gerlach, E.} and {Geyer, R.} and {Giacobbe, P.} and {Gilmore, G.} and {Girona, S.} and {Giuffrida, G.} and {Gomel, R.} and {Gomez, A.} and {Gonzalez-Santamaria, I.} and {González-Vidal, J. J.} and {Granvik, M.} and {Gutiérrez-Sánchez, R.} and {Guy, L. P.} and {Hauser, M.} and {Haywood, M.} and {Helmi, A.} and {Hidalgo, S. L.} and {Hilger, T.} and {Hładczuk, N.} and {Hobbs, D.} and {Holland, G.} and {Huckle, H. E.} and {Jasniewicz, G.} and {Jonker, P. G.} and {Juaristi Campillo, J.} and {Julbe, F.} and {Karbevska, L.} and {Kervella, P.} and {Khanna, S.} and {Kochoska, A.} and {Kontizas, M.} and {Kordopatis, G.} and {Korn, A. J.} and {Kostrzewa-Rutkowska, Z.} and {Kruszyńska, K.} and {Lambert, S.} and {Lanza, A. F.} and {Lasne, Y.} and {Le Campion, J.-F.} and {Le Fustec, Y.} and {Lebreton, Y.} and {Lebzelter, T.} and {Leccia, S.} and {Leclerc, N.} and {Lecoeur-Taibi, I.} and {Liao, S.} and {Licata, E.} and {Lindstrøm, E. P.} and {Lister, T. A.} and {Livanou, E.} and {Lobel, A.} and {Madrero Pardo, P.} and {Managau, S.} and {Mann, R. G.} and {Marchant, J. M.} and {Marconi, M.} and {Marcos Santos, M. M. S.} and {Marinoni, S.} and {Marocco, F.} and {Marshall, D. J.} and {Martin Polo, L.} and {Martín-Fleitas, J. M.} and {Masip, A.} and {Massari, D.} and {Mastrobuono-Battisti, A.} and {Mazeh, T.} and {McMillan, P. J.} and {Messina, S.} and {Michalik, D.} and {Millar, N. R.} and {Mints, A.} and {Molina, D.} and {Molinaro, R.} and {Molnár, L.} and {Montegriffo, P.} and {Mor, R.} and {Morbidelli, R.} and {Morel, T.} and {Morris, D.} and {Mulone, A. F.} and {Munoz, D.} and {Muraveva, T.} and {Murphy, C. P.} and {Musella, I.} and {Noval, L.} and {Ordénovic, C.} and {Orrù, G.} and {Osinde, J.} and {Pagani, C.} and {Pagano, I.} and {Palaversa, L.} and {Palicio, P. A.} and {Panahi, A.} and {Pawlak, M.} and {Peñalosa Esteller, X.} and {Penttilä, A.} and {Piersimoni, A. M.} and {Pineau, F.-X.} and {Plachy, E.} and {Plum, G.} and {Poggio, E.} and {Poretti, E.} and {Poujoulet, E.} and {Prša, A.} and {Pulone, L.} and {Racero, E.} and {Ragaini, S.} and {Rainer, M.} and {Raiteri, C. M.} and {Rambaux, N.} and {Ramos, P.} and {Ramos-Lerate, M.} and {Re Fiorentin, P.} and {Regibo, S.} and {Reylé, C.} and {Ripepi, V.} and {Riva, A.} and {Rixon, G.} and {Robichon, N.} and {Robin, C.} and {Roelens, M.} and {Rohrbasser, L.} and {Romero-Gómez, M.} and {Rowell, N.} and {Royer, F.} and {Rybicki, K. A.} and {Sadowski, G.} and {Sagristà Sellés, A.} and {Sahlmann, J.} and {Salgado, J.} and {Salguero, E.} and {Samaras, N.} and {Sanchez Gimenez, V.} and {Sanna, N.} and {Santoveña, R.} and {Sarasso, M.} and {Schultheis, M.} and {Sciacca, E.} and {Segol, M.} and {Segovia, J. C.} and {Ségransan, D.} and {Semeux, D.} and {Shahaf, S.} and {Siddiqui, H. I.} and {Siebert, A.} and {Siltala, L.} and {Slezak, E.} and {Smart, R. L.} and {Solano, E.} and {Solitro, F.} and {Souami, D.} and {Souchay, J.} and {Spagna, A.} and {Spoto, F.} and {Steele, I. A.} and {Steidelmüller, H.} and {Stephenson, C. A.} and {Süveges, M.} and {Szabados, L.} and {Szegedi-Elek, E.} and {Taris, F.} and {Tauran, G.} and {Taylor, M. B.} and {Teixeira, R.} and {Thuillot, W.} and {Tonello, N.} and {Torra, F.} and {Torra, J.} and {Turon, C.} and {Unger, N.} and {Vaillant, M.} and {van Dillen, E.} and {Vanel, O.} and {Vecchiato, A.} and {Viala, Y.} and {Vicente, D.} and {Voutsinas, S.} and {Weiler, M.} and {Wevers, T.} and {Wyrzykowski, Ł.} and {Yoldas, A.} and {Yvard, P.} and {Zhao, H.} and {Zorec, J.} and {Zucker, S.} and {Zurbach, C.} and {Zwitter, T.}},
	title = {Gaia Early Data Release 3 - Summary of the contents and survey properties},
	DOI= "10.1051/0004-6361/202039657",
	url= "https://doi.org/10.1051/0004-6361/202039657",
	journal = {A\&A},
	year = 2021,
	volume = 649,
	pages = "A1",
}

@article{GaiaDR2,
	author = {{Gaia Collaboration} and {Brown, A. G. A.} and {Vallenari, A.} and {Prusti, T.} and {de Bruijne, J. H. J.} and {Babusiaux, C.} and {Bailer-Jones, C. A. L.} and {Biermann, M.} and {Evans, D. W.} and {Eyer, L.} and {Jansen, F.} and {Jordi, C.} and {Klioner, S. A.} and {Lammers, U.} and {Lindegren, L.} and {Luri, X.} and {Mignard, F.} and {Panem, C.} and {Pourbaix, D.} and {Randich, S.} and {Sartoretti, P.} and {Siddiqui, H. I.} and {Soubiran, C.} and {van Leeuwen, F.} and {Walton, N. A.} and {Arenou, F.} and {Bastian, U.} and {Cropper, M.} and {Drimmel, R.} and {Katz, D.} and {Lattanzi, M. G.} and {Bakker, J.} and {Cacciari, C.} and {Castañeda, J.} and {Chaoul, L.} and {Cheek, N.} and {De Angeli, F.} and {Fabricius, C.} and {Guerra, R.} and {Holl, B.} and {Masana, E.} and {Messineo, R.} and {Mowlavi, N.} and {Nienartowicz, K.} and {Panuzzo, P.} and {Portell, J.} and {Riello, M.} and {Seabroke, G. M.} and {Tanga, P.} and {Thévenin, F.} and {Gracia-Abril, G.} and {Comoretto, G.} and {Garcia-Reinaldos, M.} and {Teyssier, D.} and {Altmann, M.} and {Andrae, R.} and {Audard, M.} and {Bellas-Velidis, I.} and {Benson, K.} and {Berthier, J.} and {Blomme, R.} and {Burgess, P.} and {Busso, G.} and {Carry, B.} and {Cellino, A.} and {Clementini, G.} and {Clotet, M.} and {Creevey, O.} and {Davidson, M.} and {De Ridder, J.} and {Delchambre, L.} and {Dell’Oro, A.} and {Ducourant, C.} and {Fernández-Hernández, J.} and {Fouesneau, M.} and {Frémat, Y.} and {Galluccio, L.} and {García-Torres, M.} and {González-Núñez, J.} and {González-Vidal, J. J.} and {Gosset, E.} and {Guy, L. P.} and {Halbwachs, J.-L.} and {Hambly, N. C.} and {Harrison, D. L.} and {Hernández, J.} and {Hestroffer, D.} and {Hodgkin, S. T.} and {Hutton, A.} and {Jasniewicz, G.} and {Jean-Antoine-Piccolo, A.} and {Jordan, S.} and {Korn, A. J.} and {Krone-Martins, A.} and {Lanzafame, A. C.} and {Lebzelter, T.} and {Löffler, W.} and {Manteiga, M.} and {Marrese, P. M.} and {Martín-Fleitas, J. M.} and {Moitinho, A.} and {Mora, A.} and {Muinonen, K.} and {Osinde, J.} and {Pancino, E.} and {Pauwels, T.} and {Petit, J.-M.} and {Recio-Blanco, A.} and {Richards, P. J.} and {Rimoldini, L.} and {Robin, A. C.} and {Sarro, L. M.} and {Siopis, C.} and {Smith, M.} and {Sozzetti, A.} and {Süveges, M.} and {Torra, J.} and {van Reeven, W.} and {Abbas, U.} and {Abreu Aramburu, A.} and {Accart, S.} and {Aerts, C.} and {Altavilla, G.} and {Álvarez, M. A.} and {Alvarez, R.} and {Alves, J.} and {Anderson, R. I.} and {Andrei, A. H.} and {Anglada Varela, E.} and {Antiche, E.} and {Antoja, T.} and {Arcay, B.} and {Astraatmadja, T. L.} and {Bach, N.} and {Baker, S. G.} and {Balaguer-Núñez, L.} and {Balm, P.} and {Barache, C.} and {Barata, C.} and {Barbato, D.} and {Barblan, F.} and {Barklem, P. S.} and {Barrado, D.} and {Barros, M.} and {Barstow, M. A.} and {Bartholomé Muñoz, S.} and {Bassilana, J.-L.} and {Becciani, U.} and {Bellazzini, M.} and {Berihuete, A.} and {Bertone, S.} and {Bianchi, L.} and {Bienaymé, O.} and {Blanco-Cuaresma, S.} and {Boch, T.} and {Boeche, C.} and {Bombrun, A.} and {Borrachero, R.} and {Bossini, D.} and {Bouquillon, S.} and {Bourda, G.} and {Bragaglia, A.} and {Bramante, L.} and {Breddels, M. A.} and {Bressan, A.} and {Brouillet, N.} and {Brüsemeister, T.} and {Brugaletta, E.} and {Bucciarelli, B.} and {Burlacu, A.} and {Busonero, D.} and {Butkevich, A. G.} and {Buzzi, R.} and {Caffau, E.} and {Cancelliere, R.} and {Cannizzaro, G.} and {Cantat-Gaudin, T.} and {Carballo, R.} and {Carlucci, T.} and {Carrasco, J. M.} and {Casamiquela, L.} and {Castellani, M.} and {Castro-Ginard, A.} and {Charlot, P.} and {Chemin, L.} and {Chiavassa, A.} and {Cocozza, G.} and {Costigan, G.} and {Cowell, S.} and {Crifo, F.} and {Crosta, M.} and {Crowley, C.} and {Cuypers†, J.} and {Dafonte, C.} and {Damerdji, Y.} and {Dapergolas, A.} and {David, P.} and {David, M.} and {de Laverny, P.} and {De Luise, F.} and {De March, R.} and {de Martino, D.} and {de Souza, R.} and {de Torres, A.} and {Debosscher, J.} and {del Pozo, E.} and {Delbo, M.} and {Delgado, A.} and {Delgado, H. E.} and {Di Matteo, P.} and {Diakite, S.} and {Diener, C.} and {Distefano, E.} and {Dolding, C.} and {Drazinos, P.} and {Durán, J.} and {Edvardsson, B.} and {Enke, H.} and {Eriksson, K.} and {Esquej, P.} and {Eynard Bontemps, G.} and {Fabre, C.} and {Fabrizio, M.} and {Faigler, S.} and {Falcão, A. J.} and {Farràs Casas, M.} and {Federici, L.} and {Fedorets, G.} and {Fernique, P.} and {Figueras, F.} and {Filippi, F.} and {Findeisen, K.} and {Fonti, A.} and {Fraile, E.} and {Fraser, M.} and {Frézouls, B.} and {Gai, M.} and {Galleti, S.} and {Garabato, D.} and {García-Sedano, F.} and {Garofalo, A.} and {Garralda, N.} and {Gavel, A.} and {Gavras, P.} and {Gerssen, J.} and {Geyer, R.} and {Giacobbe, P.} and {Gilmore, G.} and {Girona, S.} and {Giuffrida, G.} and {Glass, F.} and {Gomes, M.} and {Granvik, M.} and {Gueguen, A.} and {Guerrier, A.} and {Guiraud, J.} and {Gutiérrez-Sánchez, R.} and {Haigron, R.} and {Hatzidimitriou, D.} and {Hauser, M.} and {Haywood, M.} and {Heiter, U.} and {Helmi, A.} and {Heu, J.} and {Hilger, T.} and {Hobbs, D.} and {Hofmann, W.} and {Holland, G.} and {Huckle, H. E.} and {Hypki, A.} and {Icardi, V.} and {Janßen, K.} and {Jevardat de Fombelle, G.} and {Jonker, P. G.} and {Juhász, Á. L.} and {Julbe, F.} and {Karampelas, A.} and {Kewley, A.} and {Klar, J.} and {Kochoska, A.} and {Kohley, R.} and {Kolenberg, K.} and {Kontizas, M.} and {Kontizas, E.} and {Koposov, S. E.} and {Kordopatis, G.} and {Kostrzewa-Rutkowska, Z.} and {Koubsky, P.} and {Lambert, S.} and {Lanza, A. F.} and {Lasne, Y.} and {Lavigne, J.-B.} and {Le Fustec, Y.} and {Le Poncin-Lafitte, C.} and {Lebreton, Y.} and {Leccia, S.} and {Leclerc, N.} and {Lecoeur-Taibi, I.} and {Lenhardt, H.} and {Leroux, F.} and {Liao, S.} and {Licata, E.} and {Lindstrøm, H. E. P.} and {Lister, T. A.} and {Livanou, E.} and {Lobel, A.} and {López, M.} and {Managau, S.} and {Mann, R. G.} and {Mantelet, G.} and {Marchal, O.} and {Marchant, J. M.} and {Marconi, M.} and {Marinoni, S.} and {Marschalkó, G.} and {Marshall, D. J.} and {Martino, M.} and {Marton, G.} and {Mary, N.} and {Massari, D.} and {Matijevič, G.} and {Mazeh, T.} and {McMillan, P. J.} and {Messina, S.} and {Michalik, D.} and {Millar, N. R.} and {Molina, D.} and {Molinaro, R.} and {Molnár, L.} and {Montegriffo, P.} and {Mor, R.} and {Morbidelli, R.} and {Morel, T.} and {Morris, D.} and {Mulone, A. F.} and {Muraveva, T.} and {Musella, I.} and {Nelemans, G.} and {Nicastro, L.} and {Noval, L.} and {O’Mullane, W.} and {Ordénovic, C.} and {Ordóñez-Blanco, D.} and {Osborne, P.} and {Pagani, C.} and {Pagano, I.} and {Pailler, F.} and {Palacin, H.} and {Palaversa, L.} and {Panahi, A.} and {Pawlak, M.} and {Piersimoni, A. M.} and {Pineau, F.-X.} and {Plachy, E.} and {Plum, G.} and {Poggio, E.} and {Poujoulet, E.} and {Prša, A.} and {Pulone, L.} and {Racero, E.} and {Ragaini, S.} and {Rambaux, N.} and {Ramos-Lerate, M.} and {Regibo, S.} and {Reylé, C.} and {Riclet, F.} and {Ripepi, V.} and {Riva, A.} and {Rivard, A.} and {Rixon, G.} and {Roegiers, T.} and {Roelens, M.} and {Romero-Gómez, M.} and {Rowell, N.} and {Royer, F.} and {Ruiz-Dern, L.} and {Sadowski, G.} and {Sagristà Sellés, T.} and {Sahlmann, J.} and {Salgado, J.} and {Salguero, E.} and {Sanna, N.} and {Santana-Ros, T.} and {Sarasso, M.} and {Savietto, H.} and {Schultheis, M.} and {Sciacca, E.} and {Segol, M.} and {Segovia, J. C.} and {Ségransan, D.} and {Shih, I-C.} and {Siltala, L.} and {Silva, A. F.} and {Smart, R. L.} and {Smith, K. W.} and {Solano, E.} and {Solitro, F.} and {Sordo, R.} and {Soria Nieto, S.} and {Souchay, J.} and {Spagna, A.} and {Spoto, F.} and {Stampa, U.} and {Steele, I. A.} and {Steidelmüller, H.} and {Stephenson, C. A.} and {Stoev, H.} and {Suess, F. F.} and {Surdej, J.} and {Szabados, L.} and {Szegedi-Elek, E.} and {Tapiador, D.} and {Taris, F.} and {Tauran, G.} and {Taylor, M. B.} and {Teixeira, R.} and {Terrett, D.} and {Teyssandier, P.} and {Thuillot, W.} and {Titarenko, A.} and {Torra Clotet, F.} and {Turon, C.} and {Ulla, A.} and {Utrilla, E.} and {Uzzi, S.} and {Vaillant, M.} and {Valentini, G.} and {Valette, V.} and {van Elteren, A.} and {Van Hemelryck, E.} and {van Leeuwen, M.} and {Vaschetto, M.} and {Vecchiato, A.} and {Veljanoski, J.} and {Viala, Y.} and {Vicente, D.} and {Vogt, S.} and {von Essen, C.} and {Voss, H.} and {Votruba, V.} and {Voutsinas, S.} and {Walmsley, G.} and {Weiler, M.} and {Wertz, O.} and {Wevers, T.} and {Wyrzykowski, Ł.} and {Yoldas, A.} and {Žerjal, M.} and {Ziaeepour, H.} and {Zorec, J.} and {Zschocke, S.} and {Zucker, S.} and {Zurbach, C.} and {Zwitter, T.}},
	title = {Gaia Data Release 2 - Summary of the contents and survey properties},
	DOI= "10.1051/0004-6361/201833051",
	url= "https://doi.org/10.1051/0004-6361/201833051",
	journal = {A\&A},
	year = 2018,
	volume = 616,
	pages = "A1",
}

@article{GaiaDR3RVs,
	author = {{Katz, D.} and {Sartoretti, P.} and {Guerrier, A.} and {Panuzzo, P.} and {Seabroke, G. M.} and {Thévenin, F.} and {Cropper, M.} and {Benson, K.} and {Blomme, R.} and {Haigron, R.} and {Marchal, O.} and {Smith, M.} and {Baker, S.} and {Chemin, L.} and {Damerdji, Y.} and {David, M.} and {Dolding, C.} and {Frémat, Y.} and {Gosset, E.} and {Janßen, K.} and {Jasniewicz, G.} and {Lobel, A.} and {Plum, G.} and {Samaras, N.} and {Snaith, O.} and {Soubiran, C.} and {Vanel, O.} and {Zwitter, T.} and {Antoja, T.} and {Arenou, F.} and {Babusiaux, C.} and {Brouillet, N.} and {Caffau, E.} and {Di Matteo, P.} and {Fabre, C.} and {Fabricius, C.} and {Fragkoudi, F.} and {Haywood, M.} and {Huckle, H. E.} and {Hottier, C.} and {Lasne, Y.} and {Leclerc, N.} and {Mastrobuono-Battisti, A.} and {Royer, F.} and {Teyssier, D.} and {Zorec, J.} and {Crifo, F.} and {Jean-Antoine Piccolo, A.} and {Turon, C.} and {Viala, Y.}},
	title = {Gaia Data Release 3 - Properties and validation of the radial velocities},
	DOI= "10.1051/0004-6361/202244220",
	url= "https://doi.org/10.1051/0004-6361/202244220",
	journal = {A\&A},
	year = 2023,
	volume = 674,
	pages = "A5",
}

@ARTICLE{Majewski17APOGEE,
       author = {{Majewski}, Steven R. and {Schiavon}, Ricardo P. and {Frinchaboy}, Peter M. and {Allende Prieto}, Carlos and {Barkhouser}, Robert and {Bizyaev}, Dmitry and {Blank}, Basil and {Brunner}, Sophia and {Burton}, Adam and {Carrera}, Ricardo and {Chojnowski}, S. Drew and {Cunha}, K{\'a}tia and {Epstein}, Courtney and {Fitzgerald}, Greg and {Garc{\'\i}a P{\'e}rez}, Ana E. and {Hearty}, Fred R. and {Henderson}, Chuck and {Holtzman}, Jon A. and {Johnson}, Jennifer A. and {Lam}, Charles R. and {Lawler}, James E. and {Maseman}, Paul and {M{\'e}sz{\'a}ros}, Szabolcs and {Nelson}, Matthew and {Nguyen}, Duy Coung and {Nidever}, David L. and {Pinsonneault}, Marc and {Shetrone}, Matthew and {Smee}, Stephen and {Smith}, Verne V. and {Stolberg}, Todd and {Skrutskie}, Michael F. and {Walker}, Eric and {Wilson}, John C. and {Zasowski}, Gail and {Anders}, Friedrich and {Basu}, Sarbani and {Beland}, Stephane and {Blanton}, Michael R. and {Bovy}, Jo and {Brownstein}, Joel R. and {Carlberg}, Joleen and {Chaplin}, William and {Chiappini}, Cristina and {Eisenstein}, Daniel J. and {Elsworth}, Yvonne and {Feuillet}, Diane and {Fleming}, Scott W. and {Galbraith-Frew}, Jessica and {Garc{\'\i}a}, Rafael A. and {Garc{\'\i}a-Hern{\'a}ndez}, D. An{\'\i}bal and {Gillespie}, Bruce A. and {Girardi}, L{\'e}o and {Gunn}, James E. and {Hasselquist}, Sten and {Hayden}, Michael R. and {Hekker}, Saskia and {Ivans}, Inese and {Kinemuchi}, Karen and {Klaene}, Mark and {Mahadevan}, Suvrath and {Mathur}, Savita and {Mosser}, Beno{\^\i}t and {Muna}, Demitri and {Munn}, Jeffrey A. and {Nichol}, Robert C. and {O'Connell}, Robert W. and {Parejko}, John K. and {Robin}, A.~C. and {Rocha-Pinto}, Helio and {Schultheis}, Matthias and {Serenelli}, Aldo M. and {Shane}, Neville and {Silva Aguirre}, Victor and {Sobeck}, Jennifer S. and {Thompson}, Benjamin and {Troup}, Nicholas W. and {Weinberg}, David H. and {Zamora}, Olga},
        title = "{The Apache Point Observatory Galactic Evolution Experiment (APOGEE)}",
      journal = {\aj},
         year = 2017,
        month = sep,
       volume = {154},
       number = {3},
          eid = {94},
        pages = {94},
          doi = {10.3847/1538-3881/aa784d},
archivePrefix = {arXiv},
       eprint = {1509.05420},
 primaryClass = {astro-ph.IM},
       adsurl = {https://ui.adsabs.harvard.edu/abs/2017AJ....154...94M}
}

@ARTICLE{Kawata18,
       author = {{Kawata}, Daisuke and {Baba}, Junichi and {Ciuc{\v{a}}}, Ioana and {Cropper}, Mark and {Grand}, Robert J.~J. and {Hunt}, Jason A.~S. and {Seabroke}, George},
        title = "{Radial distribution of stellar motions in Gaia DR2}",
      journal = {\mnras},
         year = 2018,
        month = sep,
       volume = {479},
       number = {1},
        pages = {L108-L112},
          doi = {10.1093/mnrasl/sly107},
archivePrefix = {arXiv},
       eprint = {1804.10175},
 primaryClass = {astro-ph.GA},
       adsurl = {https://ui.adsabs.harvard.edu/abs/2018MNRAS.479L.108K}
}

@ARTICLE{BinneySchoenrich18,
       author = {{Binney}, James and {Sch{\"o}nrich}, Ralph},
        title = "{The origin of the Gaia phase-plane spiral}",
      journal = {\mnras},
         year = 2018,
        month = dec,
       volume = {481},
       number = {2},
        pages = {1501-1506},
          doi = {10.1093/mnras/sty2378},
archivePrefix = {arXiv},
       eprint = {1807.09819},
 primaryClass = {astro-ph.GA},
       adsurl = {https://ui.adsabs.harvard.edu/abs/2018MNRAS.481.1501B}
}

@ARTICLE{ConsidereAthanassoula82,
       author = {{Considere}, S. and {Athanassoula}, E.},
        title = "{The distribution of H II regions in external galaxies. I}",
      journal = {\aap},
         year = 1982,
        month = jul,
       volume = {111},
       number = {1},
        pages = {28-42},
       adsurl = {https://ui.adsabs.harvard.edu/abs/1982A&A...111...28C}
}

@ARTICLE{StarkBrand89,
       author = {{Stark}, Antony A. and {Brand}, Jan},
        title = "{Kinematics of Molecular Clouds. II. New Data on Nearby Giant Molecular Clouds}",
      journal = {\apj},
         year = 1989,
        month = apr,
       volume = {339},
        pages = {763},
          doi = {10.1086/167334},
       adsurl = {https://ui.adsabs.harvard.edu/abs/1989ApJ...339..763S}
}

@ARTICLE{Antoja09,
       author = {{Antoja}, T. and {Valenzuela}, O. and {Pichardo}, B. and {Moreno}, E. and {Figueras}, F. and {Fern{\'a}ndez}, D.},
        title = "{Stellar Kinematic Constraints on Galactic Structure Models Revisited: Bar and Spiral Arm Resonances}",
      journal = {\apjl},
         year = 2009,
        month = aug,
       volume = {700},
       number = {2},
        pages = {L78-L82},
          doi = {10.1088/0004-637X/700/2/L78},
archivePrefix = {arXiv},
       eprint = {0906.4682},
 primaryClass = {astro-ph.GA},
       adsurl = {https://ui.adsabs.harvard.edu/abs/2009ApJ...700L..78A}
}

@ARTICLE{Khachaturyants22,
       author = {{Khachaturyants}, Tigran and {Beraldo e Silva}, Leandro and {Debattista}, Victor P. and {Daniel}, Kathryne J.},
        title = "{Bending waves excited by irregular gas inflow along warps}",
      journal = {\mnras},
         year = 2022,
        month = may,
       volume = {512},
       number = {3},
        pages = {3500-3519},
          doi = {10.1093/mnras/stac606},
archivePrefix = {arXiv},
       eprint = {2203.03741},
 primaryClass = {astro-ph.GA},
       adsurl = {https://ui.adsabs.harvard.edu/abs/2022MNRAS.512.3500K}
}

@ARTICLE{Wang26,
       author = {{Wang}, Shuyu and {Brown}, Anthony G.~A. and {Debattista}, Victor P. and {Khachaturyants}, Tigran},
        title = "{Phase spirals induced by the gas warp}",
      journal = {arXiv e-prints},
         year = 2026,
        month = apr,
          eid = {arXiv:2604.06862},
        pages = {arXiv:2604.06862},
          doi = {10.48550/arXiv.2604.06862},
archivePrefix = {arXiv},
       eprint = {2604.06862},
 primaryClass = {astro-ph.GA},
       adsurl = {https://ui.adsabs.harvard.edu/abs/2026arXiv260406862W}
}

@ARTICLE{Debattista25,
       author = {{Debattista}, Victor P. and {Khachaturyants}, Tigran and {Amarante}, Jo{\~a}o A.~S. and {Carr}, Christopher and {Beraldo e Silva}, Leandro and {Laporte}, Chervin F.~P.},
        title = "{Azimuthal metallicity variations, spiral structure, and the failure of radial actions based on assuming axisymmetry}",
      journal = {\mnras},
         year = 2025,
        month = feb,
       volume = {537},
       number = {2},
        pages = {1620-1645},
          doi = {10.1093/mnras/staf035},
archivePrefix = {arXiv},
       eprint = {2402.08356},
 primaryClass = {astro-ph.GA},
       adsurl = {https://ui.adsabs.harvard.edu/abs/2025MNRAS.537.1620D}
}

@ARTICLE{Kuhn19,
       author = {{Kuhn}, Michael A. and {Hillenbrand}, Lynne A. and {Sills}, Alison and {Feigelson}, Eric D. and {Getman}, Konstantin V.},
        title = "{Kinematics in Young Star Clusters and Associations with Gaia DR2}",
      journal = {\apj},
         year = 2019,
        month = jan,
       volume = {870},
       number = {1},
          eid = {32},
        pages = {32},
          doi = {10.3847/1538-4357/aaef8c},
archivePrefix = {arXiv},
       eprint = {1807.02115},
 primaryClass = {astro-ph.GA},
       adsurl = {https://ui.adsabs.harvard.edu/abs/2019ApJ...870...32K}
}

@ARTICLE{Alves20,
       author = {{Alves}, Jo{\~a}o and {Zucker}, Catherine and {Goodman}, Alyssa A. and {Speagle}, Joshua S. and {Meingast}, Stefan and {Robitaille}, Thomas and {Finkbeiner}, Douglas P. and {Schlafly}, Edward F. and {Green}, Gregory M.},
        title = "{A Galactic-scale gas wave in the solar neighbourhood}",
      journal = {\nat},
         year = 2020,
        month = feb,
       volume = {578},
       number = {7794},
        pages = {237-239},
          doi = {10.1038/s41586-019-1874-z},
archivePrefix = {arXiv},
       eprint = {2001.08748},
 primaryClass = {astro-ph.GA},
       adsurl = {https://ui.adsabs.harvard.edu/abs/2020Natur.578..237A}
}

@ARTICLE{Banik23,
       author = {{Banik}, Uddipan and {van den Bosch}, Frank C. and {Weinberg}, Martin D.},
        title = "{A Comprehensive Perturbative Formalism for Phase Mixing in Perturbed Disks. II. Phase Spirals in an Inhomogeneous Disk Galaxy with a Nonresponsive Dark Matter Halo}",
      journal = {\apj},
         year = 2023,
        month = jul,
       volume = {952},
       number = {1},
          eid = {65},
        pages = {65},
          doi = {10.3847/1538-4357/acd641},
archivePrefix = {arXiv},
       eprint = {2303.00034},
 primaryClass = {astro-ph.GA},
       adsurl = {https://ui.adsabs.harvard.edu/abs/2023ApJ...952...65B}
}

@ARTICLE{Quillen11,
       author = {{Quillen}, Alice C. and {Dougherty}, Jamie and {Bagley}, Micaela B. and {Minchev}, Ivan and {Comparetta}, Justin},
        title = "{Structure in phase space associated with spiral and bar density waves in an N-body hybrid galactic disc}",
      journal = {\mnras},
         year = 2011,
        month = oct,
       volume = {417},
       number = {1},
        pages = {762-784},
          doi = {10.1111/j.1365-2966.2011.19349.x},
archivePrefix = {arXiv},
       eprint = {1010.5745},
 primaryClass = {astro-ph.GA},
       adsurl = {https://ui.adsabs.harvard.edu/abs/2011MNRAS.417..762Q}
}

@ARTICLE{Quillen18,
       author = {{Quillen}, Alice C. and {Carrillo}, Ismael and {Anders}, Friedrich and {McMillan}, Paul and {Hilmi}, Tariq and {Monari}, Giacomo and {Minchev}, Ivan and {Chiappini}, Cristina and {Khalatyan}, Arman and {Steinmetz}, Matthias},
        title = "{Spiral arm crossings inferred from ridges in Gaia stellar velocity distributions}",
      journal = {\mnras},
         year = 2018,
        month = nov,
       volume = {480},
       number = {3},
        pages = {3132-3139},
          doi = {10.1093/mnras/sty2077},
archivePrefix = {arXiv},
       eprint = {1805.10236},
 primaryClass = {astro-ph.GA},
       adsurl = {https://ui.adsabs.harvard.edu/abs/2018MNRAS.480.3132Q}
}

@ARTICLE{BW67,
       author = {{Barbanis}, B. and {Woltjer}, L.},
        title = "{Orbits in Spiral Galaxies and the Velocity Dispersion of Population i Stars}",
      journal = {\apj},
         year = 1967,
        month = nov,
       volume = {150},
        pages = {461},
          doi = {10.1086/149349},
       adsurl = {https://ui.adsabs.harvard.edu/abs/1967ApJ...150..461B}
}

@ARTICLE{Monari18,
       author = {{Monari}, G. and {Famaey}, B. and {Minchev}, I. and {Antoja}, T. and {Bienaym{\'e}}, O. and {Gibson}, B.~K. and {Grebel}, E.~K. and {Kordopatis}, G. and {McMillan}, P. and {Navarro}, J. and {Parker}, Q.~A. and {Quillen}, A.~C. and {Reid}, W. and {Seabroke}, G. and {Siebert}, A. and {Steinmetz}, M. and {Wyse}, R.~F.~G. and {Zwitter}, T.},
        title = "{Coma Berenices: The First Evidence for Incomplete Vertical Phase-mixing in Local Velocity Space with RAVE{\textemdash}Confirmed with Gaia DR2}",
      journal = {Research Notes of the American Astronomical Society},
         year = 2018,
        month = may,
       volume = {2},
       number = {2},
          eid = {32},
        pages = {32},
          doi = {10.3847/2515-5172/aac38e},
archivePrefix = {arXiv},
       eprint = {1804.07767},
 primaryClass = {astro-ph.GA},
       adsurl = {https://ui.adsabs.harvard.edu/abs/2018RNAAS...2...32M}
}

@ARTICLE{SchoenrichDehnen18,
       author = {{Sch{\"o}nrich}, Ralph and {Dehnen}, Walter},
        title = "{Warp, waves, and wrinkles in the Milky Way}",
      journal = {\mnras},
         year = 2018,
        month = aug,
       volume = {478},
       number = {3},
        pages = {3809-3824},
          doi = {10.1093/mnras/sty1256},
archivePrefix = {arXiv},
       eprint = {1712.06616},
 primaryClass = {astro-ph.GA},
       adsurl = {https://ui.adsabs.harvard.edu/abs/2018MNRAS.478.3809S}
}

@ARTICLE{Koppelman18,
       author = {{Koppelman}, Helmer and {Helmi}, Amina and {Veljanoski}, Jovan},
        title = "{One Large Blob and Many Streams Frosting the nearby Stellar Halo in Gaia DR2}",
      journal = {\apjl},
         year = 2018,
        month = jun,
       volume = {860},
       number = {1},
          eid = {L11},
        pages = {L11},
          doi = {10.3847/2041-8213/aac882},
archivePrefix = {arXiv},
       eprint = {1804.11347},
 primaryClass = {astro-ph.GA},
       adsurl = {https://ui.adsabs.harvard.edu/abs/2018ApJ...860L..11K}
}

@ARTICLE{Bland-Hawthorn19,
       author = {{Bland-Hawthorn}, Joss and {Sharma}, Sanjib and {Tepper-Garcia}, Thor and {Binney}, James and {Freeman}, Ken C. and {Hayden}, Michael R. and {Kos}, Janez and {De Silva}, Gayandhi M. and {Ellis}, Simon and {Lewis}, Geraint F. and {Asplund}, Martin and {Buder}, Sven and {Casey}, Andrew R. and {D'Orazi}, Valentina and {Duong}, Ly and {Khanna}, Shourya and {Lin}, Jane and {Lind}, Karin and {Martell}, Sarah L. and {Ness}, Melissa K. and {Simpson}, Jeffrey D. and {Zucker}, Daniel B. and {Zwitter}, Toma{\v{z}} and {Kafle}, Prajwal R. and {Quillen}, Alice C. and {Ting}, Yuan-Sen and {Wyse}, Rosemary F.~G.},
        title = "{The GALAH survey and Gaia DR2: dissecting the stellar disc's phase space by age, action, chemistry, and location}",
      journal = {\mnras},
         year = 2019,
        month = jun,
       volume = {486},
       number = {1},
        pages = {1167-1191},
          doi = {10.1093/mnras/stz217},
archivePrefix = {arXiv},
       eprint = {1809.02658},
 primaryClass = {astro-ph.GA},
       adsurl = {https://ui.adsabs.harvard.edu/abs/2019MNRAS.486.1167B}
}

@ARTICLE{Chakrabarty2007,
       author = {{Chakrabarty}, D.},
        title = "{Phase space structure in the solar neighbourhood}",
      journal = {\aap},
         year = 2007,
        month = may,
       volume = {467},
       number = {1},
        pages = {145-162},
          doi = {10.1051/0004-6361:20066677},
archivePrefix = {arXiv},
       eprint = {astro-ph/0703242},
 primaryClass = {astro-ph},
       adsurl = {https://ui.adsabs.harvard.edu/abs/2007A&A...467..145C}
}

@ARTICLE{Raboud98,
       author = {{Raboud}, D. and {Grenon}, M. and {Martinet}, L. and {Fux}, R. and {Udry}, S.},
        title = "{Evidence for a signature of the galactic bar in the solar neighbourhood}",
      journal = {\aap},
         year = 1998,
        month = jul,
       volume = {335},
        pages = {L61-L64},
          doi = {10.48550/arXiv.astro-ph/9802266},
archivePrefix = {arXiv},
       eprint = {astro-ph/9802266},
 primaryClass = {astro-ph},
       adsurl = {https://ui.adsabs.harvard.edu/abs/1998A&A...335L..61R}
}

@ARTICLE{Fragkoudi19,
       author = {{Fragkoudi}, F. and {Katz}, D. and {Trick}, W. and {White}, S.~D.~M. and {Di Matteo}, P. and {Sormani}, M.~C. and {Khoperskov}, S. and {Haywood}, M. and {Hall{\'e}}, A. and {G{\'o}mez}, A.},
        title = "{On the ridges, undulations, and streams in Gaia DR2: linking the topography of phase space to the orbital structure of an N-body bar}",
      journal = {\mnras},
         year = 2019,
        month = sep,
       volume = {488},
       number = {3},
        pages = {3324-3339},
          doi = {10.1093/mnras/stz1875},
archivePrefix = {arXiv},
       eprint = {1901.07568},
 primaryClass = {astro-ph.GA},
       adsurl = {https://ui.adsabs.harvard.edu/abs/2019MNRAS.488.3324F}
}

@ARTICLE{Khoperskov19,
       author = {{Khoperskov}, Sergey and {Di Matteo}, Paola and {Gerhard}, Ortwin and {Katz}, David and {Haywood}, Misha and {Combes}, Fran{\c{c}}oise and {Berczik}, Peter and {Gomez}, Ana},
        title = "{The echo of the bar buckling: Phase-space spirals in Gaia Data Release 2}",
      journal = {\aap},
         year = 2019,
        month = feb,
       volume = {622},
          eid = {L6},
        pages = {L6},
          doi = {10.1051/0004-6361/201834707},
archivePrefix = {arXiv},
       eprint = {1811.09205},
 primaryClass = {astro-ph.GA},
       adsurl = {https://ui.adsabs.harvard.edu/abs/2019A&A...622L...6K}
}

@ARTICLE{Khanna19,
       author = {{Khanna}, Shourya and {Sharma}, Sanjib and {Tepper-Garcia}, Thor and {Bland-Hawthorn}, Joss and {Hayden}, Michael and {Asplund}, Martin and {Buder}, Sven and {Chen}, Boquan and {De Silva}, Gayandhi M. and {Freeman}, Ken C. and {Kos}, Janez and {Lewis}, Geraint F. and {Lin}, Jane and {Martell}, Sarah L. and {Simpson}, Jeffrey D. and {Nordlander}, Thomas and {Stello}, Dennis and {Ting}, Yuan-Sen and {Zucker}, Daniel B. and {Zwitter}, Toma{\v{z}}},
        title = "{The GALAH survey and Gaia DR2: Linking ridges, arches, and vertical waves in the kinematics of the Milky Way}",
      journal = {\mnras},
         year = 2019,
        month = nov,
       volume = {489},
       number = {4},
        pages = {4962-4979},
          doi = {10.1093/mnras/stz2462},
archivePrefix = {arXiv},
       eprint = {1902.10113},
 primaryClass = {astro-ph.GA},
       adsurl = {https://ui.adsabs.harvard.edu/abs/2019MNRAS.489.4962K}
}

@ARTICLE{Minchev09,
       author = {{Minchev}, I. and {Quillen}, A.~C. and {Williams}, M. and {Freeman}, K.~C. and {Nordhaus}, J. and {Siebert}, A. and {Bienaym{\'e}}, O.},
        title = "{Is the Milky Way ringing? The hunt for high-velocity streams}",
      journal = {\mnras},
         year = 2009,
        month = jun,
       volume = {396},
       number = {1},
        pages = {L56-L60},
          doi = {10.1111/j.1745-3933.2009.00661.x},
archivePrefix = {arXiv},
       eprint = {0902.1531},
 primaryClass = {astro-ph.GA},
       adsurl = {https://ui.adsabs.harvard.edu/abs/2009MNRAS.396L..56M}
}

@ARTICLE{Gomez12,
       author = {{G{\'o}mez}, Facundo A. and {Minchev}, Ivan and {Villalobos}, {\'A}lvaro and {O'Shea}, Brian W. and {Williams}, Mary E.~K.},
        title = "{Signatures of minor mergers in Milky Way like disc kinematics: ringing revisited}",
      journal = {\mnras},
         year = 2012,
        month = jan,
       volume = {419},
       number = {3},
        pages = {2163-2172},
          doi = {10.1111/j.1365-2966.2011.19867.x},
archivePrefix = {arXiv},
       eprint = {1105.4231},
 primaryClass = {astro-ph.GA},
       adsurl = {https://ui.adsabs.harvard.edu/abs/2012MNRAS.419.2163G}
}

@ARTICLE{Helmi99,
       author = {{Helmi}, Amina and {White}, Simon D.~M. and {de Zeeuw}, P. Tim and {Zhao}, Hongsheng},
        title = "{Debris streams in the solar neighbourhood as relicts from the formation of the Milky Way}",
      journal = {\nat},
         year = 1999,
        month = nov,
       volume = {402},
       number = {6757},
        pages = {53-55},
          doi = {10.1038/46980},
archivePrefix = {arXiv},
       eprint = {astro-ph/9911041},
 primaryClass = {astro-ph},
       adsurl = {https://ui.adsabs.harvard.edu/abs/1999Natur.402...53H}
}

@ARTICLE{Laporte19,
       author = {{Laporte}, Chervin F.~P. and {Minchev}, Ivan and {Johnston}, Kathryn V. and {G{\'o}mez}, Facundo A.},
        title = "{Footprints of the Sagittarius dwarf galaxy in the Gaia data set}",
      journal = {\mnras},
         year = 2019,
        month = may,
       volume = {485},
       number = {3},
        pages = {3134-3152},
          doi = {10.1093/mnras/stz583},
archivePrefix = {arXiv},
       eprint = {1808.00451},
 primaryClass = {astro-ph.GA},
       adsurl = {https://ui.adsabs.harvard.edu/abs/2019MNRAS.485.3134L}
}

@ARTICLE{LBK72,
       author = {{Lynden-Bell}, D. and {Kalnajs}, A.~J.},
        title = "{On the generating mechanism of spiral structure}",
      journal = {\mnras},
         year = 1972,
        month = jan,
       volume = {157},
        pages = {1},
          doi = {10.1093/mnras/157.1.1},
       adsurl = {https://ui.adsabs.harvard.edu/abs/1972MNRAS.157....1L}
}

@ARTICLE{SellwoodMasters22,
       author = {{Sellwood}, J.~A. and {Masters}, Karen L.},
        title = "{Spirals in Galaxies}",
      journal = {\araa},
         year = 2022,
        month = aug,
       volume = {60},
          doi = {10.1146/annurev-astro-052920-104505},
archivePrefix = {arXiv},
       eprint = {2110.05615},
 primaryClass = {astro-ph.GA},
       adsurl = {https://ui.adsabs.harvard.edu/abs/2022ARA&A..60...73S}
}

@ARTICLE{Kazantzidis08,
       author = {{Kazantzidis}, Stelios and {Bullock}, James S. and {Zentner}, Andrew R. and {Kravtsov}, Andrey V. and {Moustakas}, Leonidas A.},
        title = "{Cold Dark Matter Substructure and Galactic Disks. I. Morphological Signatures of Hierarchical Satellite Accretion}",
      journal = {\apj},
         year = 2008,
        month = nov,
       volume = {688},
       number = {1},
        pages = {254-276},
          doi = {10.1086/591958},
archivePrefix = {arXiv},
       eprint = {0708.1949},
 primaryClass = {astro-ph},
       adsurl = {https://ui.adsabs.harvard.edu/abs/2008ApJ...688..254K}
}

@ARTICLE{Roskar08b,
       author = {{Ro{\v{s}}kar}, Rok and {Debattista}, Victor P. and {Quinn}, Thomas R. and {Stinson}, Gregory S. and {Wadsley}, James},
        title = "{Riding the Spiral Waves: Implications of Stellar Migration for the Properties of Galactic Disks}",
      journal = {\apjl},
         year = 2008,
        month = sep,
       volume = {684},
       number = {2},
        pages = {L79},
          doi = {10.1086/592231},
archivePrefix = {arXiv},
       eprint = {0808.0206},
 primaryClass = {astro-ph},
       adsurl = {https://ui.adsabs.harvard.edu/abs/2008ApJ...684L..79R}
}

@ARTICLE{Mark74,
       author = {{Mark}, J.~W. -K.},
        title = "{On density waves in galaxies. I. Source terms and action conservation.}",
      journal = {\apj},
         year = 1974,
        month = nov,
       volume = {193},
        pages = {539-559},
          doi = {10.1086/153192},
       adsurl = {https://ui.adsabs.harvard.edu/abs/1974ApJ...193..539M}
}

@ARTICLE{JB90,
       author = {{Jenkins}, Adrian and {Binney}, James},
        title = "{Spiral heating of galactic discs}",
      journal = {\mnras},
         year = 1990,
        month = jul,
       volume = {245},
        pages = {305-317},
       adsurl = {https://ui.adsabs.harvard.edu/abs/1990MNRAS.245..305J}
}

@ARTICLE{Beane18,
       author = {{Beane}, Angus and {Ness}, Melissa K. and {Bedell}, Megan},
        title = "{Actions Are Weak Stellar Age Indicators in the Milky Way Disk}",
      journal = {\apj},
         year = 2018,
        month = nov,
       volume = {867},
       number = {1},
          eid = {31},
        pages = {31},
          doi = {10.3847/1538-4357/aae07f},
archivePrefix = {arXiv},
       eprint = {1807.05986},
 primaryClass = {astro-ph.SR},
       adsurl = {https://ui.adsabs.harvard.edu/abs/2018ApJ...867...31B}
}

@ARTICLE{Dehnen2000,
       author = {{Dehnen}, Walter},
        title = "{The Effect of the Outer Lindblad Resonance of the Galactic Bar on the Local Stellar Velocity Distribution}",
      journal = {\aj},
         year = 2000,
        month = feb,
       volume = {119},
       number = {2},
        pages = {800-812},
          doi = {10.1086/301226},
archivePrefix = {arXiv},
       eprint = {astro-ph/9911161},
 primaryClass = {astro-ph},
       adsurl = {https://ui.adsabs.harvard.edu/abs/2000AJ....119..800D}
}

@ARTICLE{Fux01,
       author = {{Fux}, R.},
        title = "{Order and chaos in the local disc stellar kinematics induced by the Galactic bar}",
      journal = {\aap},
         year = 2001,
        month = jul,
       volume = {373},
        pages = {511-535},
          doi = {10.1051/0004-6361:20010561},
archivePrefix = {arXiv},
       eprint = {astro-ph/0105398},
 primaryClass = {astro-ph},
       adsurl = {https://ui.adsabs.harvard.edu/abs/2001A&A...373..511F}
}

@ARTICLE{Stromberg46,
       author = {{Str{\"o}mberg}, Gustaf},
        title = "{The Motions of the Stars Within 20 Parsecs of the Sun.}",
      journal = {\apj},
         year = 1946,
        month = jul,
       volume = {104},
        pages = {12},
          doi = {10.1086/144830},
       adsurl = {https://ui.adsabs.harvard.edu/abs/1946ApJ...104...12S}
}

@ARTICLE{Roman50a,
       author = {{Roman}, Nancy G.},
        title = "{Some characteristics of the spectra of F-, G-, and K-type stars.}",
      journal = {\aj},
         year = 1950,
        month = oct,
       volume = {55},
        pages = {182},
          doi = {10.1086/106403},
       adsurl = {https://ui.adsabs.harvard.edu/abs/1950AJ.....55..182R}
}

@ARTICLE{Roman50b,
       author = {{Roman}, Nancy G.},
        title = "{A Correlation Between the Spectroscopic and Dynamical Characteristics of the Late F - and Early G - Type Stars.}",
      journal = {\apj},
         year = 1950,
        month = nov,
       volume = {112},
        pages = {554},
          doi = {10.1086/145367},
       adsurl = {https://ui.adsabs.harvard.edu/abs/1950ApJ...112..554R}
}

@ARTICLE{SeabrokeGilmore07,
       author = {{Seabroke}, G.~M. and {Gilmore}, G.},
        title = "{Revisiting the relations: Galactic thin disc age-velocity dispersion relation}",
      journal = {\mnras},
         year = 2007,
        month = oct,
       volume = {380},
       number = {4},
        pages = {1348-1368},
          doi = {10.1111/j.1365-2966.2007.12210.x},
archivePrefix = {arXiv},
       eprint = {0707.1027},
 primaryClass = {astro-ph},
       adsurl = {https://ui.adsabs.harvard.edu/abs/2007MNRAS.380.1348S}
}

@ARTICLE{Weinberg94,
       author = {{Weinberg}, Martin D.},
        title = "{Kinematic Signature of a Rotating Bar near a Resonance}",
      journal = {\apj},
         year = 1994,
        month = jan,
       volume = {420},
        pages = {597},
          doi = {10.1086/173589},
archivePrefix = {arXiv},
       eprint = {astro-ph/9304026},
 primaryClass = {astro-ph},
       adsurl = {https://ui.adsabs.harvard.edu/abs/1994ApJ...420..597W}
}

@ARTICLE{Bird21,
       author = {{Bird}, Jonathan C. and {Loebman}, Sarah R. and {Weinberg}, David H. and {Brooks}, Alyson M. and {Quinn}, Thomas R. and {Christensen}, Charlotte R.},
        title = "{Inside out and upside-down: The roles of gas cooling and dynamical heating in shaping the stellar age-velocity relation}",
      journal = {\mnras},
         year = 2021,
        month = may,
       volume = {503},
       number = {2},
        pages = {1815-1827},
          doi = {10.1093/mnras/stab289},
archivePrefix = {arXiv},
       eprint = {2005.12948},
 primaryClass = {astro-ph.GA},
       adsurl = {https://ui.adsabs.harvard.edu/abs/2021MNRAS.503.1815B}
}

@ARTICLE{Lacey84,
       author = {{Lacey}, C.~G.},
        title = "{The influence of massive gas clouds on stellar velocity dispersions in galactic discs}",
      journal = {\mnras},
         year = 1984,
        month = jun,
       volume = {208},
        pages = {687-707},
          doi = {10.1093/mnras/208.4.687},
       adsurl = {https://ui.adsabs.harvard.edu/abs/1984MNRAS.208..687L}
}

@ARTICLE{Weinberg04,
       author = {{Weinberg}, Martin D.},
        title = "{Time-dependent secular evolution in galaxies}",
      journal = {arXiv e-prints},
         year = 2004,
        month = apr,
          eid = {astro-ph/0404169},
        pages = {astro-ph/0404169},
          doi = {10.48550/arXiv.astro-ph/0404169},
archivePrefix = {arXiv},
       eprint = {astro-ph/0404169},
 primaryClass = {astro-ph},
       adsurl = {https://ui.adsabs.harvard.edu/abs/2004astro.ph..4169W}
}

@ARTICLE{Bird12,
       author = {{Bird}, Jonathan C. and {Kazantzidis}, Stelios and {Weinberg}, David H.},
        title = "{Radial mixing in galactic discs: the effects of disc structure and satellite bombardment}",
      journal = {\mnras},
         year = 2012,
        month = feb,
       volume = {420},
       number = {2},
        pages = {913-925},
          doi = {10.1111/j.1365-2966.2011.19728.x},
archivePrefix = {arXiv},
       eprint = {1104.0933},
 primaryClass = {astro-ph.GA},
       adsurl = {https://ui.adsabs.harvard.edu/abs/2012MNRAS.420..913B}
}

@ARTICLE{McMillan11a,
       author = {{McMillan}, Paul J.},
        title = "{Mass models of the Milky Way}",
      journal = {\mnras},
         year = 2011,
        month = jul,
       volume = {414},
       number = {3},
        pages = {2446-2457},
          doi = {10.1111/j.1365-2966.2011.18564.x},
archivePrefix = {arXiv},
       eprint = {1102.4340},
 primaryClass = {astro-ph.GA},
       adsurl = {https://ui.adsabs.harvard.edu/abs/2011MNRAS.414.2446M}
}

@ARTICLE{SC84,
       author = {{Sellwood}, J.~A. and {Carlberg}, R.~G.},
        title = "{Spiral instabilities provoked by accretion and star formation}",
      journal = {\apj},
         year = 1984,
        month = jul,
       volume = {282},
        pages = {61-74},
          doi = {10.1086/162176},
       adsurl = {https://ui.adsabs.harvard.edu/abs/1984ApJ...282...61S}
}

@ARTICLE{AumerBinney09,
       author = {{Aumer}, Michael and {Binney}, James J.},
        title = "{Kinematics and history of the solar neighbourhood revisited}",
      journal = {\mnras},
         year = 2009,
        month = aug,
       volume = {397},
       number = {3},
        pages = {1286-1301},
          doi = {10.1111/j.1365-2966.2009.15053.x},
archivePrefix = {arXiv},
       eprint = {0905.2512},
 primaryClass = {astro-ph.GA},
       adsurl = {https://ui.adsabs.harvard.edu/abs/2009MNRAS.397.1286A}
}

@ARTICLE{Binney10,
       author = {{Binney}, James},
        title = "{Distribution functions for the Milky Way}",
      journal = {\mnras},
         year = 2010,
        month = feb,
       volume = {401},
       number = {4},
        pages = {2318-2330},
          doi = {10.1111/j.1365-2966.2009.15845.x},
archivePrefix = {arXiv},
       eprint = {0910.1512},
 primaryClass = {astro-ph.GA},
       adsurl = {https://ui.adsabs.harvard.edu/abs/2010MNRAS.401.2318B}
}

@ARTICLE{BM11,
       author = {{Binney}, James and {McMillan}, Paul},
        title = "{Models of our Galaxy - II}",
      journal = {\mnras},
         year = 2011,
        month = may,
       volume = {413},
       number = {3},
        pages = {1889-1898},
          doi = {10.1111/j.1365-2966.2011.18268.x},
archivePrefix = {arXiv},
       eprint = {1101.0747},
 primaryClass = {astro-ph.GA},
       adsurl = {https://ui.adsabs.harvard.edu/abs/2011MNRAS.413.1889B}
}

@ARTICLE{deSilva15GALAH,
       author = {{De Silva}, G.~M. and {Freeman}, K.~C. and {Bland-Hawthorn}, J. and
         {Martell}, S. and {de Boer}, E. Wylie and {Asplund}, M. and
         {Keller}, S. and {Sharma}, S. and {Zucker}, D.~B. and {Zwitter}, T. and
         {Anguiano}, B. and {Bacigalupo}, C. and {Bayliss}, D. and
         {Beavis}, M.~A. and {Bergemann}, M. and {Campbell}, S. and
         {Cannon}, R. and {Carollo}, D. and {Casagrande}, L. and {Casey}, A.~R. and
         {Da Costa}, G. and {D'Orazi}, V. and {Dotter}, A. and {Duong}, L. and
         {Heger}, A. and {Ireland}, M.~J. and {Kafle}, P.~R. and {Kos}, J. and
         {Lattanzio}, J. and {Lewis}, G.~F. and {Lin}, J. and {Lind}, K. and
         {Munari}, U. and {Nataf}, D.~M. and {O'Toole}, S. and {Parker}, Q. and
         {Reid}, W. and {Schlesinger}, K.~J. and {Sheinis}, A. and
         {Simpson}, J.~D. and {Stello}, D. and {Ting}, Y. -S. and {Traven}, G. and
         {Watson}, F. and {Wittenmyer}, R. and {Yong}, D. and {{\v{Z}}erjal}, M.},
        title = "{The GALAH survey: scientific motivation}",
      journal = {\mnras},
         year = 2015,
        month = may,
       volume = {449},
       number = {3},
        pages = {2604-2617},
          doi = {10.1093/mnras/stv327},
archivePrefix = {arXiv},
       eprint = {1502.04767},
 primaryClass = {astro-ph.GA},
       adsurl = {https://ui.adsabs.harvard.edu/abs/2015MNRAS.449.2604D}
}

@BOOK{Chandrasekhar42Book,
       author = {{Chandrasekhar}, Subrahmanyan},
        title = "{Principles of stellar dynamics}",
         year = 1942,
       adsurl = {https://ui.adsabs.harvard.edu/abs/1942psd..book.....C}
}

@ARTICLE{Jeans1915,
       author = {{Jeans}, J.~H.},
        title = "{On the theory of star-streaming and the structure of the universe}",
      journal = {\mnras},
         year = 1915,
        month = dec,
       volume = {76},
        pages = {70-84},
          doi = {10.1093/mnras/76.2.70},
       adsurl = {https://ui.adsabs.harvard.edu/abs/1915MNRAS..76...70J}
}

@ARTICLE{Shu77,
       author = {{Shu}, F.~H.},
        title = "{Self-similar collapse of isothermal spheres and star formation.}",
      journal = {\apj},
         year = 1977,
        month = jun,
       volume = {214},
        pages = {488-497},
          doi = {10.1086/155274},
       adsurl = {https://ui.adsabs.harvard.edu/abs/1977ApJ...214..488S}
}

@ARTICLE{Sellwood11,
       author = {{Sellwood}, J.~A.},
        title = "{The lifetimes of spiral patterns in disc galaxies}",
      journal = {\mnras},
         year = 2011,
        month = jan,
       volume = {410},
       number = {3},
        pages = {1637-1646},
          doi = {10.1111/j.1365-2966.2010.17545.x},
archivePrefix = {arXiv},
       eprint = {1008.2737},
 primaryClass = {astro-ph.CO},
       adsurl = {https://ui.adsabs.harvard.edu/abs/2011MNRAS.410.1637S}
}

@ARTICLE{Sellwood12,
       author = {{Sellwood}, J.~A.},
        title = "{Spiral Instabilities in N-body Simulations. I. Emergence from Noise}",
      journal = {\apj},
         year = 2012,
        month = may,
       volume = {751},
       number = {1},
          eid = {44},
        pages = {44},
          doi = {10.1088/0004-637X/751/1/44},
archivePrefix = {arXiv},
       eprint = {1203.0444},
 primaryClass = {astro-ph.GA},
       adsurl = {https://ui.adsabs.harvard.edu/abs/2012ApJ...751...44S}
}

@ARTICLE{FouvryPichon15b,
       author = {{Fouvry}, Jean-Baptiste and {Pichon}, Christophe},
        title = "{Secular resonant dressed orbital diffusion - II. Application to an isolated self-similar tepid galactic disc}",
      journal = {\mnras},
         year = 2015,
        month = may,
       volume = {449},
       number = {2},
        pages = {1982-1995},
          doi = {10.1093/mnras/stv360},
archivePrefix = {arXiv},
       eprint = {1504.04831},
 primaryClass = {astro-ph.GA},
       adsurl = {https://ui.adsabs.harvard.edu/abs/2015MNRAS.449.1982F}
}

@misc{McAuley23, 
    author = {McAuley, Olivia}, 
    title = {Unpublished master's thesis: "Tracing the Resonances in Barred Galaxies", Bryn Mawr College}, 
    year = {2023}
}

@ARTICLE{Frankel20,
       author = {{Frankel}, Neige and {Sanders}, Jason and {Ting}, Yuan-Sen and
         {Rix}, Hans-Walter},
        title = "{Keeping It Cool: Much Orbit Migration, yet Little Heating, in the Galactic Disk}",
      journal = {\apj},
         year = 2020,
        month = jun,
       volume = {896},
       number = {1},
          eid = {15},
        pages = {15},
          doi = {10.3847/1538-4357/ab910c},
archivePrefix = {arXiv},
       eprint = {2002.04622},
 primaryClass = {astro-ph.GA},
       adsurl = {https://ui.adsabs.harvard.edu/abs/2020ApJ...896...15F}
}

@ARTICLE{Ricker15,
       author = {{Ricker}, George R. and {Winn}, Joshua N. and {Vanderspek}, Roland and
         {Latham}, David W. and {Bakos}, G{\'a}sp{\'a}r {\'A}. and
         {Bean}, Jacob L. and {Berta-Thompson}, Zachory K. and
         {Brown}, Timothy M. and {Buchhave}, Lars and {Butler}, Nathaniel R.},
        title = "{Transiting Exoplanet Survey Satellite (TESS)}",
      journal = {Journal of Astronomical Telescopes, Instruments, and Systems},
         year = "2015",
        month = "Jan",
       volume = {1},
          eid = {014003},
        pages = {014003},
          doi = {10.1117/1.JATIS.1.1.014003},
       adsurl = {https://ui.adsabs.harvard.edu/abs/2015JATIS...1a4003R}
}

@ARTICLE{QuillenMinchev05,
       author = {{Quillen}, A.~C. and {Minchev}, Ivan},
        title = "{The Effect of Spiral Structure on the Stellar Velocity Distribution in the Solar Neighborhood}",
      journal = {\aj},
         year = 2005,
        month = aug,
       volume = {130},
       number = {2},
        pages = {576-585},
          doi = {10.1086/430885},
archivePrefix = {arXiv},
       eprint = {astro-ph/0502205},
 primaryClass = {astro-ph},
       adsurl = {https://ui.adsabs.harvard.edu/abs/2005AJ....130..576Q}
}

@ARTICLE{Sellwood10b,
       author = {{Sellwood}, J.~A.},
        title = "{A recent Lindblad resonance in the solar neighbourhood}",
      journal = {\mnras},
         year = "2010",
        month = "Nov",
       volume = {409},
       number = {1},
        pages = {145-155},
          doi = {10.1111/j.1365-2966.2010.17305.x},
archivePrefix = {arXiv},
       eprint = {1001.5197},
 primaryClass = {astro-ph.GA},
       adsurl = {https://ui.adsabs.harvard.edu/abs/2010MNRAS.409..145S}
}

@ARTICLE{Wisz26,
    author = {{Wisz}, M.E. and {Masters}, Karen L. and {Daniel} Kathryne J. and {Stark}, David V. and {Belfiore}, Francesco},
    title = "{The Impact of Bars, Spirals and Bulge-Size on Gas-Phase Metallicity Gradients in MaNGA Galaxies}",
    journal = {in preparation},
    note = {in preparation},
    year = 2026
}

@ARTICLE{Dehnen98,
   author = {{Dehnen}, W.},
    title = "{The Distribution of Nearby Stars in Velocity Space Inferred from HIPPARCOS Data}",
  journal = {AJ},
   eprint = {astro-ph/9803110},
     year = 1998,
    month = jun,
   volume = 115,
    pages = {2384-2396},
      doi = {10.1086/300364},
   adsurl = {http://adsabs.harvard.edu/abs/1998AJ....115.2384D}
}

@ARTICLE{Hunt19,
       author = {{Hunt}, Jason A.~S. and {Bub}, Mathew W. and {Bovy}, Jo and {Mackereth}, J. Ted and {Trick}, Wilma H. and {Kawata}, Daisuke},
        title = "{Signatures of resonance and phase mixing in the Galactic disc}",
      journal = {\mnras},
         year = 2019,
        month = nov,
       volume = {490},
       number = {1},
        pages = {1026-1043},
          doi = {10.1093/mnras/stz2667},
archivePrefix = {arXiv},
       eprint = {1904.10968},
 primaryClass = {astro-ph.GA},
       adsurl = {https://ui.adsabs.harvard.edu/abs/2019MNRAS.490.1026H}
}

@ARTICLE{HuntBovy18,
       author = {{Hunt}, Jason A.~S. and {Bovy}, Jo},
        title = "{The 4:1 outer Lindblad resonance of a long-slow bar as an explanation for the Hercules stream}",
      journal = {\mnras},
         year = 2018,
        month = jul,
       volume = {477},
       number = {3},
        pages = {3945-3953},
          doi = {10.1093/mnras/sty921},
archivePrefix = {arXiv},
       eprint = {1803.02358},
 primaryClass = {astro-ph.GA},
       adsurl = {https://ui.adsabs.harvard.edu/abs/2018MNRAS.477.3945H}
}

@ARTICLE{DanielWyse18,
       author = {{Daniel}, Kathryne J. and {Wyse}, Rosemary F.~G.},
        title = "{Constraints on radial migration in spiral galaxies - II. Angular momentum distribution and preferential migration}",
      journal = {\mnras},
         year = 2018,
        month = may,
       volume = {476},
       number = {2},
        pages = {1561-1580},
          doi = {10.1093/mnras/sty199},
archivePrefix = {arXiv},
       eprint = {1801.08455},
 primaryClass = {astro-ph.GA},
       adsurl = {https://ui.adsabs.harvard.edu/abs/2018MNRAS.476.1561D}
}

@ARTICLE{Khrapov21,
       author = {{Khrapov}, Sergey and {Khoperskov}, Alexander and {Korchagin}, Vladimir},
        title = "{Modeling of Spiral Structure in a Multi-Component Milky Way-Like Galaxy}",
      journal = {Galaxies},
         year = 2021,
        month = apr,
       volume = {9},
       number = {2},
          eid = {29},
        pages = {29},
          doi = {10.3390/galaxies9020029},
archivePrefix = {arXiv},
       eprint = {2105.03198},
 primaryClass = {astro-ph.GA},
       adsurl = {https://ui.adsabs.harvard.edu/abs/2021Galax...9...29K}
}

@ARTICLE{Hunt18b,
       author = {{Hunt}, Jason A.~S. and {Hong}, Jack and {Bovy}, Jo and
         {Kawata}, Daisuke and {Grand}, Robert J.~J.},
        title = "{Transient spiral structure and the disc velocity substructure in Gaia DR2}",
      journal = {\mnras},
         year = "2018",
        month = "Dec",
       volume = {481},
       number = {3},
        pages = {3794-3803},
          doi = {10.1093/mnras/sty2532},
archivePrefix = {arXiv},
       eprint = {1806.02832},
 primaryClass = {astro-ph.GA},
       adsurl = {https://ui.adsabs.harvard.edu/abs/2018MNRAS.481.3794H}
}

@ARTICLE{DeSimone04,
       author = {{De Simone}, Richard and {Wu}, Xiaoan and {Tremaine}, Scott},
        title = "{The stellar velocity distribution in the solar neighbourhood}",
      journal = {\mnras},
         year = 2004,
        month = may,
       volume = {350},
       number = {2},
        pages = {627-643},
          doi = {10.1111/j.1365-2966.2004.07675.x},
archivePrefix = {arXiv},
       eprint = {astro-ph/0310906},
 primaryClass = {astro-ph},
       adsurl = {https://ui.adsabs.harvard.edu/abs/2004MNRAS.350..627D}
}

@ARTICLE{Grand12a,
       author = {{Grand}, Robert J.~J. and {Kawata}, Daisuke and {Cropper}, Mark},
        title = "{The dynamics of stars around spiral arms}",
      journal = {\mnras},
         year = 2012,
        month = apr,
       volume = {421},
       number = {2},
        pages = {1529-1538},
          doi = {10.1111/j.1365-2966.2012.20411.x},
archivePrefix = {arXiv},
       eprint = {1112.0019},
 primaryClass = {astro-ph.GA},
       adsurl = {https://ui.adsabs.harvard.edu/abs/2012MNRAS.421.1529G}
}

@ARTICLE{MiyamotoNagai75,
       author = {{Miyamoto}, M. and {Nagai}, R.},
        title = "{Three-dimensional models for the distribution of mass in galaxies.}",
      journal = {\pasj},
         year = 1975,
        month = jan,
       volume = {27},
        pages = {533-543},
       adsurl = {https://ui.adsabs.harvard.edu/abs/1975PASJ...27..533M}
}

@ARTICLE{DehnenBinney98b,
       author = {{Dehnen}, Walter and {Binney}, James J.},
        title = "{Local stellar kinematics from HIPPARCOS data}",
      journal = {\mnras},
         year = 1998,
        month = aug,
       volume = {298},
       number = {2},
        pages = {387-394},
          doi = {10.1046/j.1365-8711.1998.01600.x},
archivePrefix = {arXiv},
       eprint = {astro-ph/9710077},
 primaryClass = {astro-ph},
       adsurl = {https://ui.adsabs.harvard.edu/abs/1998MNRAS.298..387D}
}

@ARTICLE{CoxGomez02,
       author = {{Cox}, Donald P. and {G{\'o}mez}, Gilberto C.},
        title = "{Analytical Expressions for Spiral Arm Gravitational Potential and Density}",
      journal = {\apjs},
         year = 2002,
        month = oct,
       volume = {142},
       number = {2},
        pages = {261-267},
          doi = {10.1086/341946},
archivePrefix = {arXiv},
       eprint = {astro-ph/0207635},
 primaryClass = {astro-ph},
       adsurl = {https://ui.adsabs.harvard.edu/abs/2002ApJS..142..261C}
}

@ARTICLE{Grand15,
       author = {{Grand}, Robert J.~J. and {Bovy}, Jo and {Kawata}, Daisuke and {Hunt}, Jason A.~S. and {Famaey}, Benoit and {Siebert}, Arnaud and {Monari}, Giacomo and {Cropper}, Mark},
        title = "{Spiral- and bar-driven peculiar velocities in Milky Way-sized galaxy simulations}",
      journal = {\mnras},
         year = 2015,
        month = oct,
       volume = {453},
       number = {2},
        pages = {1867-1878},
          doi = {10.1093/mnras/stv1785},
archivePrefix = {arXiv},
       eprint = {1506.02668},
 primaryClass = {astro-ph.GA},
       adsurl = {https://ui.adsabs.harvard.edu/abs/2015MNRAS.453.1867G}
}

@ARTICLE{NFW96,
       author = {{Navarro}, Julio F. and {Frenk}, Carlos S. and {White}, Simon D.~M.},
        title = "{The Structure of Cold Dark Matter Halos}",
      journal = {\apj},
         year = 1996,
        month = may,
       volume = {462},
        pages = {563},
          doi = {10.1086/177173},
archivePrefix = {arXiv},
       eprint = {astro-ph/9508025},
 primaryClass = {astro-ph},
       adsurl = {https://ui.adsabs.harvard.edu/abs/1996ApJ...462..563N}
}

@ARTICLE{Kennicutt81,
       author = {{Kennicutt}, R.~C., Jr.},
        title = "{The shapes of spiral arms along the Hubble sequence.}",
      journal = {\aj},
         year = 1981,
        month = dec,
       volume = {86},
        pages = {1847-1858},
          doi = {10.1086/113064},
       adsurl = {https://ui.adsabs.harvard.edu/abs/1981AJ.....86.1847K}
}

@ARTICLE{YuHo20,
       author = {{Yu}, Si-Yue and {Ho}, Luis C.},
        title = "{The Statistical Properties of Spiral Arms in Nearby Disk Galaxies}",
      journal = {\apj},
         year = 2020,
        month = sep,
       volume = {900},
       number = {2},
          eid = {150},
        pages = {150},
          doi = {10.3847/1538-4357/abac5b},
       adsurl = {https://ui.adsabs.harvard.edu/abs/2020ApJ...900..150Y}
}

@ARTICLE{Diaz-Garcia19,
       author = {{D{\'\i}az-Garc{\'\i}a}, S. and {Salo}, H. and {Knapen}, J.~H. and {Herrera-Endoqui}, M.},
        title = "{The shapes of spiral arms in the S$^{4}$G survey and their connection with stellar bars}",
      journal = {\aap},
         year = 2019,
        month = nov,
       volume = {631},
          eid = {A94},
        pages = {A94},
          doi = {10.1051/0004-6361/201936000},
archivePrefix = {arXiv},
       eprint = {1908.04246},
 primaryClass = {astro-ph.GA},
       adsurl = {https://ui.adsabs.harvard.edu/abs/2019A&A...631A..94D}
}

@ARTICLE{Masters19,
       author = {{Masters}, Karen L. and {Lintott}, Chris J. and {Hart}, Ross E. and {Kruk}, Sandor J. and {Smethurst}, Rebecca J. and {Casteels}, Kevin V. and {Keel}, William C. and {Simmons}, Brooke D. and {Stanescu}, Dennis O. and {Tate}, Jean and {Tomi}, Satoshi},
        title = "{Galaxy Zoo: unwinding the winding problem - observations of spiral bulge prominence and arm pitch angles suggest local spiral galaxies are winding}",
      journal = {\mnras},
         year = 2019,
        month = aug,
       volume = {487},
       number = {2},
        pages = {1808-1820},
          doi = {10.1093/mnras/stz1153},
archivePrefix = {arXiv},
       eprint = {1904.11436},
 primaryClass = {astro-ph.GA},
       adsurl = {https://ui.adsabs.harvard.edu/abs/2019MNRAS.487.1808M}
}

@ARTICLE{BovyRix13,
       author = {{Bovy}, Jo and {Rix}, Hans-Walter},
        title = "{A Direct Dynamical Measurement of the Milky Way's Disk Surface Density Profile, Disk Scale Length, and Dark Matter Profile at 4 kpc <\raisebox{-0.5ex}\textasciitilde R <\raisebox{-0.5ex}\textasciitilde 9 kpc}",
      journal = {\apj},
         year = 2013,
        month = dec,
       volume = {779},
       number = {2},
          eid = {115},
        pages = {115},
          doi = {10.1088/0004-637X/779/2/115},
archivePrefix = {arXiv},
       eprint = {1309.0809},
 primaryClass = {astro-ph.GA},
       adsurl = {https://ui.adsabs.harvard.edu/abs/2013ApJ...779..115B}
}

@ARTICLE{SellwoodTrick19,
       author = {{Sellwood}, J.~A. and {Trick}, Wilma H. and {Carlberg}, R.~G. and
         {Coronado}, Johanna and {Rix}, Hans-Walter},
        title = "{Discriminating among theories of spiral structure using Gaia DR2}",
      journal = {\mnras},
         year = "2019",
        month = "Apr",
       volume = {484},
       number = {3},
        pages = {3154-3167},
          doi = {10.1093/mnras/stz140},
archivePrefix = {arXiv},
       eprint = {1810.03325},
 primaryClass = {astro-ph.GA},
       adsurl = {https://ui.adsabs.harvard.edu/abs/2019MNRAS.484.3154S}
}

@ARTICLE{Quillen03,
       author = {{Quillen}, A.~C.},
        title = "{Chaos Caused by Resonance Overlap in the Solar Neighborhood: Spiral Structure at the Bar's Outer Lindblad Resonance}",
      journal = {\aj},
         year = 2003,
        month = feb,
       volume = {125},
       number = {2},
        pages = {785-793},
          doi = {10.1086/345725},
archivePrefix = {arXiv},
       eprint = {astro-ph/0204040},
 primaryClass = {astro-ph},
       adsurl = {https://ui.adsabs.harvard.edu/abs/2003AJ....125..785Q}
}

@ARTICLE{Fujii19,
       author = {{Fujii}, M.~S. and {B{\'e}dorf}, J. and {Baba}, J. and {Portegies Zwart}, S.},
        title = "{Modelling the Milky Way as a dry Galaxy}",
      journal = {\mnras},
         year = 2019,
        month = jan,
       volume = {482},
       number = {2},
        pages = {1983-2015},
          doi = {10.1093/mnras/sty2747},
archivePrefix = {arXiv},
       eprint = {1807.10019},
 primaryClass = {astro-ph.GA},
       adsurl = {https://ui.adsabs.harvard.edu/abs/2019MNRAS.482.1983F}
}
\bibliographystyle{aasjournal}

\end{document}